\documentclass[preprintnumbers, floatfix, letterpaper,aps,twocolumn,prd,epsfig,nofootinbib,natbib,longbibliography]{revtex4-1}
\usepackage{graphicx}
\usepackage{epstopdf}
\usepackage{latexsym}
\usepackage{amssymb}
\usepackage{amsmath}
\usepackage{color}
\usepackage{mathrsfs}
\usepackage{multirow}
\usepackage{tabularx,booktabs}
\newcolumntype{Y}{>{\centering\arraybackslash}X}

\usepackage{subfigure}
\usepackage{float}
\usepackage{epsfig}
\usepackage{epstopdf}
\usepackage{amsmath}
\usepackage{hyperref}
\hypersetup{colorlinks=True,citecolor=blue}
\usepackage{amsfonts}
\usepackage{amssymb}
\usepackage{float}
\usepackage{bm}
\usepackage{graphicx}
\usepackage{color}
\usepackage{verbatim}
\usepackage{makecell}
\setcellgapes{4pt}

\begin{document}

 \newcommand{\bq}{\begin{equation}}
 \newcommand{\eq}{\end{equation}}
 \newcommand{\bqn}{\begin{eqnarray}}
 \newcommand{\eqn}{\end{eqnarray}}
 \newcommand{\nb}{\nonumber}
 \newcommand{\lb}{\label}
\newcommand{\PRL}{Phys. Rev. Lett.}
\newcommand{\PL}{Phys. Lett.}
\newcommand{\PR}{Phys. Rev.}
\newcommand{\CQG}{Class. Quantum Grav.}

\title{Genericness of pre-inflationary  dynamics and probability of the desired slow-roll inflation  in modified loop quantum cosmologies}

\author{Bao-Fei Li $^{1,2}$}
\email{ Bao-Fei$\_$Li@baylor.edu}
\author{Parampreet Singh$^3$}
\email{psingh@lsu.edu}
\author{Anzhong Wang$^{1, 2}$\footnote{The corresponding author}}
\email{Anzhong$\_$Wang@baylor.edu}
\affiliation{$^{1}$Institute for Theoretical Physics $\&$ Cosmology,
Zhejiang University of Technology, Hangzhou, 310032, China\\
$^2$GCAP-CASPER, Department of Physics, Baylor University, Waco, TX, 76798-7316, USA\\
$^3$ Department of Physics and Astronomy, $\&$ Center for Computation and Technology, Louisiana State University, Baton Rouge, LA 70803, USA}


\begin{abstract}
We study the evolution of spatially flat Friedmann-Lema\^{i}tre-Robertson-Walker universe for chaotic and Starobinsky  potentials in the framework of modified 
loop quantum cosmologies. These models result in a non-singular bounce as in loop quantum cosmology, but with far more complex modified Friedmann 
dynamics with higher order than quadratic terms in energy density. For the kinetic energy dominated bounce, we obtain analytical solutions using different 
approximations and compare with numerical evolution for various physical variables. The relative error turns out to be less than $0.3\%$ in the bounce regime 
for both of the potentials. Generic features of dynamics, shared with loop quantum cosmology, are established using analytical and numerical solutions. Detailed 
properties of three distinct phases in dynamics separating bounce regime, transition stage and inflationary phase are studied. For the potential energy dominated bounce, we qualitatively describe its generic features and confirm by simulations that they all lead to the desired slow-roll phase in the chaotic inflation. However, in the Starobinsky potential, the potential energy dominated bounce cannot  give rise to any inflationary phase. Finally, we compute the probability for the desired slow-roll
inflation to occur in the chaotic inflation and as in loop quantum cosmology, find a very large probability for the universe to undergo inflation. 

\end{abstract}

\maketitle

\section{Introduction}
\label{Intro}
\renewcommand{\theequation}{1.\arabic{equation}}\setcounter{equation}{0}
 
The inflationary paradigm  helps to resolve several long-standing problems of the standard big-bang cosmology  by assuming that the universe experiences an exponential expansion in its very early stage \cite{ag1981}. But, inflationary spacetime are past-incomplete and a  
spacetime singularity inevitably occurs \cite{BV94}. At the  singularity, spacetime curvature becomes infinite, and hence general relativity (GR) breaks down.  Therefore, it is not clear how to impose physical initial conditions at this point, 
instead, the initial conditions  are usually imposed at the onset of inflation \cite{DB09}. This leads to a fundamental question about the phase before inflation, and the way new physics at Planck scale affects inflationary predictions. Such questions can only be addressed in a framework which addresses the problem of cosmological singularities.

In the last few decades, as a background independent quantum theory of gravity,  loop quantum gravity (LQG)  has been extensively developed and  applied  to different contexts. When the techniques of LQG are applied to the cosmology where the usual homogeneous and isotropic spacetime is assumed, one is led to 
 loop quantum cosmology (LQC) \cite{review1,review2}.  In LQC, the big-bang singularity is generically replaced by a quantum bounce \cite{aps3, aps, slqc}, at which the energy density of the universe achieves its maximum value.    
 The existence of the  bounce is a  consequence of  pure quantum geometric effects, irrelevant to the matter content,  and thus a robust feature against the quantization ambiguities in constructing the effective Hamiltonian from the LQG. Interestingly, LQC permits an effective spacetime description derived from an effective Hamiltonian \cite{VT08}, turning out to be an excellent approximation to the underlying quantum dynamics for isotropic \cite{numlsu-2} as well as anisotropic spacetimes \cite{numlsu-4},  using which singularity resolution has been shown to be a generic result in isotropic and anisotropic spacetimes \cite{generic-singularity}, and various phenomenological implications for inflation and CMB have been studied \cite{agullo-singh}. 
 
Given this success of LQC, it is pertinent to understand how various quantization ambiguities affect the physical predictions. This issue becomes more relevant when we note that so far there is no systematic derivation of LQC as a cosmological sector of LQG. In this direction, different paths to obtain LQC-like Hamiltonian from LQG have been followed (see for eg. \cite{YDM09,DL17,DL18,ABBMS18,BM19}). Our focus in this manuscript is on modified loop quantum cosmologies which differ from standard LQC in the way Lorentzian term in the Hamiltonian constraint is treated \cite{YDM09,DL17,DL18}. This exercise, first carried out in  \cite{YDM09}, led to two modifications of LQC which have been recently studied in more detail  \cite{adlp, lsw2018, IA19, GQMM19, lsw2018b, SS18, SS19}. In particular,  big-bang singularity is replaced by a quantum bounce in these two models, and slow-roll inflation is an attractor for various potentials \cite{lsw2018b}.

The first modified loop quantum cosmology (mLQC-I) is derived by treating the Lorentzian term in the Hamiltonian constraint separately and using Thiemann's regularization in the full LQG \cite{thiemann}. To the leading order, the resulting effective Hamiltonian is identical to the one recently obtained   from  the expectation values of the Hamiltonian operator in  LQG  with the help of complexifier coherent states in the homogeneous and isotropic flat Friedmann-Lema\^{i}tre-Robertson-Walker (FLRW) spacetime \cite{DL17,DL18}.  Even for a massless scalar field, the evolution of the universe in this model is asymmetric about the bounce, in contrast to the standard LQC \cite{adlp,lsw2018}. In addition, the pre-bounce branch results in an emergent Planckian size cosmological constant \cite{adlp}, similar to a quantization of the Schwarzschild interior in LQC \cite{djs},  and a rescaled Newton's constant \cite{lsw2018}. The post-bounce branch results in a classical universe at late times where GR is an excellent approximation. In the Planck regime, mLQC-I shows departures from standard LQC which are captured by a higher than $\rho^2$ modified Friedmann dynamics.  

If the proportionality of Ashtekar's connection  and extrinsic curvature, valid for spatially flat models, is used along with Thiemann's regularization, one is led to mLQC-II which is studied in detail in \cite{lsw2018b}. Unlike mLQC-I, the bounce in this model is perfectly symmetric for a massless scalar field. Though in this sense mLQC-II is similar to LQC, but like mLQC-I the modified Friedmann dynamics in the Planck regime is far more non-trivial with higher than $\rho^2$ terms which characterize LQC dynamics. 
  
For the inflationary spacetimes, investigations in LQC have focussed on two issues. First to understand the way Planck scale physics affects the pre-inflationary dynamics resulting in potential phenomenological signatures, and second to establish the naturalness of inflation. Given that modified loop quantum cosmologies differ significantly from LQC in the Planck regime, it is pertinent to ask whether the results obtained in LQC hold qualitative validity in modified loop quantum cosmologies. Our goal in this manuscript is to answer this question by investigating these two issues:  genericness of the pre-inflationary dynamics   and the probability of the desired slow-roll inflation in above two modified cosmological models. In LQC, the former was studied in   \cite{ZWCKS16,ZWCKS17,SSWW17,SSW18a,Sha18,SSWW18b,JMZ18,bcl2018,SZW19}, while the latter in \cite{{SVV06,ZL07,as2011,CK11,corichi-sloan,bedic,LB13,CZ15}}. For mLQC-I and mLQC-II, some aspects of pre-inflationary dynamics were noted by authors in \cite{lsw2018} and \cite{lsw2018b}. In this paper, we start with an overview of effective dynamics of mLQC-I and mLQC-II in Sec. II. This is followed in Sec. III by a numerical study of the dynamics of the mLQCs for two different potentials, the chaotic and Starobinsky potentials. First, we  discuss the case in which the kinetic energy (KE) of the inflaton dominates the evolution of the universe at the bounce. A remarkable feature emerges from such studies: The evolution of the universe can be generically divided into three distinctive phases -- {\em bouncing, transition, slow-roll inflation}, quite similar to the case of LQC \cite{ZWCKS16,ZWCKS17}. This division does not depend on the choice of inflationary potentials or their initial conditions, as long as 
\bq
\lb{1.1}
\frac{1}{2}\dot\phi^2(t_B) \gg V\left(\phi_B\right), 
\eq
holds, where $t_B$ is the moment that the quantum bounce occurs, and $\phi_B \equiv \phi(t_B)$. Then we  move onto the PE dominated bounce which, although devoid of three distinctive phases as in the KE dominated case,  can also lead to the desired slow-roll inflation in the chaotic inflation as confirmed by our numeric simulations. However, there is no slow-roll phase with the Starobinsky potential when the bounce is PE dominated, similar to the LQC case \cite{bonga}. In Sec. IV, we obtain analytical solutions for the KE dominated bounce by using different approximations in each of these three phases.  Our comparison of analytical solutions with numerical evolution shows that the former are reliable approximations to extract phenomenological implications, especially in the bounce regime. This result simplifies understanding Planck scale physics of mLQC-I and mLQC-II which is comparatively more non-trivial than LQC. In particular, the solutions in the bouncing phase have generic features shared with LQC. This is because, the KE  dominates the evolution of the universe during this whole phase, once it dominates at the beginning (the bounce).  Again, this is quite
similar to the LQC case \cite{ZWCKS16,ZWCKS17}, because the corrections of the modified LQCs are high-orders and restricted to the bounce regime \cite{YDM09,lsw2018,lsw2018b}. Thus, the generic features obtained in \cite{ZWCKS16,ZWCKS17} for LQC turn out to be robust for mLQC-I and mLQC-II as long as the bounce is dominated by the KE of the scalar field \footnote{It is interesting to note that this robust feature was also shown to be true generically for the universe  emerging from an initial singularity in a non-inflating state where the kinetic energy of the inflaton well dominates its potential energy \cite{HBLH14,H3L18a,H3L18b}.}.  In Sec. V, we compute the probability of the occurrence of the desired slow-roll inflation using Liouville measure \cite{as2011},  consistent with current observations \cite{Planck2015,Planck2018}. For the chaotic inflation, the probability for a desired slow-roll inflation
 to {\em not} occur is $ P^{\mathrm{I}}(\text{not realized})\lesssim 1.12\times 10^{-5}$ for mLQC-I and  $P^{ \mathrm{II}}(\text{not realized})\lesssim 2.62\times 10^{-6}$ for mLQC-II, respectively. In comparison, for LQC one gets $P (\text{not realized})\lesssim2.74\times 10^{-6}$ \cite{as2011}. Thus, the result of probability of inflation to occur holds for mLQC-I and mLQC-II. We conclude with a summary of results in Sec. VI. In an Appendix, we summarize some relevant details for slow-roll inflationary models. In this manuscript, we will work in the  units $\hbar=c=1$ and keep the Planck mass $m_{\mathrm{Pl}} $ ($\equiv1/\sqrt{G}$) explicit.

\section{Effective dynamics  of modified loop quantum cosmological models: a brief overview}
\label{Section2}
\renewcommand{\theequation}{2.\arabic{equation}}\setcounter{equation}{0}
In this section, we provide an overview of the modified Friedmann dynamics for mLQC-I and mLQC-II. Our notation follows the one in \cite{lsw2018,lsw2018b}. The modified Friedmann dynamics can be obtained from the effective Hamiltonian constraint using Hamilton's equations. The validity of effective spacetime description for LQC has been tested in various works \cite{ps-kv,aps3,numlsu-2,numlsu-4}, where an excellent agreement with the underlying quantum dynamics is seen. In this manuscript, we will assume the validity of the effective dynamics in mLQC-I and mLQC-II.

\subsection{mLQC-I}
For the spatially-flat FLRW universe 
\bq
\lb{bg}
ds^2 = - dt^2 + a^2(t)\left(dx^2 + dy^2 + dz^2\right),
\eq
the effective Hamiltonian for the first modified loop quantum cosmological model (mLQC-I) reads \cite{YDM09,DL17},  
\bqn
\lb{ham}
\mathcal {H}^{\mathrm{I}} &=&\frac{3v}{8\pi G\lambda^2}\left\{\sin^2(\lambda b)-\frac{(\gamma^2+1)\sin^2(2\lambda b)}{4\gamma^2}\right\}\nb\\
&& ~~~~~~~~~~~~ +\mathcal{H}_M, 
\eqn
where $v \equiv v_o a^3$,  $v_o$ is the volume of fiducial cell in $\mathbb{R}^3$ spatial manifold.   
The variable $b$, which is given by $b = \gamma H$  in the classical limit, satisfies the canonical  relation 
\bq
\lb{bv}
\{b,v\}=4\pi G\gamma, 
\eq
where $H$ denotes the Hubble rate $H \equiv \dot{a}/a = \dot{v}/(3v)$.  Here, 
 $\gamma$ is the Barbero-Immirzi parameter whose value is set to $\gamma \approx 0.2375$ using 
black hole thermodynamics in LQG \cite{Mei04}, while $\lambda$ is defined by  $ \lambda^2 \equiv   4\sqrt{3}\pi\gamma \ell_{\mathrm{Pl}}^2$.  
Then,  the Hamilton's equations for the basic variables $b$ and $v$ take the  form,
\bqn
\lb{eomA}
\dot v&=&\Big\{v, \mathcal H^{\mathrm{I}}\Big\}=\frac{3v\sin(2\lambda b)}{2\gamma \lambda}\Big\{(\gamma^2+1)\cos(2\lambda b)-\gamma^2\Big\}, \nb\\
\\
\lb{eomB}
\dot b&=&\Big\{b, \mathcal H^{\mathrm{I}}\Big\}=\frac{3\sin^2(\lambda b)}{2\gamma \lambda^2}\Big\{\gamma^2\sin^2(\lambda b)-\cos^2(\lambda b)\Big\}\nb\\
&& ~~~~~~~~~~~~~~ -4\pi G\gamma P,
\eqn
where $P\equiv -{\partial \mathcal{H}_M}/{\partial v}$ is the pressure.   
Once the matter Hamiltonian $\mathcal{H}_M$ is specified,  the vanishing of the 
Hamiltonian constraint,
$ \mathcal{C}^{\mathrm{I}} := 16 \pi G{\cal H}^{\mathrm{I}} \approx 0$ 
and Eqs. (\ref{eomA}) and (\ref{eomB}) uniquely determine the evolution of the universe. In this paper, we shall consider
only the case in which the matter is described by a single scalar field $\phi$ with potential $V(\phi)$, for which  $\mathcal{H}_M$  is given by
\bqn
\lb{hamA}
 \mathcal{H}_M = \frac{1}{2}v\left(\frac{p_{\phi}^2}{v^2} + 2V(\phi)\right), 
\eqn
where  $p_{\phi}$ is the momentum of $\phi$. As a result, the Hamilton's equations of the matter sector  read
\begin{equation}
\lb{2.9}
\dot \phi = \frac{p_\phi}{v}, ~~~ 
\dot p_\phi=-vV_{,\phi} ,
\end{equation}
where $V_{,\phi} \equiv \partial V(\phi)/\partial \phi$.  

To consider the dynamics of the system (\ref{eomA}) and (\ref{eomB}), it is often more convenient and transparent to write them in terms of $H, \; \rho$ and $P$, from which we can easily find 
their classical correspondence, where $\rho \equiv  \mathcal{H}_M/v$. In doing so, it was found that the generalized Friedmann-Raychaudhuri (FR) equations are divided into two branches \cite{lsw2018}, 
referred to, respectively, as the $b_{\pm}$ branches. In the $b_{-}$ branch, the FR equations take the form, 
\bqn
\lb{1.4a}
&&H^2 =\frac{8\pi G \rho}{3}\left(1-\frac{\rho}{\rho^{\scriptscriptstyle{\mathrm{I}}}_c}\right)\Bigg[1 \nb \\ &&~~~~~~~~~~~~~~~~~~~~+\frac{\gamma^2 \rho/\rho^{\scriptscriptstyle{\mathrm{I}}}_c}{\left(\gamma^2+1\right)\left(1+\sqrt{1-\rho/\rho^{\scriptscriptstyle{\mathrm{I}}}_c}\right)^2}\Bigg], \\
\lb{1.4b}
&&\frac{\ddot a}{a}= \nonumber -\frac{4\pi G}{3}\left(\rho + 3P\right) \\
&& ~~+ \frac{4\pi G \rho^2}{3\rho^{\scriptscriptstyle{\mathrm{I}}}_c}\left[\frac{\left(7\gamma^2+ 8\right) -4\rho/\rho^{\scriptscriptstyle{\mathrm{I}}}_c+\left(5\gamma^2 +8\right)\sqrt{1-\rho/\rho^{\scriptscriptstyle{\mathrm{I}}}_c}}{(\gamma^2 +1)\left(1+\sqrt{1-\rho/\rho^{\scriptscriptstyle{\mathrm{I}}}_c}\right)^2}\right]\nb\\
  &&  ~~+ 4\pi G P \left[\frac{3\gamma^2+2+2\sqrt{1-\rho/\rho^{\scriptscriptstyle{\mathrm{I}}}_c}}{(\gamma^2+1)\left(1+\sqrt{1-\rho/\rho^{\scriptscriptstyle{\mathrm{I}}}_c}\right)}\right]\frac{\rho}{\rho^{\scriptscriptstyle{\mathrm{I}}}_c},
\eqn
where 
\bq
\lb{rhocI}
\rho^{\scriptscriptstyle{\mathrm{I}}}_c \equiv \frac{\rho_c}{4(1+\gamma^2)}, \quad \rho_c \equiv \frac{3}{8\pi \lambda^2\gamma^2 G}.
\eq
From the above equations, it can be shown that  the energy conservation law 
\bq
\lb{ecl}
\dot \rho+3H(\rho +P)=0,
\eq
still holds, while in terms of $\rho$ and $P$, we always have 
\bq
\lb{1.3}
\dot b= -4\pi G  \gamma (\rho+P) = -4\pi G  \gamma \dot\phi^2.
\eq
Note that in writing the last step, we had used the fact,
\bqn
\lb{rhoP}
\rho &\equiv& \frac{1}{v}{\mathcal H_M}\left(b, v, \phi, p_{\phi}\right) = \frac{1}{2}\dot\phi^2 + V(\phi),\nb\\
P&\equiv& -\frac{\partial}{\partial v} {\mathcal H_M}\left(b, v, \phi, p_{\phi}\right) = \frac{1}{2}\dot\phi^2 - V(\phi),
\eqn
for $\mathcal{H}_M$ given by Eq.(\ref{hamA}).

In the $b_{+}$ branch, the FR equations take the form \cite{lsw2018}, 
\bqn
\lb{1.6a}
&&H^2=\frac{8\pi G\alpha  \rho_\Lambda}{3}\left(1-\frac{\rho}{\rho^{\scriptscriptstyle{\mathrm{I}}}_c}\right)\Bigg[1\nb\\&&
~~~~~~~~~~~~~~~~~~~~~~~~+\frac{\rho\left(1-2\gamma^2+\sqrt{1-\rho/\rho^{\scriptscriptstyle{\mathrm{I}}}_c}\right)}{4\gamma^2\rho^{\scriptscriptstyle{\mathrm{I}}}_c\left(1+\sqrt{1-\rho/\rho^{\scriptscriptstyle{\mathrm{I}}}_c}\right)}\Bigg],\\
\lb{1.6b}
&&\frac{\ddot a}{a}=- \frac{4\pi \alpha G}{3}\left(\rho + 3P - 2\rho_\Lambda \right) \nb\\&&~~ +4\pi G\alpha P\left(\frac{2-3\gamma^2 +2\sqrt{1-\rho/\rho^{\scriptscriptstyle{\mathrm{I}}}_c}}{(1-5\gamma^2)\left(1+\sqrt{1-\rho/\rho^{\scriptscriptstyle{\mathrm{I}}}_c}\right)}\right)\frac{\rho}{\rho^{\scriptscriptstyle{\mathrm{I}}}_c}\nb\\
&&~~- \frac{4\pi G\alpha \rho^2 \left(7\gamma^2-8+4\rho/\rho^{\scriptscriptstyle{\mathrm{I}}}_c+\left(5\gamma^2-8\right)\sqrt{1-\rho/\rho^{\scriptscriptstyle{\mathrm{I}}}_c}\right)}{3\rho^{\scriptscriptstyle{\mathrm{I}}}_c(1-5\gamma^2)\left(1+\sqrt{1-\rho/\rho^{\scriptscriptstyle{\mathrm{I}}}_c}\right)^2},\nb\\
\eqn
where 
\bq
\lb{arhol}
\alpha\equiv \frac{1-5\gamma^2}{\gamma^2+1}, \quad
 \rho_\Lambda \equiv \frac{3}{8\pi G\alpha \lambda^2(1+\gamma^2)^2}.
 \eq
From Eqs.(\ref{1.4a})-(\ref{1.4b}) we can see that a quantum bounce always happens at $\rho = \rho^{\scriptscriptstyle{\mathrm{I}}}_c \simeq 0.24 \rho_c$,  that is, the big bang singularity is replaced by a quantum bounce, similar to LQC \cite{review2}.\footnote{In this paper,  the maximum or critical density in LQC is always  denoted by $\rho_c$, and the critical density in mLQC-I  (mLQC-II)  is denoted 
by $\rho_c^{{\scriptscriptstyle{\mathrm{I}}}}$ ($\rho_c^{{\scriptscriptstyle{\mathrm{II}}}}$).} But, in contrast to LQC now the bounce is asymmetric \cite{DL17,lsw2018}, and to be consistent with observations the $b_-$ branch must correspond to the post-bounce, while the $b_+$ branch  to the pre-bounce \cite{lsw2018}. Using above equations it is straightforward to show that the conservation law (\ref{ecl}) and Eq.(\ref{1.3}) holds also in $b_+$ branch. Moreover, similar to LQC, the quantum bounce in mLQC-I is also accompanied by a phase of SI, during which we have $\dot H > 0$. Using, 
\bqn
\lb{hdot}
\dot H&=&\frac{4 G \pi (P+\rho)}{(1+\gamma^2)}\left(2\gamma^2+2\frac{\rho}{\rho_c^{{\scriptscriptstyle{\mathrm{I}}}}}-3\gamma^2\sqrt{1-\frac{\rho}{\rho_c^{{\scriptscriptstyle{\mathrm{I}}}}}}-1\right), \nb\\
\eqn
we find that in the postbounce phase,  the SI phase lasts till  $\rho = \rho_s^{{\scriptscriptstyle{\mathrm{I}}}}$, where 
\bqn
\rho_s^{{\scriptscriptstyle{\mathrm{I}}}}=\frac{\rho_c^{{\scriptscriptstyle{\mathrm{I}}}}}{8}\left(4-8\gamma^2-9\gamma^4+3\gamma^2\sqrt{8+16\gamma^2+9\gamma^4}\right). \nb\\
\eqn 
For $\gamma \simeq 0.2375$, we find $\rho_s^{{\scriptscriptstyle{\mathrm{I}}}}\simeq 0.503\rho_c^{{\scriptscriptstyle{\mathrm{I}}}}$. 
 
\subsection{mLQC-II}
  
The effective Hamiltonian for the second modified loop quantum cosmological model  (mLQC-II) reads \cite{ YDM09}, 
\bqn
\lb{2.1}
\mathcal H^{\mathrm{II}}&=&-\frac{3v}{2\pi G\lambda^2\gamma^2}\sin^2\left(\frac{\lambda b}{2}\right)\left\{1+\gamma^2\sin^2\left(\frac{\lambda b}{2}\right)\right\}\nb\\
&& +\mathcal{H}_M,
\eqn
which yields the Hamilton equations 
\bqn
\lb{LQG2a}
\dot v&=&\Big\{v, \mathcal H^{\mathrm{II}}\Big\}=\frac{3v\sin(\lambda b)}{\gamma \lambda}\Big\{1+\gamma^2-\gamma^2\cos\left(\lambda b\right)\Big\},\nb\\
\\
\lb{LQG2b}
\dot b&=&\Big\{b, \mathcal H^{\mathrm{II}}\Big\}=-\frac{6\sin^2\left(\frac{\lambda b}{2}\right)}{\gamma \lambda^2}\Bigg\{1+\gamma^2\sin^2\left(\frac{\lambda b}{2}\right)\Bigg\}\nb\\
&&~~~~~~~~~~~~~~ ~  -4\pi G\gamma P = -4\pi G\gamma (\rho+P) .
\eqn
Then, the corresponding FR equations take the form \cite{lsw2018b},  
\bqn
\lb{2.6aa}
&&H^2
=\frac{16\pi G \rho}{3}\left(1-\frac{\rho}{\rho^{\scriptscriptstyle{\mathrm{II}}}_c}\right)\nb\\
&&~~~~\times \left(\frac{1+4\gamma^2(\gamma^2+1)\rho/\rho^{\scriptscriptstyle{\mathrm{II}}}_c}{1+2\gamma^2\rho/\rho^{\scriptscriptstyle{\mathrm{II}}}_c+\sqrt{1+4\gamma^2(1+\gamma^2)\rho/\rho^{\scriptscriptstyle{\mathrm{II}}}_c}}\right), \\
\lb{2.6bb}
&&\frac{\ddot a}{a}
=-\frac{4\pi G}{3}\left(\rho+3P\right)\nb\\&&-\frac{16\left(1+\gamma^2\right)\pi G P\rho}{\rho_c^{{\scriptscriptstyle{\mathrm{II}}}}}\left[\frac{3\gamma^2+3}{1+\sqrt{1+4\gamma^2(1+\gamma^2) \rho/\rho_c^{{\scriptscriptstyle{\mathrm{II}}}}}}-2\right] \nb\\ 
&&
-\frac{16(1+\gamma^2)\pi G \rho^2}{3\rho_c^{\scriptscriptstyle{\mathrm{II}}}}\left[\frac{7\gamma^2-16\gamma^2(1+\gamma^2)\rho/\rho_c^{{\scriptscriptstyle{\mathrm{II}}}}-1}{(1+\sqrt{1+4\gamma^2(1+\gamma^2)\rho/\rho_c^{{\scriptscriptstyle{\mathrm{II}}}}})^2}\right. \nb\\
&&\left. +\frac{(5\gamma^2-3)\sqrt{1+4\gamma^2(1+\gamma^2)\rho/\rho_c^{{\scriptscriptstyle{\mathrm{II}}}}}}{(1+\sqrt{1+4\gamma^2(1+\gamma^2)\rho/\rho_c^{{\scriptscriptstyle{\mathrm{II}}}}})^2}\right] .
\eqn
From these two equations, it is straightforward to show that the energy conservation law (\ref{ecl}) still holds.
In addition, from these equations it can be also shown that a quantum bounce  happens at 
\bq
\lb{2.8}
\rho_c^{{\scriptscriptstyle{\mathrm{II}}}} \equiv 4(\gamma^2+1)\rho_c.
\eq
Thus, the big bang singularity is also resolved in this model, similar to the case of LQC and mLQC-I. 
As in the case of these models, the  SI  phase can be identified  when $\dot H > 0$. Using, 
\bqn
\lb{2.8}
\dot H&&=\frac{4\pi G (\rho+P)}{\gamma^2}\Big[ 2\gamma^2+8\gamma^2\left(\gamma^2+1\right) \rho/\rho_c^{{\scriptscriptstyle{\mathrm{II}}}} +3\nb\\ &&
-3\left(\gamma^2+1\right)\sqrt{1+4\gamma^2\left(\gamma^2+1\right)\rho/\rho_c^{{\scriptscriptstyle{\mathrm{II}}}}}\Big],
\eqn
we find that this phase lasts till  $\rho = \rho_s^{{\scriptscriptstyle{\mathrm{II}}}}$, where 
\bqn
\rho_s^{{\scriptscriptstyle{\mathrm{II}}}}\equiv \frac{3(\gamma^2+1)\sqrt{1+2\gamma^2+9\gamma^4}+9\gamma^4+10\gamma^2-3}{32\gamma^2\left(\gamma^2+1\right)}\rho_c^{{\scriptscriptstyle{\mathrm{II}}}}.  \nb\\
\eqn 
For $\gamma \simeq 0.2375$, we find  $\rho_s^{{\scriptscriptstyle{\mathrm{II}}}} \simeq 0.513\rho_c^{{\scriptscriptstyle{\mathrm{II}}}}$. This phase plays an important role in the bouncing regime of mLQC-II.  
 
\section{Numerical analysis of modified loop quantum cosmological models}
\label{Section3}

In this section, we will carry out the numerical analysis of the pre-inflationary dynamics of the FLRW  flat universe, by paying  particular attention to their generic properties.
The background pre-inflationary dynamics in LQC has already  been  studied extensively  (see for eg. \cite{PS06,ZWCKS17,ZWCKS16,SSWW17,SSW18a,Sha18,SSWW18b,bcl2018}). Here we shall extend such studies to mLQCs, of which we shall study 
their general properties for inflationary paradigm. 
We will discuss  two representative inflationary  potentials, the chaotic and Starobinsky potentials, for a given 
set of initial conditions ($a_B, \phi_B, \dot\phi_B$) at the bounce, although our main conclusions obtained from these two models are expected to hold quite generally. Note that because of the rescaling symmetry, $a(t) \rightarrow a(t)/a_0$, 
without loss of generality, we can always set $a_B = 1$. Moreover, at the bounce we have
\bq
\lb{3.1}
\frac{1}{2}\dot\phi_B^2 + V(\phi_B) = \rho_c^{A},
\eq
where $A = \mathrm{I, II}$. Hence, once $V(\phi)$ is given, we have 
\bq
\lb{3.2}
 \dot\phi_B = \pm \sqrt{2(\rho_c^A  - V(\phi_B))},
\eq
which indicates that for any given $\phi_B$, $\dot\phi_B$ can be determined up to a sign. As a result, the parameter space at the bounce consists of ($\phi_B, \mathrm{sign}(\dot\phi_B)$).

We first recall some physical quantities relevant for our study. 

\begin{enumerate} 

\item  The equation of state of the scalar field is defined by
\bq
\lb{4.2}
w_{\phi} \equiv \frac{P(t)}{\rho(t)}=\frac{\frac{1}{2}\dot \phi^2-V}{\frac{1}{2}\dot \phi^2+V}.
\eq
It can take any value in the range of $ [-1,1]$ for a single scalar field with positive potential \cite{rs2012}. In particular, $w_{\phi} =1$ for the extreme KE dominated state and  $w_{\phi} =-1$ for the extreme PE dominated state.

\item The first-order Hubble rate and potential slow-roll parameters are defined by
\bqn
\lb{4.3}
\epsilon_H&=&-\frac{\dot H}{H^2}, \quad  \eta_H=-\frac{\ddot H}{2H \dot H},\\
\epsilon_V&=&\frac{1}{16\pi G}\left(\frac{V_{,\phi}}{V}\right)^2, \quad \eta_V=\frac{V_{,\phi\phi}}{8\pi G V}.
\eqn
These two sets of slow-roll parameters are commonly in use for different purposes \cite{DB09}. The potential slow-roll parameters labelled by a subscript `V' can be used to determine which part of the potential is able to successfully drive the inflation. The Hubble slow-roll parameters labelled with an index `H' are widely used in numerical simulations to define precisely when the slow-roll inflation begins and ends. Since, in the classical regime, the acceleration of the scale factor satisfies the relation, 
\bq
\lb{4.4}
\ddot a = aH^2 \left(1-\epsilon_H\right),
\eq
the universe experiences an accelerated expansion when $\epsilon_H<1$, while the slow-roll inflation takes place only when $\epsilon_H \ll 1$ and $\eta_H \ll 1$. In this paper,  for concreteness, we define  the onset of inflation as the time $t_i$ when  $|\eta_H|=0.03$ for the first time in the transition phase. The end of the slow-roll inflation
is defined at the time  $t_{\text{end}}$ when  $\epsilon_H$ grows to unity for the first time after $t_i$\footnote{There are no fixed rules to precisely define  the beginning and end moments of the slow-roll inflation and several definitions exist in the literature \cite{DB09,SSWW17,am2015}. However, all of them give similar results.}. 

\item The SI  phase starts at the bounce and ends when the Hubble rate attains its maximum value at $\dot H=0$. The number of e-foldings in this period are given by 
\bq
\lb{4.5}
N_S= \ln\left(\frac{a_S}{a_B}\right),
\eq  
where $a_S$ represents the scale factor at the end of the SI  phase $t_S$,  and $a_B$ is the scale factor at the bounce which is always set to unity in our numerical simulations. 

\end {enumerate}

With the relevant physical quantities introduced, we now turn to study of pre-inflationary dynamics in chaotic and Starobinsky potentials in mLQC-I and mLQC-II.

\subsection{Chaotic Inflation}

For the chaotic inflation,  the potential is given by, 
 \bq
\lb{4.8a}
V(\phi)=\frac{1}{2}m^2\phi^2,
\eq
where the mass of the scalar field is set to $m= 1.26\times 10^{-6}~m_{\mathrm{Pl}} $ [Cf. Appendix A]. In this case, since the modified Friedmann and Klein-Gordon equations are invariant under the symmetry transformation $(\phi, \dot \phi)\rightarrow (-\phi, -\dot \phi)$, it is sufficient to consider only half of the parameter space where $\dot \phi_B>0$. In this case, $\phi_B$ can take any value in the range $|\phi_B|\le\phi^A_\text{max}=\sqrt{2\rho^A_c}/m$.


The physical picture of the pre-inflationary background evolution in mLQC-I and mLQC-II  can be understood by considering climbing up and rolling down of the inflaton along the potential (\ref{4.8a}). The motion of the inflaton is, of course,  governed by the Klein-Gordon equation
\bq
\lb{KG}
\ddot \phi=-3 H \dot \phi -m^2 \phi.
\eq 

The  property of the background dynamics in mLQC-I/II depends on whether the bounce is dominated by the KE or the PE of the scalar field. In the following, we will first quantitatively analyze the KE dominated bounce in great detail, then proceed with a qualitative analysis of the PE dominated bounce.

\begin{figure}[h!] 
{
\includegraphics[width=7cm]{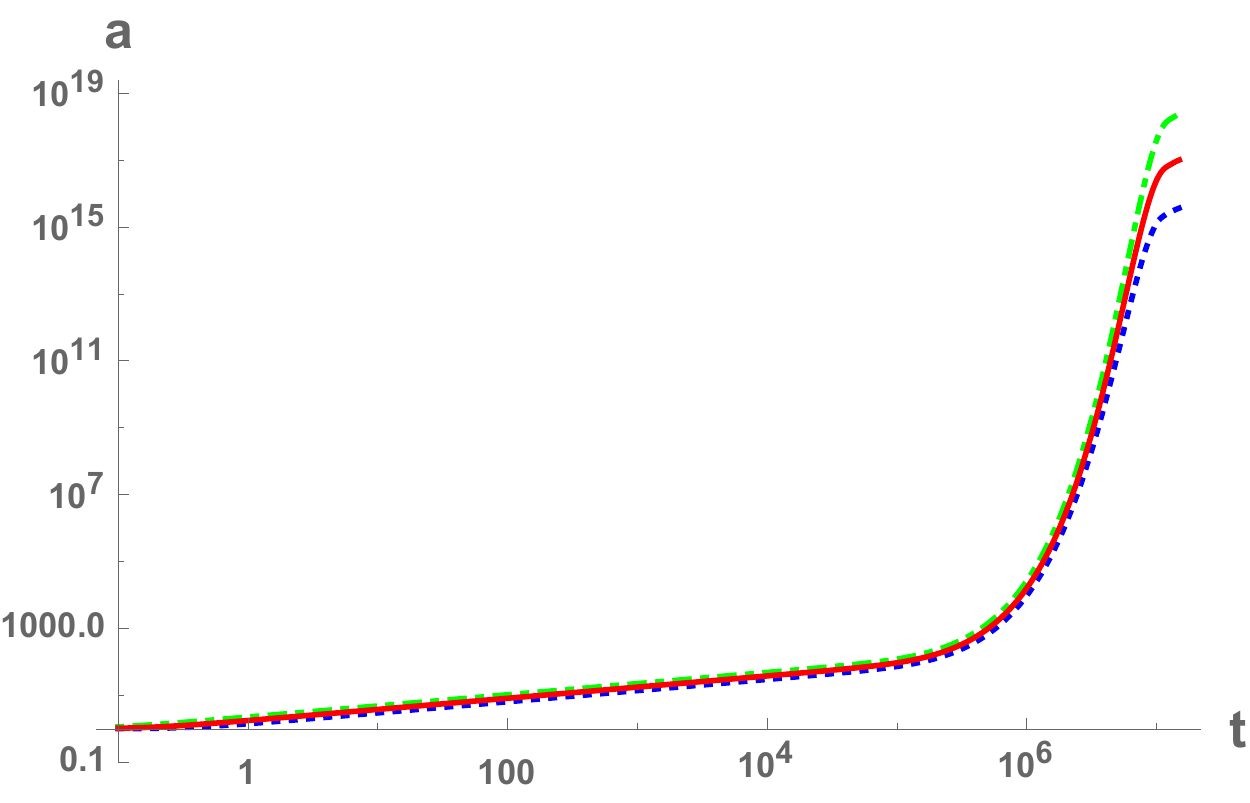}
\includegraphics[width=7cm]{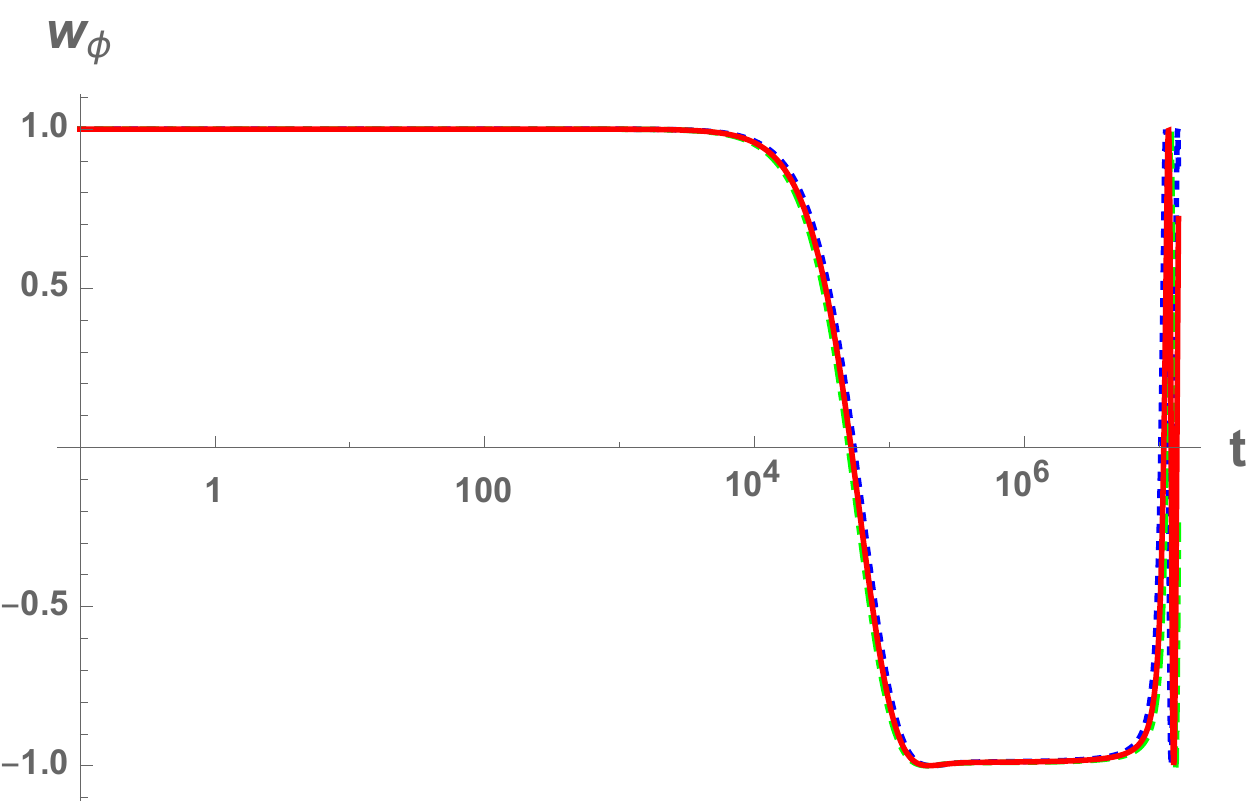}
\includegraphics[width=7cm]{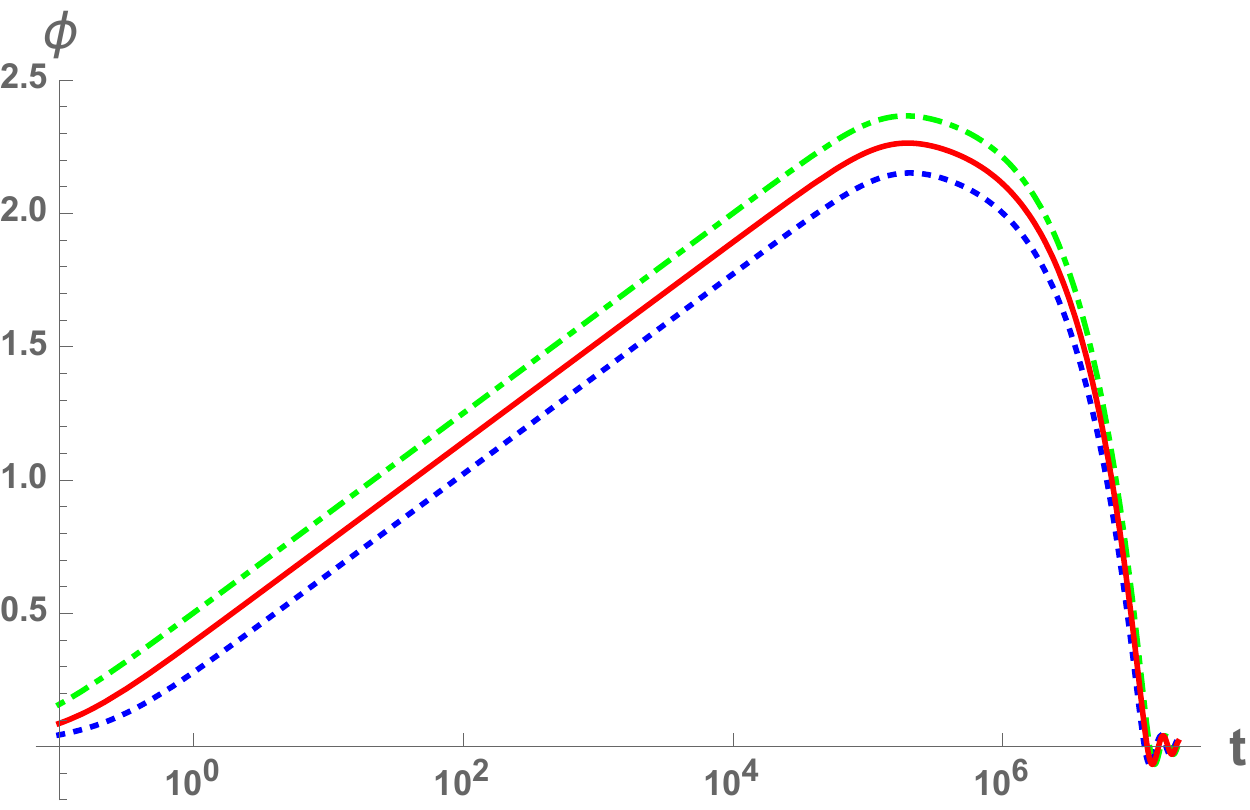}
\includegraphics[width=7cm]{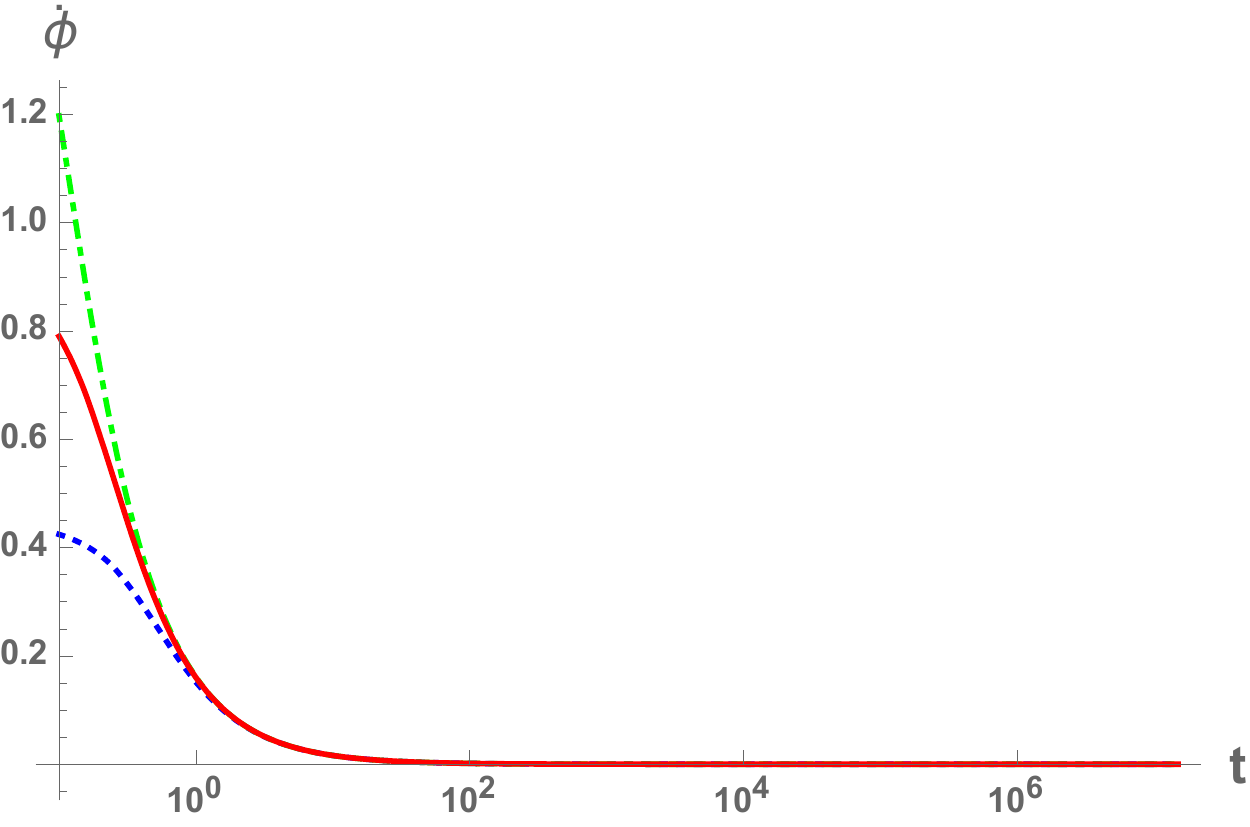}
}
\caption{The evolution of several quantities in the post-bounce phase are depicted and compared: LQC (red solid curves), mLQC-I (blue dotted curves) and mLQC-II (green dot-dashed curves) with the chaotic potential. The initial condition for the simulation is chosen at the bounce with  $\phi_B=0, \dot \phi_B>0$ so that the bounce is completely dominated by the KE  of the inflaton. The mass of the scalar field is set to $m= 1.26\times 10^{-6}~m_{\mathrm{Pl}}$.}
\label{fig3}
\end{figure}

\begin{figure}[h!] 
{
\includegraphics[width=7cm]{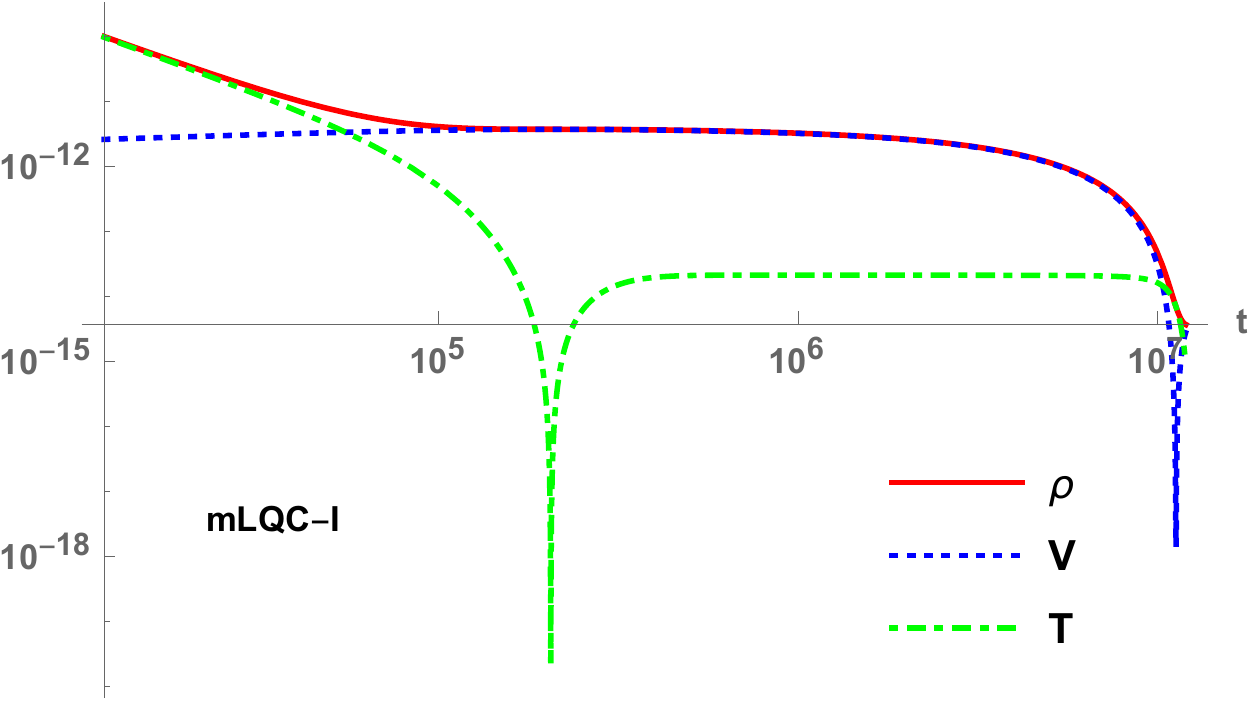}
\includegraphics[width=7cm]{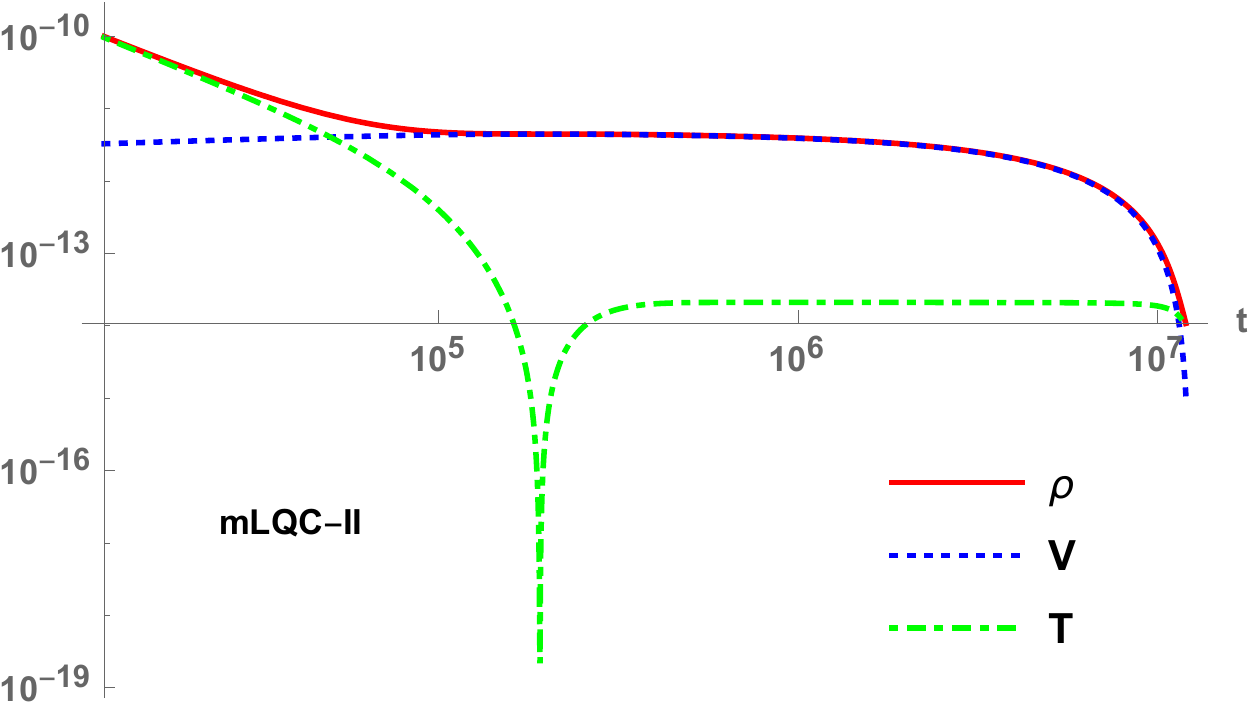}
\includegraphics[width=7cm]{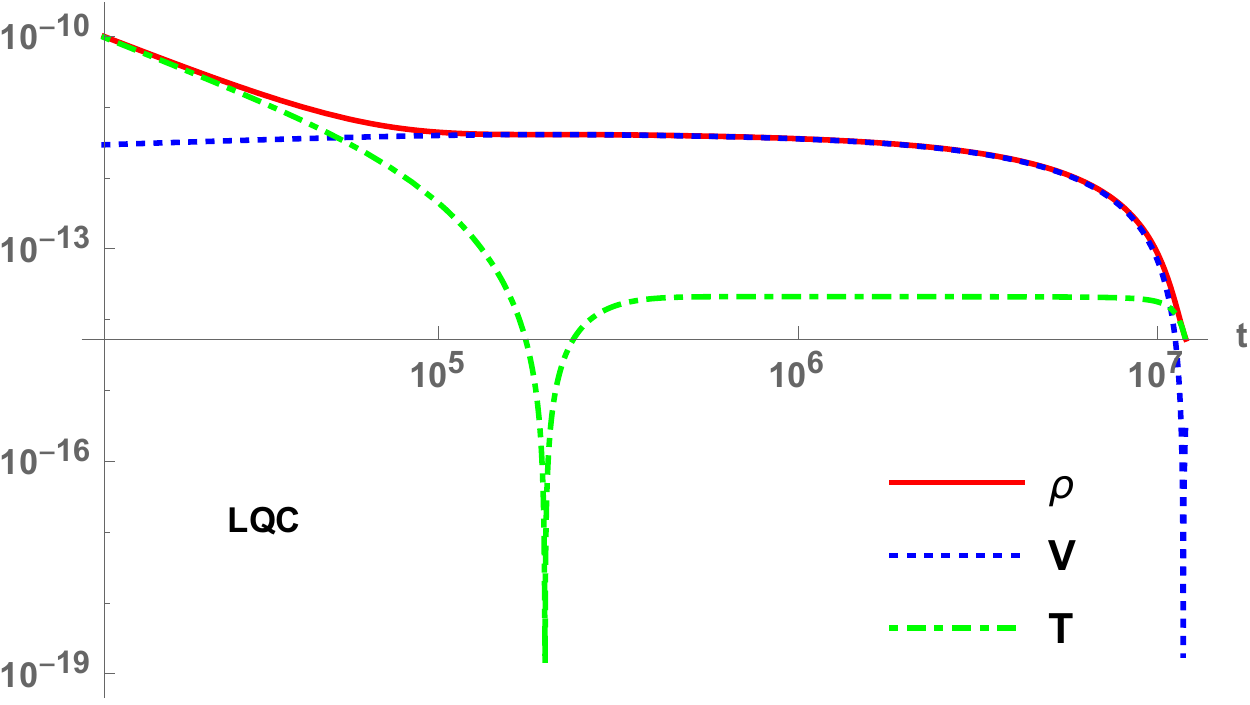}
}
\caption{With the same initial  conditions as Fig. \ref{fig3}, the change in energy density $\rho$, the KE and PE of the scalar field is shown  in each of the three models, 
LQC, mLQC-I and mLQC-II.}
\label{fig4}
\end{figure}

\subsubsection{Kinetic-Energy Dominated Case}

The generic feature of the background evolution in mLQC-I/II for the kinetic dominated bounce can be understood  from the extreme case when $\phi_B=0$ at the bounce point. In this extreme case, 
 the inflaton is initially climbing up the potential with a negative acceleration $\ddot \phi<0$. Due to the frictional term ($-3 H \dot \phi$),   
 the inflaton loses its KE rapidly,  while its PE stays almost as a constant. This behavior of the inflaton is explicitly captured by Fig. \ref{fig4} and the last two subfigures of Fig. \ref{fig3}.  
 As the total energy of the inflaton keeps decreasing, there exists a moment  $t_\text{eq}$ at which the KE  becomes equal to the PE. In Tables  \ref{t1} and \ref{t2}, $t_\text{eq}$ (KE = PE) is explicitly listed out for a number of initial conditions when the bounce is kinetic dominated.  This equilibrium point can also be seen in Fig. \ref{fig3} as the point where $\omega_\phi=0$. Shortly after $t_\text{eq}$, the speed of the  inflaton decreases to zero and the inflaton starts to roll down the potential from the turnaround point $\dot \phi=0$ (this point appears in Fig. \ref{fig3} as the peak of $\phi$).  As the inflaton rolls down the potential from the turnaround point, the driving force $(-3H\dot \phi)$ grows gradually. Finally, when the magnitude of the driving force becomes almost equal to that of the conservative force $(-m^2\phi)$, the acceleration of the inflaton becomes negligible as compared to its velocity, and, hence the slow-roll inflation takes place. 
 
In addition to the initial conditions shown in Tables \ref{t1} and \ref{t2}, we have also studied various other initial conditions, and found that, for KE well-dominated cases at the bounce,  the evolution of mLQC-I/II universe before reheating  can be generically divided into three distinctive phases, {\em bouncing, transition, slow-roll inflation}, similar to that in the LQC case \cite{ZWCKS17,ZWCKS16,SSWW17,SSW18a,Sha18,SSWW18b,bcl2018}, although some differences may appear when considering  them in details.\\

i) {\em The bouncing phase}: It  starts  from the bounce and lasts till the moment when loop quantum gravitational corrections can be safely ignored and dynamics is well approximated by classical Friedmann dynamics. For the kinetic dominated bounce, this phase ends at $t_\text{eq}$  when the KE  of the inflaton equals its PE. 
Its duration quantitatively depends on the ratio $KE/PE$. In this case, a larger $KE/PE$  implies a longer period of the bouncing phase. 
For example, in Table \ref{t1}, when $\phi_B/m_{\mathrm{Pl}} $ changes from $0$ to $0.917$ and $2$, its duration decreases from $6.24\times 10^4~m^{-1}_{\mathrm{Pl}}$, to $4.41\times 10^4~m^{-1}_\mathrm{Pl}$ and $3.26\times 10^4~m^{-1}_\mathrm{Pl}$, respectively. The analytic solutions in this phase is given by Eqs. (\ref{2.8m})
and (\ref{fittingA}) for  mLQC-I. For mLQC-II, they are given by  Eqs. (\ref{2.8mA}) and (\ref{parameter2}). Estimates from analytical solutions which are discussed in the next section are also provided in Table \ref{t1}. 

 One distinguishable stage in the bouncing phase is so-call SI phase when the Hubble parameter increases and the universe experiences a stage of super-inflation. For the KE dominated bounce, as the energy density follows the inverse power law of Eq.(\ref{2.6c}), the e-folds of SI can be simply estimated via
\bq
N_S=\frac{1}{6}\ln\left(\frac{\rho^A_c}{\rho^A_S}\right),
\eq
where $\rho^A_c$ and $\rho^A_S$ stand for the critical energy density and the energy density at which the SI  phase ends in each model. Specifically, in mLQC-I, the SI stops at $\dot H=0$ which implies $\rho_s=0.503 \rho^{\scriptscriptstyle{\mathrm{I}}}_c$ after the bounce \cite{lsw2018},  giving $N_S=0.114$,  while in mLQC-II  the SI  phase stops at $\rho_S= 0.513 \rho^{{\scriptscriptstyle{\mathrm{II}}}}_c$, from which we find $N_S=0.111$. These numbers of SI e-folds are consistent with the numeric results presented in Tables  \ref{t1} and \ref{t2}. Qualitatively, the bouncing phase, including the SI regime, is expected to produce an observable effect on the intermediate regime of the power spectra as has been discovered in LQC \cite{bgslb2015}. We will come back to this issue in our future work.

ii) {\em The transition phase}:  This phase starts from  $t_\text{eq}$ where equation of state vanishes, and lasts till the onset of slow-roll inflation. 
During this phase, the KE decreases dramatically, so that the universe soon enters into an accelerating phase, and the effective equation of state resembles a step function, as shown explicitly in Fig. \ref{fig3}.

iii) {\em The slow-roll inflationary phase}: It starts at the moment  $t_i$ when $|\eta_H| \simeq 0.03$ ($w_{\phi} \simeq -1$) for the first time after $t_\text{eq}$ and ends at the time $t_\text{end}$ when $\epsilon_H$ equals unity for the first time after $t_i$. 
It should be noted that a different value of $\eta_H$ for the slow-roll to take place does change the inflationary e-folds $N_\mathrm{SR}$. In order to study the sensitivity of $N_\mathrm{SR}$ on $\eta_H$, numeric simulations are performed with the initial conditions $\phi_B/m_\mathrm{Pl}=0, 5, 8, 10$ in mLQC-I/II and $\eta_H$ is chosen among 0.1, 0.01, 0.001 and $10^{-5}$. From the simulations we have found: $\eta_H=0.1$ gives the similar inflationary e-folds as $\eta_H=0.01$ with $\Delta N_\mathrm{SR} \le 1$, while $\eta_H=0.001$ and $\eta_H=10^{-5}$ give the similar result ($\Delta N_\mathrm{SR} \le 1$). The relative error between the inflationary e-folds when $\eta_H=0.01$ and $\eta_H=0.001$ attains its maximum at $\phi_B=0$ which is about $5\%$ ($\Delta N_\mathrm{SR} \approx 1.5$) and its minimum at $\phi_B/m_\mathrm{Pl}=10$ which is about $2\%$ ($\Delta N_\mathrm{SR} \approx 21$). For  $\phi_B/m_\mathrm{Pl}=5(8)$, the relative error is around $4\%(3\%)$. Therefore, it turns out that the inflationary e-folds does not sensitively depend on $\eta_H$.

From Tables  \ref{t1} and \ref{t2},  we find that there are three situations for which the slow-roll inflation can take place after the inflaton is `released' from the bounce point: (a) With the initial condition $(\phi_B>0, \dot \phi_B>0$) in the first three blocks of Tables \ref{t1} and \ref{t2}, the inflaton starts to climb up the potential, reaches the highest point (turnaround point) and then the slow-roll inflation occurs when it rolls down the potential. In this case, a larger $\phi_B$ always indicates a larger amount of inflationary e-folds. Intuitively, this is because a larger initial $\phi_B$ is able to help the inflaton  reach a higher turnaround point. Thus, the inflaton takes more time to roll down the potential to the point $\phi_\mathrm{end}$. 
In the simulations, $\phi_\mathrm{end} =0.201 m_{\mathrm{Pl}} $ as we have used $\epsilon_H=1$ to signify the end of the slow-roll. However, from the analytic results and  Table  \ref{appb} in Appendix A, $\phi_\mathrm{end}= 0.282 m_{\mathrm{Pl}}$ as we have used $\epsilon_V=1$ in this Table. (b) The slow-roll inflation can also take place when the initial condition is $\phi_B=-0.200 m_{\mathrm{Pl}}$. In this case, the inflaton first rolls down the left wing of the potential, losing its energy constantly, then climbs up the right wing of the potential, reaches the turnaround point and then undergoes the slow-roll inflation. In order for inflation to occur, the turnaround point must be larger than $\phi_\mathrm{end}$ ($1.96~ m_{\mathrm{Pl}} $ in mLQC-I and $2.18~m_{\mathrm{Pl}} $ in mLQC-II, both of which  are smaller than those resulting from $\phi_B =0$ and thus lead to a smaller value of the inflationary e-folds). Otherwise, the slow-roll inflation cannot take place on the right wing of the potential. (c) The slow-roll inflation takes place on the left wing of the potential when $\phi_B=-5.158 m_{\mathrm{Pl}} $ in mLQC-I and $\phi_B =-5.386 m_{\mathrm{Pl}} $ in mLQC-II. In this case, the inflaton directly rolls down the left wing of the potential and when its energy density decreases to the order of $10^{-12}\rho_\mathrm{Pl}$, the slow-roll occurs. There is no turnaround point in this process. To better understand it, we can take a look at the last block in Table  \ref{t1} as an example. The key  point is the behavior of two forces in Eq. (\ref{KG}). As $\phi<0$ and $\dot \phi>0$, $(-3H\dot \phi)$ corresponds to frictional force while ($-m^2 \phi$) is a positive conservative force. Until the end of the SI  phase, the frictional force is of the Planckian scale which is much larger than the conservative force ($\approx 10^{-12} m^3_{\mathrm{Pl}}$). Afterwards, as both of the KE  and the Hubble rate drop rapidly, at the event $\mathrm{KE}=\mathrm{PE}$, the frictional force is about $-10^{-11} m^3_{\mathrm{Pl}}$. At the onset of the slow-roll, its magnitude becomes almost equal to the conservative force \footnote{At the onset of the slow-roll, $-3H\dot \phi=-4.79\times10^{-12}~ m^3_{\mathrm{Pl}}, -m^2\phi=4.75\times10^{-12}~m^3_{\mathrm{Pl}}$.}, leading to a negligible acceleration and thus the slow-roll inflation.  

Finally, we can compare the two models directly from the case in the first blocks of Tables \ref{t1} and \ref{t2}. Here the initial conditions are set such that for both mLQC-I and mLQC-II, $\phi_B = 0$. One finds that though mLQC-I yields slightly higher e-folds at the end of SI , the total e-folds at the end of the slow-roll are much larger in mLQC-II. This is because the larger  critical  energy density in mLQC-II makes the initial velocity of inflaton greater than that in mLQC-I. Thus, the turnaround point in mLQC-II is further away from the origin, giving a larger number of total e-folds.

\begin{widetext}

\begin{table}\scriptsize
\caption{In this table, we compare the analytic and numeric results in mLQC-I with the chaotic potential. The analytic and numeric values of several observables are listed in a sequence of events, including Bounce, End of SI (superinflation), Equilibrium point when KE equals PE, the turnaround point when $\dot \phi =0$, the onset of the slow-roll inflation and EOI (end of the inflation). The subscript `A' denotes analytic results and subscript `N' denotes numeric ones. All the e-folds are counted starting from the bounce to the particular event labelled in the first column of the Table.  The Planck mass $m_{\mathrm{Pl}} $ is set to unity for conciseness. }

\begin{center}
 \begin{tabular}{|c||c|c||c|c||c|c||c|c||c|c|} 
 \hline
 Event  &\bf{$t_A$} &\bf{$t_N $}  & \bf{$\phi_A$} & \bf{$\phi_N$} &  \bf{$\dot \phi_A$}&\bf{$\dot \phi_N$}&\bf{$H_A$} &\bf{$H_N$}&$N_A$& $N_N$\\ 
 \hline
 
 Bounce & 0& 0 &0& 0&0.440&0.440&0&$1.84\times10^{-12}$&0&0\\ 
 End of SI &0.364& 0.363 & 0.141& 0.141&0.311 &0.312&0.452&0.453&0.115&0.114\\
KE=PE&$6.24\times10^4$&$5.51\times10^4$ & 2.07 &2.04 &$2.61\times10^{-6}$&$2.57\times10^{-6}$&$5.35\times10^{-6}$&$7.44\times10^{-6}$&4.01&4.01\\ 
$\dot \phi=0$&$2.23\times10^5$  &$2.05\times10^5$ &2.20&2.15&$-1.06\times10^{-22}$&$1.14\times10^{-23}$&$5.68\times10^{-6}$&$5.55\times10^{-6}$&4.91&4.88\\
Slow-Roll &$4.82\times10^5$   & $4.82\times10^5$ & 2.16 &2.11&$-2.11\times10^{-7}$&$-2.03\times10^{-7}$ &$5.57\times10^{-6}$ &$5.45\times10^{-6}$&6.37&6.40\\
EOI&$9.63\times10^6$ & $9.63\times10^6$  & 0.282&0.201&$-2.05\times10^{-7}$&$-1.79\times10^{-7}$ &$7.27\times10^{-7}$ &$6.36\times10^{-7}$&35.1 &34.7\\
 \hline
 
  Bounce &0&0& 0.917&0.917& 0.440 & 0.440& 0 &$1.84\times10^{-12}$& 0&0\\ 
 End of SI  &0.364 &0.363& 1.06 &1.06&  0.311&0.312& 0.452&0.453&0.115&0.114 \\
KE=PE& $4.41\times10^4$&$3.89\times10^4$& 2.93&2.90 & $3.70\times10^{-6}$ &$3.66\times10^{-6}$&$7.56\times10^{-6}$ &$1.06\times 10^{-5}$ &3.90&3.89 \\ 
$\dot \phi=0$& $1.72\times10^5$&$1.59\times10^5$&3.07 &3.02&$7.94\times10^{-23}$ &$-1.23\times10^{-23}$&  $7.91\times10^{-6}$&$7.78\times10^{-6}$&4.90&4.88\\
Slow-Roll &$3.63\times10^5$ &$3.53\times10^5$&3.04&2.99&$-2.10\times10^{-7}$ &$-2.03\times10^{-7}$&$7.83\times10^{-6}$ &$7.72\times10^{-6}$ &6.41&6.38\\
EOI&$1.38\times10^7$ & $1.41\times10^7$  & 0.282&0.201&$-2.05\times10^{-7}$&$-1.79\times10^{-7}$ &$7.27\times10^{-7}$  &$6.36\times10^{-7}$&63.9&63.0\\
 \hline
 
  Bounce &0&0& 2.00&2.00& 0.440 & 0.440& 0 &$1.84\times10^{-12}$& 0&0\\ 
 End of SI  &0.364&0.363& 2.14 &2.14&  0.311&0.312& 0.452&0.453&0.115&0.114 \\
KE=PE& $3.26\times10^4$&$2.87\times10^4$& 3.97  &3.93& $5.00\times10^{-6}$ &$4.96\times10^{-6}$&$1.02\times10^{-5}$ &$1.43\times 10^{-5}$&3.80&3.79 \\ 
$\dot \phi=0$& $1.37\times10^5$&$1.27\times10^5$&4.11 &4.06&$7.94\times10^{-23}$&$4.21\times10^{-23}$ &  $1.06\times10^{-5}$&$1.04\times10^{-5}$&4.89&4.87\\
Slow-Roll &$2.82\times10^5$ &$2.75\times10^5$ &4.08 &4.03&$-2.09\times10^{-7}$ &$-2.03\times10^{-7}$&$1.05\times10^{-5}$&$1.04\times10^{-5}$  &6.42&6.41\\
EOI&$1.88\times10^7$ & $1.92\times10^7$  & 0.282  &0.201 &$-2.05\times10^{-7}$ &$-1.79\times10^{-7}$ &$7.27\times10^{-7}$ &$6.36\times10^{-7}$ &111&109\\
 \hline
 
  Bounce & 0 &0 & -0.200 &-0.200 & 0.440&0.440 & 0 &$1.84\times10^{-12}$&0&0\\
 End of SI & 0.364 &0.363& $-5.87\times10^{-2}$ &$-5.89\times 10^{-2}$ & 0.311 &0.312& 0.452 &0.453&0.115&0.114\\
KE=PE&$6.85\times10^4$ &$6.05\times10^4$ & 1.89 &1.86 & $2.38\times10^{-6}$ &$2.34\times10^{-6}$ &$4.87\times10^{-6}$ &$6.77\times10^{-6}$&4.04&4.04\\ 
$\dot \phi=0$&$2.38\times10^5$ &$2.19\times10^5$ & 2.01 &1.96 &0 &$-2.41\times10^{-23}$ & $5.19\times10^{-6}$ &$5.06\times10^{-6}$ &4.91&4.88\\
Slow-Roll &$5.18\times10^5$ &$5.31\times10^5$ & 1.97  &1.91 &$-2.12\times10^{-7}$ &$-2.03\times10^{-7}$ & $5.08\times10^{-6}$ &$4.95\times10^{-6}$&6.35&6.45\\
EOI&$8.73\times10^6$  & $9.08\times10^6$  & 0.282 &0.201&$-2.05\times10^{-7}$ &$-1.79\times10^{-7}$&$7.27\times10^{-7}$&$6.36\times10^{-7}$ &30.2 &29.9\\
 \hline
 
 Bounce &0&0& -5.16 &-5.16 & 0.440 & 0.440 & 0 &$1.84\times10^{-12}$ & 0&0\\ 
 End of SI &0.364 &0.363 & -5.02 &-5.02 &  0.311 &0.312 &  0.452&0.453 & 0.115 &0.114\\
KE=PE&$4.10\times10^4$ & $4.10\times10^4$ &-3.15&-3.19 & $3.97\times10^{-6}$ &$4.01\times10^{-6}$ &$8.13\times10^{-6}$ &$1.16\times 10^{-5}$ &3.87 &3.88\\ 
Slow-Roll &$3.69\times10^5$ & $3.41\times10^5$ &-2.93 &-2.99 &$1.93\times 10^{-7}$ &$2.07\times10^{-7}$ &$7.57\times 10^{-6}$ &$7.71\times10^{-6}$ &6.40&6.34\\
EOI&$1.33\times10^7$ & $1.41\times10^7$   & -0.282 &-0.201 &$2.05\times10^{-7}$&$1.79\times10^{-7}$ &$7.27\times10^{-7}$ &$6.36\times10^{-7}$ &60.0&62.9\\
 \hline
 \hline

\end{tabular}
\end{center}
\label{t1}
\end{table}

\begin{figure}[h!] 
{
\includegraphics[width=8cm]{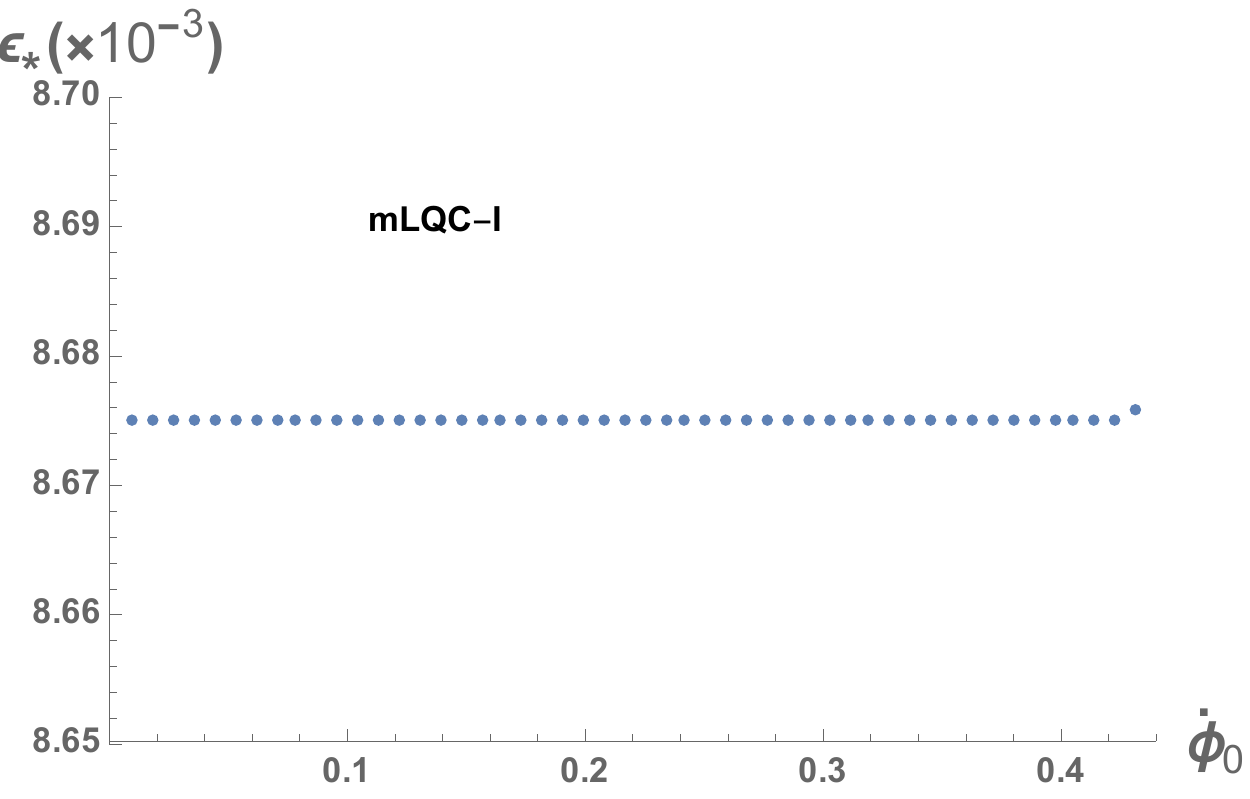}
\includegraphics[width=8cm]{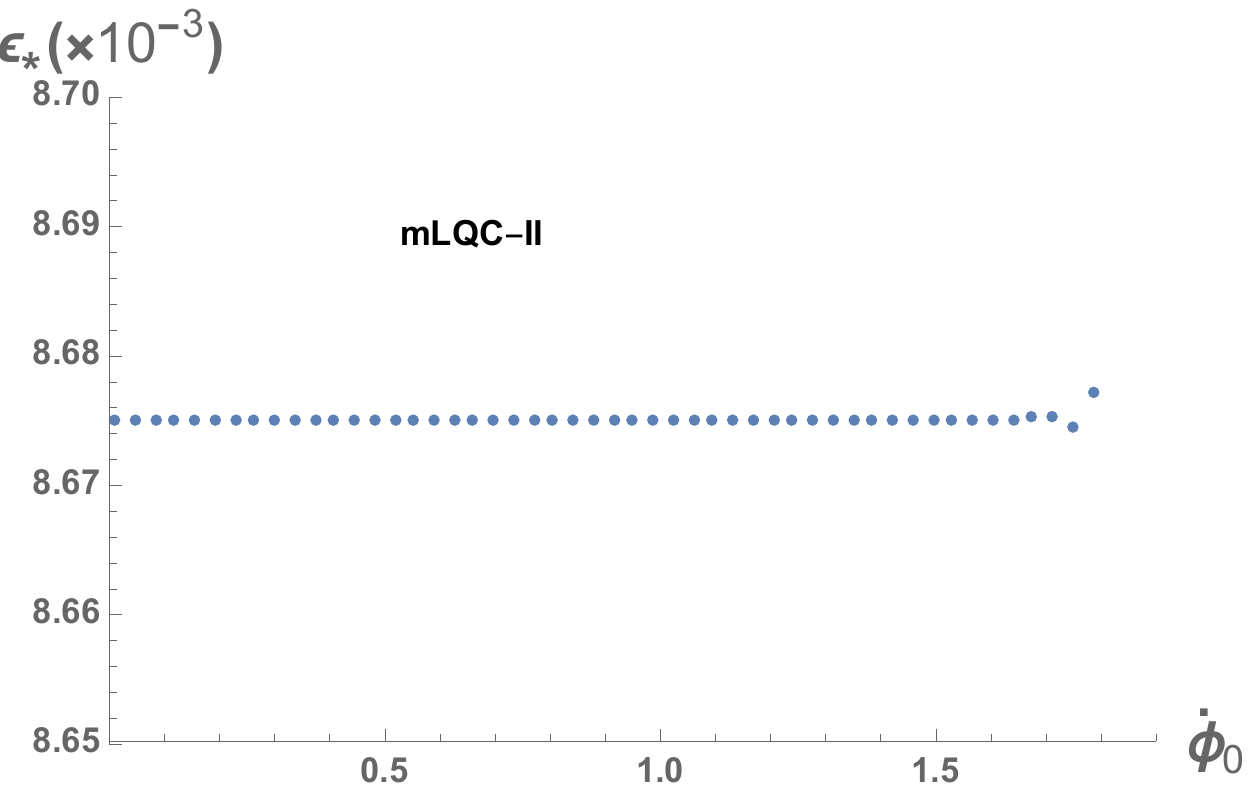}
}
\caption{The conditions of the numeric simulations in the figure are imposed at $\phi_0=-5.20~m_\mathrm{Pl}$ for mLQC-I and $\phi_0=-5.40~m_\mathrm{Pl}$ for mLQC-II while the velocity is chosen from the interval $\dot \phi_0 \in \left(0, \sqrt{2 \rho^A_S}\right)$ with  uniformly distributed data points. Each dot in the figure represents the result of $\epsilon_*$ for one particular set of ($\phi_0, \dot{\phi}_0$). When the pivot mode exits the horizon during the slow-roll, $\phi_*=-3.02~ m_\mathrm{Pl}$. Thus, in mLQC-I/II, the Universe is evolved from $\phi_0$ to $\phi_*$. The $\epsilon_*$($=8.675\times10^{-3}$) at the horizon-crossing as shown in the figure turns out to be constant for all permissible initial conditions.}
\label{pedom}
\end{figure}

\begin{figure}[h!] 
{
\includegraphics[width=8cm]{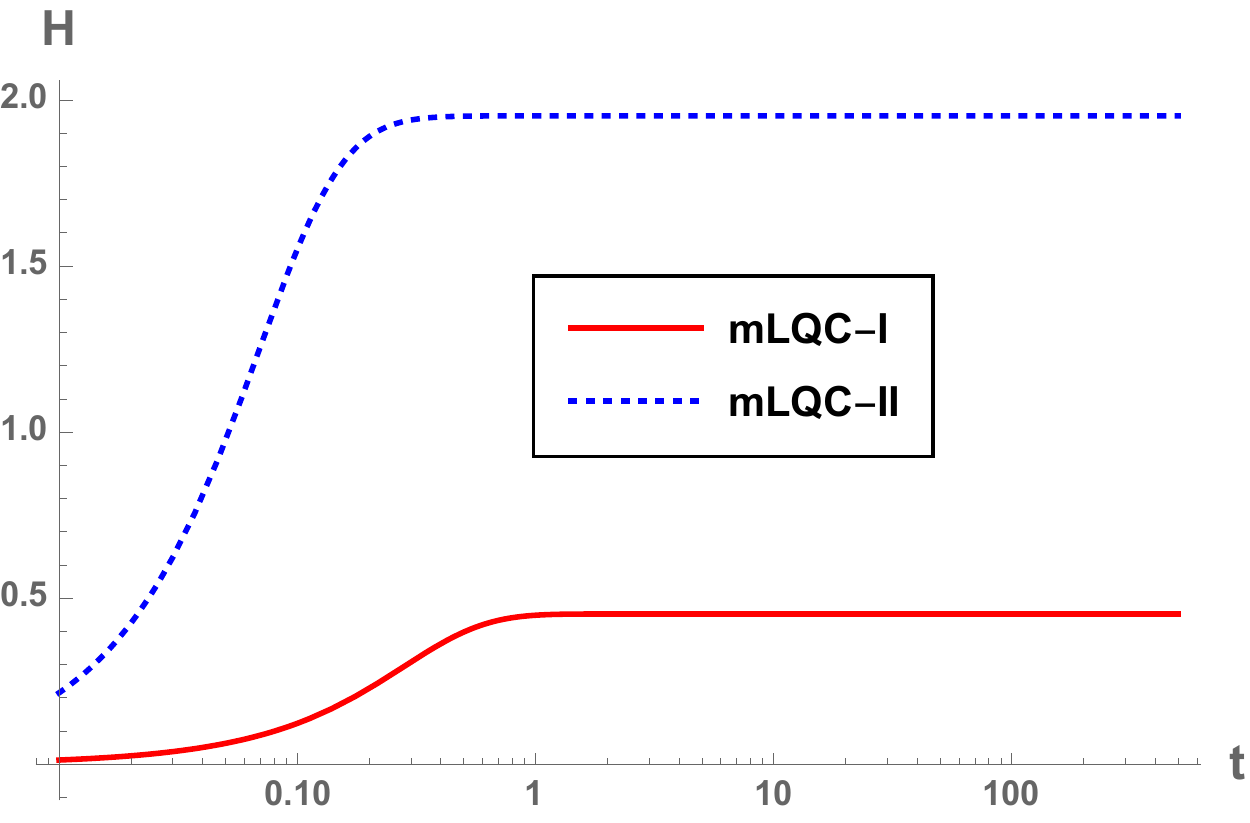}
\includegraphics[width=8cm]{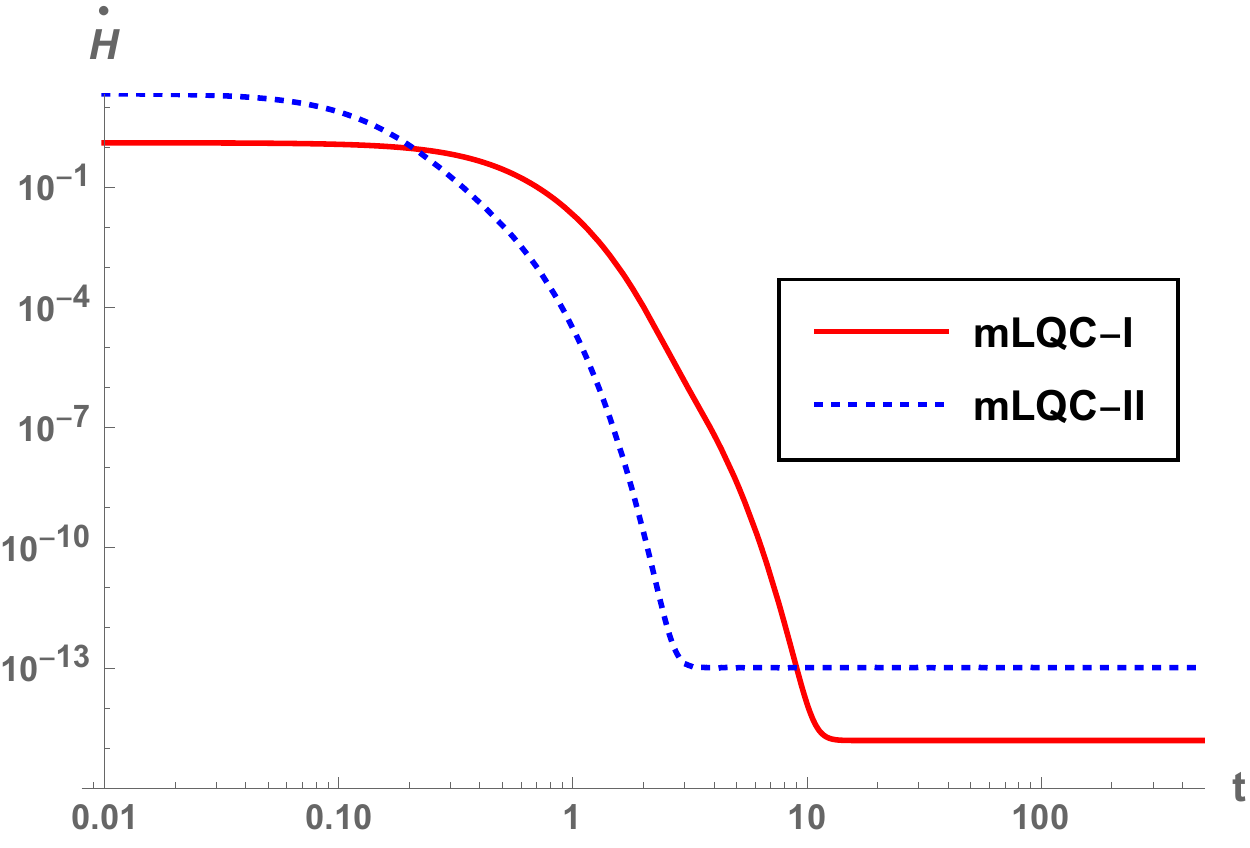}
\includegraphics[width=8cm]{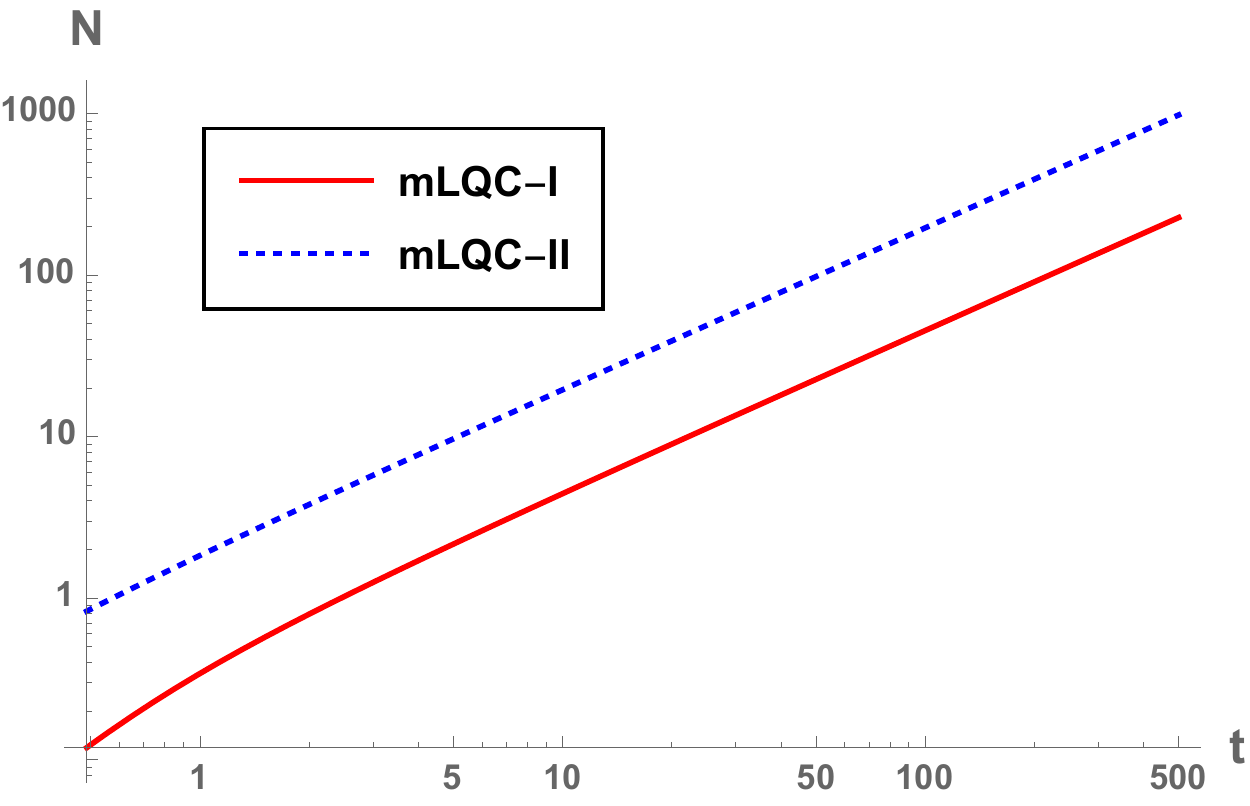}
\includegraphics[width=8cm]{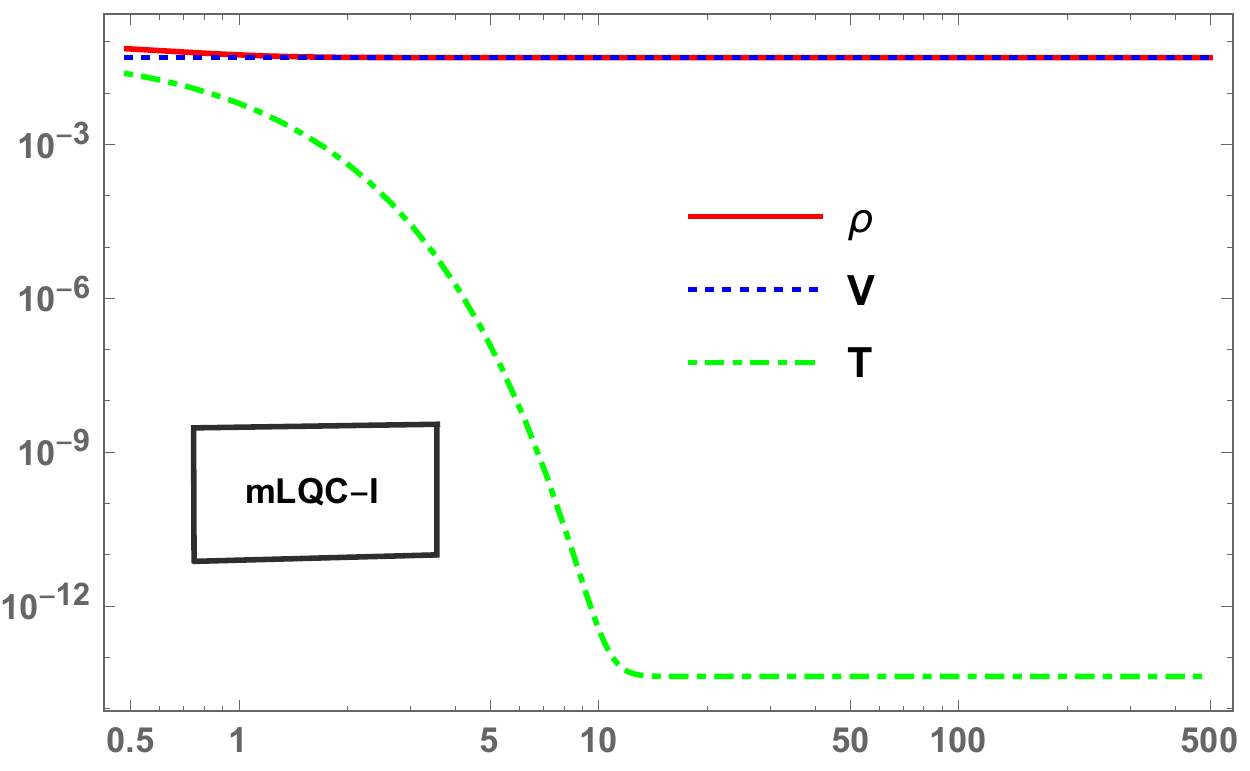}
}
\caption{Two examples when the bounce is dominated by the PE in mLQC-I/II are shown in the figure. The evolution of the Hubble parameter (top left), the rate of the Hubble parameter (top right), the e-folds (bottom left) and energy density (bottom right) are depicted in each model, except for the last panel in which only the change of the KE, PE and energy density in mLQC-I is plotted out as the shapes of these curves are similar in mLQC-II.   For mLQC-I,  $\phi_B$ at the bounce is set to $-2.48\times 10^5~m_\mathrm{Pl}$ while for mLQC-II, $\phi_B= -1.01\times 10^6 ~m_\mathrm{Pl}$. With these initial conditions, there exists in both models a long period of SI phase during which the slow-roll inflation takes place.}
\label{pedom2}
\end{figure}

\end{widetext}

\subsubsection{Potential-Energy Dominated Case}

When $ \mathrm{PE} \gtrsim \mathrm{KE}$ at the bounce point, the kinetic energy of the inflaton decays so rapidly that the Universe enters into the slow-roll phase at  $t \approx 10 \sim 100 ~m^{-1}_\mathrm{Pl}$  right after the bounce. Moreover, when $\mathrm{PE} \ge \rho^A_S\approx 0.5 \rho^A_c$ at the bounce point, the duration of the SI phase is substantially extended. This is because the slow-roll occurs shortly after the bounce when the energy density exceeds $\rho^A_S$. As a result,  the SI phase partly overlaps with the slow-roll phase. By using Eq. (\ref{app2}), it is straightforward to estimate the SI e-folds which  turns out to be 
$$
N_S\approx \frac{4\pi}{m^2} \left(PE_B-\rho^A_S\right), 
$$
where $PE_B$ denotes the potential energy at the bounce point and the formula is only valid when $\mathrm{PE}_B \ge \rho^A_S$. In Fig. \ref{pedom2}, we show explicitly two simulations in mLQC-I/II when the bounce is PE dominated. In mLQC-I, the slow-roll takes place at $t_i=13.9 ~t_\mathrm{Pl}$ and the e-folds from the bounce to the onset of inflation is $N_i=6.16$ while in mLQC-II, the slow-roll takes place at $t_i=3.44 ~t_\mathrm{Pl}$, $N_i=6.60$. Therefore, although the slow-roll takes place shortly after the bounce for the PE dominated bounce, as the universe is in a state of superinflation, the e-folds at the onset of the slow-roll is still comparable to that in the KE dominated case (see Tables. I and II). Since the slow-roll partly overlaps with the SI phase, the Hubble parameter is of the Planckian scale as shown in the top-left panel of Fig. \ref{pedom2} while $\dot H$ remains a very tiny positive quantity (top right panel). The e-folds of the slow-roll phase also blows up rapidly (bottom left). The change in the KE, PE and the total energy density is depicted in mLQC-I in the bottom right panel. The slow-roll takes place when the KE drops down to $10^{-12} \rho_\mathrm{Pl}$. 

For PE dominated bounce, the lower bound on the slow-roll inflationary e-folds can also be roughly estimated as $N_\mathrm{SR} \gtrsim 10^{11}$. Such large amount e-folds imposes mathematically serious challenges for simulating the whole process from the bounce to the end of the slow-roll. In addition, in such cases any quantum effects will be washed out at the current time.
A convenient way to show that all the initial conditions satisfying  $\mathrm{PE}\gtrsim \mathrm{KE}$ can lead to the desired slow-roll inflation is first proposed in \cite{as2011}. Due to the reflection symmetry $\left(\phi, \dot \phi\right)\rightarrow \left(-\phi, -\dot \phi\right)$, it is sufficient  to  focus only on the left wing of the potential where $\phi<0$. Now, it should be noted that, given an arbitrary $\phi_0(<0)$ which lies in the intervals given by Eqs. (\ref{infefold1})-(\ref{infefold2}), all the trajectories with the conditions $\phi_B<\phi_0$ will definitely pass through $\phi_0$ at some finite time as proved in \cite{as2011}. Thus, numeric simulations can be simply performed in the interval $\left(\phi_0, \phi_*\right)$ where $\phi_*=-3.02 ~m_\mathrm{Pl}$ is the value of the scalar field when the pivot mode exits the horizon during the slow-roll. In order to exhaust all possible conditions  at the bounce point,  the velocity of the scalar field at $\phi_0$ should be chosen from the interval $\dot \phi_0 \in \left(0, \sqrt{2 \rho^A_S}\right)$. For $\dot \phi_0 \approx 0$, the bounce is dominated by the PE while $\dot \phi_0 \approx \sqrt{2 \rho^A_S}$ represents the case when the bounce is kinetic dominated. The key point is to show all the possible initial conditions at $\phi_0$ will result in the same $\epsilon_*$ when the inflaton rolls down to the pivot value $\phi_*$. Afterwards, the succeeding evolution from $\phi_*$ to $\phi_\mathrm{end}$ is the desired slow-roll phase. In Fig. \ref{pedom}, we show the results of our simulations for mLQC-I/II. In mLQC-I, $\phi_0$ is set to $-5.20 ~m_\mathrm{Pl}$ and in mLQC-II, $\phi_0$ is set to $-5.40 ~m_\mathrm{Pl}$. It can be seen clearly from the figure that in both cases, for almost all $\dot \phi_0$, $\epsilon_*\approx 8.675\times 10^{-3}$ which implies at $\phi=\phi_*$, the Universe is already in the slow-roll phase and also the pivot mode exits the horizon right at this moment. As a result, all the initial conditions with $\phi_B<\phi_0$ at the bounce, including those dominated by the PE,  will lead to the desired slow-roll phase. Similar analysis also applies to the right wing of the potential when $\phi>0$ due to the reflection symmetry. 

\subsubsection{Concluding Remarks}
So far, we have discussed the generic properties of the pre-inflationary dynamics in mLQC-I and mLQC-II in details. Based on the scalar power spectral amplitude and scalar spectral index from 2015 Planck CMB data \footnote{Planck2018 CMB data  is quite similar to the 2015 one, so no significant difference is obtained from the newly released data \cite{Planck2018}.}, the number of e-folds from the horizon exit to  the end of inflation for the pivot mode is $N_*=56.6$. In order to be consistent with  this observation, the total number of inflationary e-folds must be larger than $N_*$. According to our numeric simulations (see also Tables  \ref{t1} and \ref{t2}) and the analysis of both KE and PE dominated bounce in the above subsections, the range of $\phi_B$ which allows for at least $N_*$ inflationary e-folds in each model is given by  
\bqn
\lb{infefold1}
&&  \phi^{\mathrm{I}}_B \in (-\phi^{{\scriptscriptstyle{\mathrm{I}}}}_\text{max}, -5.518~m_{\mathrm{Pl}} )\cup(0.917~m_{\mathrm{Pl}} ,\phi^{{\scriptscriptstyle{\mathrm{I}}}}_\text{max}), ~~~~~~ \\
\lb{infefold2}
&& \phi^{\mathrm{II}}_B \in (-\phi^{{\scriptscriptstyle{\mathrm{II}}}}_\text{max}, -5.386~m_{\mathrm{Pl}} )\cup(0.689~m_{\mathrm{Pl}} ,\phi^{{\scriptscriptstyle{\mathrm{II}}}}_\text{max}), 
\eqn
where
\bqn
\lb{Max}
\phi^{{\scriptscriptstyle{\mathrm{I}}}}_\text{max}= 3.49\times10^5~m_{\mathrm{Pl}},\quad
\phi^{{\scriptscriptstyle{\mathrm{II}}}}_\text{max}= 1.48\times10^6~m_{\mathrm{Pl}}. ~~~~~~~~
\eqn
Noting that in the extreme PE dominated state, $\phi=\pm \phi^A_\mathrm{max}$ while $\phi=0$ corresponds to the extreme  KE dominated state. In Sec. IV, we will talk about a well-defined physical measure available at the bounce point, which combined with Eqs. (\ref{infefold1})-(\ref{infefold2}) gives the probability of the desired slow-roll in the chaotic inflation.

Moreover, by employing the physical measure in Sec. IV, the ratio of the KE dominated bounce over the PE dominated bounce in mLQC-I and mLQC-II can be computed as 
\bqn
\frac{\int^{\phi_\mathrm{mid}}_0 d\omega^A}{\int^{\phi_\mathrm{max}}_{\phi_\mathrm{mid}} d\omega^A} =7.07:2.93,
\eqn
where $d\omega^A$ represents the measure given by Eq. (\ref{measure1}) and (\ref{measure2}) in mLQC-I and mLQC-II, respectively, $\phi_\mathrm{mid}=\sqrt{\rho^A_c}/m$ and $\phi_\mathrm{max}=\sqrt{2\rho^A_c}/m$.

\begin{widetext}

\begin{table} \scriptsize
\caption{Similar to Table \ref{t1}, the analytic and numeric results are compared in mLQC-II with the chaotic potential.  All the notations in this table are the same as those in Table \ref{t1}.}
\begin{center}
 \begin{tabular}{|c||c|c||c|c||c|c||c|c||c|c|} 
 \hline
 Event  &\bf{$t_A$} &\bf{$t_N $}  & \bf{$\phi_A$} & \bf{$\phi_N$} & \bf{$\dot \phi_A$} & \bf{$\dot \phi_N$}&\bf{$H_A$} &\bf{$H_N$} &$N_A$& $N_N$\\ 
 \hline
 
 Bounce & 0 &0& 0  &0& 1.86 &1.86 & 0 &$2.96\times10^{-16}$&0&0\\ 
 End of SI &$8.22\times10^{-2}$ &$8.17\times10^{-2}$& $0.135$ &0.134& 1.33  &1.33& 1.96 &1.96&0.112&0.111\\
KE=PE&$5.64\times10^4$ &$5.00\times10^4$& 2.29 &2.25& $2.89\times10^{-6}$&$2.84\times10^{-6}$ &$5.91\times10^{-6}$ &$8.21\times10^{-6}$&4.46&4.45\\ 
$\dot \phi=0$&$2.07\times10^5$ &$1.91\times10^5$ &2.42 &2.37 &$1.06\times10^{-22}$&$2.31\times10^{-23}$&$6.25\times10^{-6}$&$6.10\times10^{-6}$ &5.39&5.36\\
Slow-Roll &$4.45\times10^5$  & $4.39\times10^5$ & 2.38  &2.33 &$-2.11\times10^{-7}$  &$-2.03\times10^{-7}$ &$6.14\times10^{-6}$  &$6.02\times10^{-6}$ &6.86&6.86\\
EOI&$1.07\times10^7$& $1.10\times10^7$   & 0.282 &0.201 &$-2.05\times10^{-7}$ &$-1.79\times10^{-7}$ &$7.27\times10^{-7}$ &$6.36\times10^{-7}$&42.0 &41.3\\
 \hline
 
  Bounce &0 &0& 0.689&0.689& 1.86  & 1.86& 0 &$2.96\times10^{-16}$ & 0&0\\ 
 End of SI  &$8.22\times10^{-2}$ &$8.17\times10^{-2}$& 0.824 &0.823 &  1.33 &1.33& 1.96&1.96&0.112&0.111 \\
KE=PE& $4.40\times10^4$ &$3.89\times10^4$ & 2.94&2.90 & $3.70\times10^{-6}$ &$3.66\times10^{-6}$ &$7.58\times10^{-6}$ &$1.06\times 10^{-5}$  &4.38&4.37 \\ 
$\dot \phi=0$& $1.72\times10^5$ &$1.59\times10^5$&3.08 &3.02 &$-7.94\times10^{-23}$ &$6.91\times10^{-23}$ &  $7.93\times10^{-6}$ &$7.78\times10^{-6}$&5.38&5.36\\
Slow-Roll &$3.63\times10^5$ &$3.55\times10^5$  &3.04&2.99 &$-2.10\times10^{-7}$ &$-2.03\times10^{-7}$ &$7.85\times10^{-6}$ &$7.72\times10^{-6}$ &6.89&6.87\\
EOI&$1.38\times10^7$ & $1.41\times10^7$  & 0.282  &0.201 &$-2.05\times10^{-7}$ &$-1.79\times10^{-7}$ &$7.27\times10^{-7}$ &$6.36\times10^{-7}$ &64.6&63.5\\
 \hline
 
  Bounce &0 &0& 1.90 &1.90& 1.86 & 1.86 & 0 &$1.48\times10^{-16}$ & 0&0\\ 
 End of SI  &$8.22\times10^{-2}$ &$8.17\times10^{-2}$& 2.04 &2.03 &  1.33 &1.33 & 1.96 &1.96&0.112&0.111 \\
KE=PE& $3.16\times10^4$ &$2.79\times10^4$ & 4.10 &4.06 & $5.16\times10^{-6}$ &$5.11\times10^{-6}$ &$1.06\times10^{-5}$ &$1.48\times 10^{-5}$ &4.26&4.26 \\ 
$\dot \phi=0$& $1.34\times10^5$ &$1.24\times10^5$&4.24 &4.18 &$2.65\times10^{-23}$ &$1.18\times10^{-23}$ &  $1.09\times10^{-5}$ &$1.08\times10^{-5}$&5.37&5.35\\
Slow-Roll &$2.74\times10^5$ &$2.70\times10^5$ &4.21&4.16&$-2.09\times10^{-7}$ &$-2.03\times10^{-7}$&$1.09\times10^{-5}$ &$1.07\times10^{-5}$ &6.90&6.92\\
EOI&$1.94\times10^7$ & $1.98\times10^7$  & 0.282 &0.201&$-2.05\times10^{-7}$ &$-1.79\times10^{-7}$ &$7.27\times10^{-7}$  &$6.36\times10^{-7}$&118&116\\
 \hline
 
  Bounce & 0 &0& -0.200 &-0.200& 1.86  &1.86& 0 &$2.96\times10^{-16}$&0&0\\
 End of SI &$8.22\times10^{-2}$ &$8.17\times10^{-2}$ & $-6.47\times10^{-2}$ &$-6.55\times 10^{-2}$& 1.33 &1.33& 1.96  &1.96&0.112&0.111\\
KE=PE&$6.14\times10^4$ &$5.44\times10^4$& 2.11 &2.07 & $2.65\times10^{-6}$ &$2.60\times10^{-6}$&$5.43\times10^{-6}$ &$7.53\times10^{-6}$ &4.49&4.48\\ 
$\dot \phi=0$&$2.20\times10^5$ &$2.03\times10^5$& 2.23 &2.18&$-7.94\times10^{-23}$ &$-3.44\times10^{-23}$ & $5.76\times10^{-6}$ &$5.62\times10^{-6}$&5.39&5.36\\
Slow-Roll &$4.76\times10^5$&$4.79\times10^5$ & 2.19 &2.13&$-2.11\times10^{-7}$ &$-2.03\times10^{-7}$& $5.65\times10^{-6}$ &$5.51\times10^{-6}$&6.85&6.90\\
EOI&$9.78\times10^6$   & $1.01\times10^7$ & 0.282 &0.201&$-2.05\times10^{-7}$  &$-1.79\times10^{-7}$ &$7.27\times10^{-7}$ &$6.36\times10^{-7}$&36.5&35.9\\
 \hline
 
  Bounce &0 &0& -5.39&-5.39& 1.86 & 1.86 & 0 &$2.96\times10^{-16}$& 0&0\\ 
 End of SI &$8.22\times10^{-2}$ &$8.17\times10^{-2}$ & -5.25 &-5.25 &  1.33&1.33&  1.96&1.96& 0.112&0.111\\
KE=PE&$4.11\times10^4$ & $3.61\times10^4$&-3.15 &-3.19& $3.96\times10^{-6}$ &$4.01\times10^{-6}$&$8.11\times10^{-6}$ &$1.16\times 10^{-5}$&4.35 &4.36\\ 
Slow-Roll &$3.70\times10^5$ & $3.45\times10^5$&-2.93 &-2.99&$1.93\times 10^{-7}$&$2.07\times10^{-7}$ &$7.55\times 10^{-6}$&$7.71\times10^{-6}$ &6.88&6.85\\
EOI&$1.33\times10^7$   & $1.41\times10^7$ & -0.282 &-0.201&$2.05\times10^{-7}$ &$1.79\times10^{-7}$ &$7.27\times10^{-7}$ &$6.36\times10^{-7}$&60.2 &63.4\\
 \hline
 \hline
\end{tabular}
\label{t2}
\end{center}
\end{table}
\end{widetext}

  \begin{figure}[h!] 
{
\includegraphics[width=8cm]{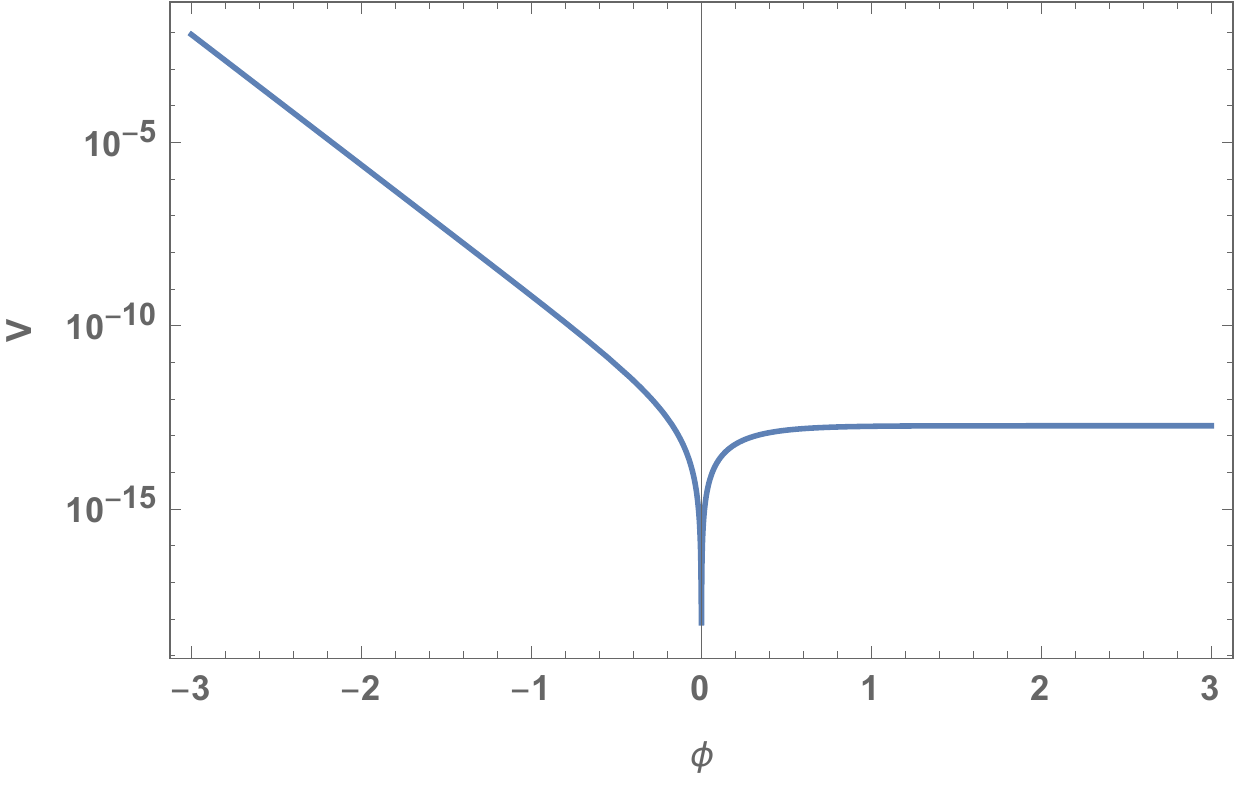}
}
\caption{A plot of the Starobinsky potential with the mass set to $2.49 \times 10^{-6} ~m_{\mathrm{Pl}} $ which shows the energy scale of the PE   in the range of $\phi/m_{\mathrm{Pl}} \in(-3.00, 3.00)$. For the desired slow-roll inflation to occur, the bounce point must be dominated by the KE  of the inflaton as  discussed in the text.}
\label{starpotential}
\end{figure}

\begin{figure}[h!] 
{
\includegraphics[width=7cm]{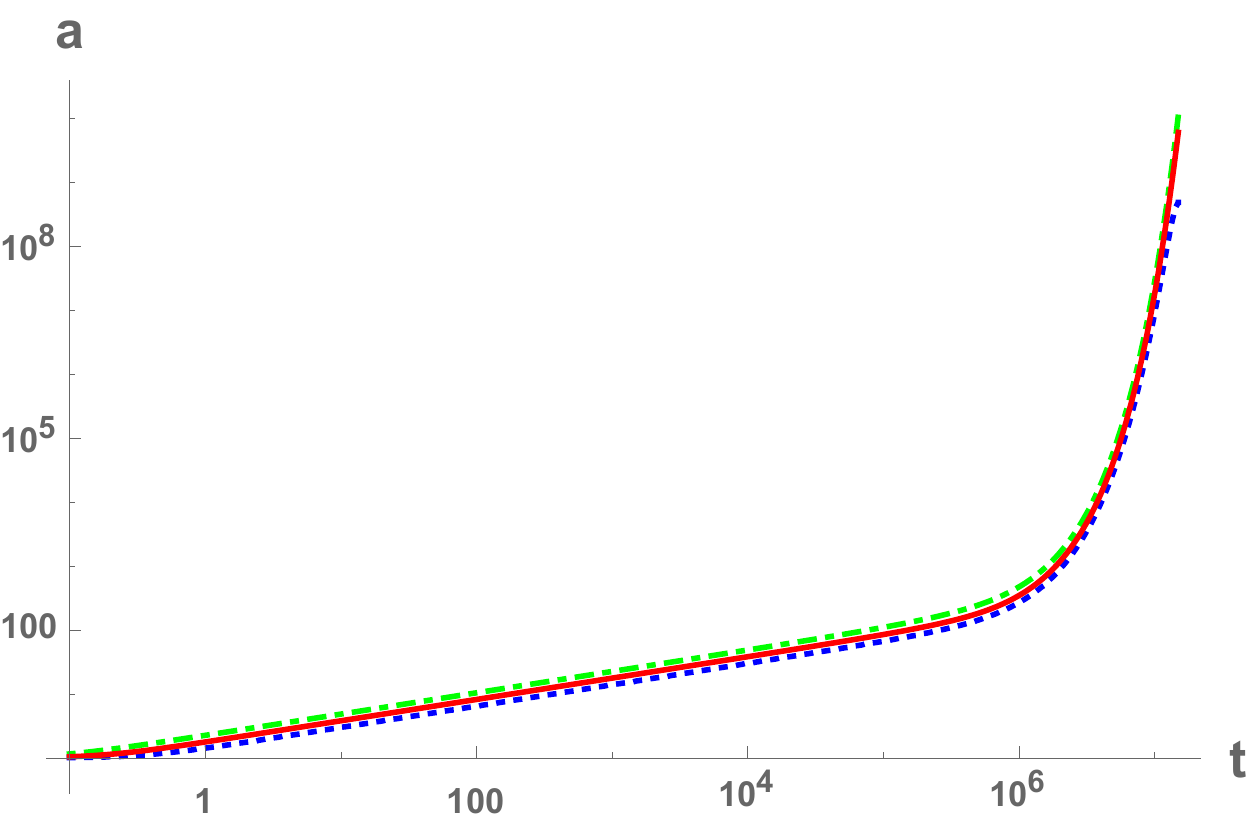}
\includegraphics[width=7cm]{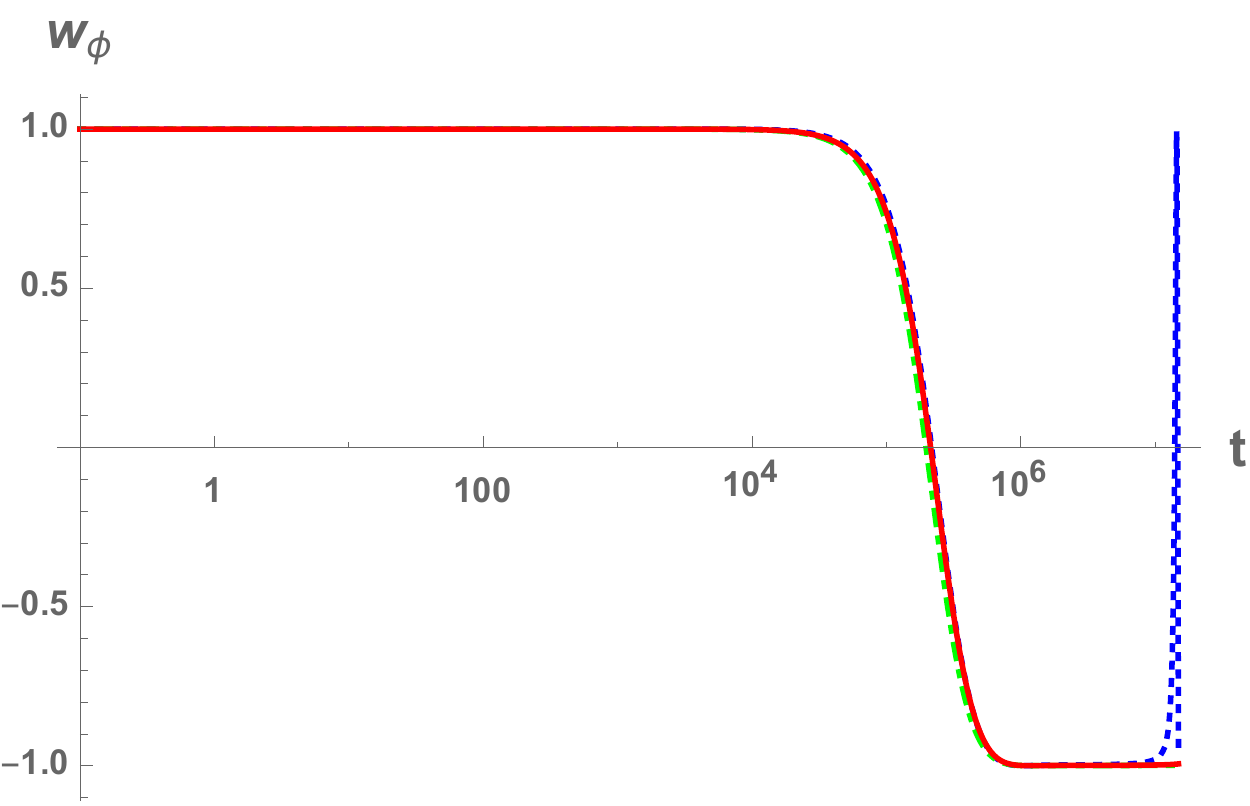}
\includegraphics[width=7cm]{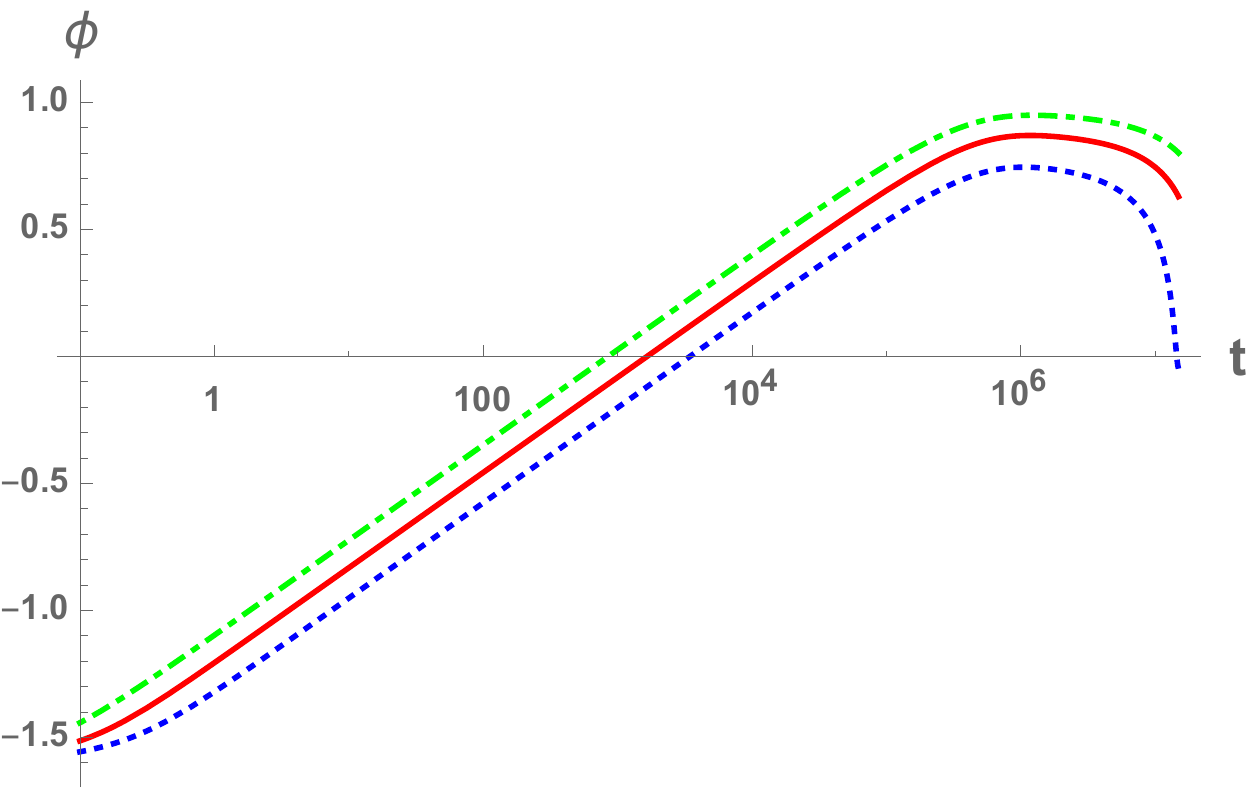}
}
\caption{The evolution of several quantities in the post-bounce phase are depicted and compared  in LQC (red solid curves), mLQC-I (blue dotted curves) and mLQC-II (green dot-dashed curves) with the Starobinsky potential.The initial condition for the simulation is chosen at the bounce with  $\phi_B=-1.6~m_{\mathrm{Pl}} , \dot \phi_B>0$. }
\label{fig10}
\end{figure} 

\subsection{Starobinsky potential}
 
In GR, Starobinsky inflation results from adding a $R^2$ term in the Einstein-Hilbert action, which results in the following effective potential 
\bq
\lb{star}
V=\frac{3m^2}{32\pi G}\left(1-e^{-\sqrt{16\pi G/3} \phi}\right)^2,
\eq
where the mass is set to $2.49 \times 10^{-6} ~m_{\mathrm{Pl}} $ (Appendix A). Since loop cosmological models are not based on canonical quantization, one assumes the above potential to hold as is in effective dynamics to understand phenomenological implications \cite{bonga,lsw2018b}.\footnote{Note that in LQC, the covariant action exists \cite{olmo-singh} but it is of infinite order in curvature terms in the Palatini framework. Starobinsky inflation for such a Palatini action has not been implemented so far.} Unlike the $m^2 \phi^2$ potential discussed above, the slow-roll inflation can  only take place on the right wing of the Starobinsky potential where $\phi>\phi_\text{end}=0.188~m_{\mathrm{Pl}} $.  The flatness of the right wing makes the tensor-to-scalar ratio rather small as compared with the chaotic potential. Again,  the initial data for numeric simulations is chosen at the bounce point where the parameter space consists of  $\phi_B$ and the sign of $\dot \phi_B$. Due to the asymmetry of the potential, both $\dot \phi_B>0$ and $\dot \phi_B<0$  should be considered separately.   In mLQC-I/II, the critical energy density sets up a lower bound on the range of  $\phi_B$ which is explicitly given by 
\bq
\phi^A_\text{min} = -\sqrt{\frac{3}{16\pi G}}\ln\left(1+\sqrt{\frac{32 \pi G \rho^A_c}{3m^2}}\right).
\eq
More specifically, $\phi^{{\scriptscriptstyle{\mathrm{I}}}}_\text{min}=-3.30~m_{\mathrm{Pl}} $ and $\phi^{{\scriptscriptstyle{\mathrm{II}}}}_\text{min}=-3.65~m_{\mathrm{Pl}} $. For the Starobinsky potential, the number of e-folds of the desired slow-roll inflation is $N_*=55.0$ based on the 2015 Planck CMB observations  (Appendix A). Depending on the sign of the initial velocity of the inflaton field,  there exist two situations that can result in an inflationary phase with an e-folds larger than $N_*$. First, if the initial velocity of the inflaton is positive, the initial value of the inflaton field at the bounce  can be no less than $-1.349~m_{\mathrm{Pl}} $ ($-1.577~m_{\mathrm{Pl}} $)  for mLQC-I (mLQC-II). The values of some important observables at different events when the evolution is initiated with these critical values of the scalar field($\phi_B=-1.349/1.577~m_{\mathrm{Pl}} $ for mLQC-I/II) are explicitly given in Tables \ref{star1}-\ref{star2}. As depicted in Fig. \ref{fig10}, in this case the inflaton first rolls down the left wing of the potential then climbs up the hill on the right and reaches a turnaround point. Shortly after it reverses its direction of motion, the slow-roll phase takes place and continues until  the inflaton reaches $\phi_\text{end}$. Therefore, for inflation to be possible, one necessary condition is that the value of the inflaton at the turnaround point has to be larger than $\phi_\text{end}$. Otherwise, as depicted in Fig. \ref{fig5}, when the inflaton starts from the top of the left wing of the potential $\phi_B/m_{\mathrm{Pl}}  = -3.00$, as it experiences the frictional force and covers a longer distance while rolling down the hill, it turns around at a point $\phi \lesssim \phi_\text{end}$ which directly leads to the missing of the slow-roll phase. As a result, in the Starobinsky potential, the slow-roll inflation does not exist for the PE dominated bounce in both mLQC-I and mLQC-II. Second, when the initial velocity of the inflaton is negative, the initial value of the inflaton field at the bounce  must be larger than $3.494~m_{\mathrm{Pl}} $ (or $3.722~m_{\mathrm{Pl}} $) for mLQC-I (or mLQC-II). In this case, with the negative initial velocity, the inflaton rolls down the right uphill and directly triggers the slow-roll phase in the same manner as in the chaotic inflation  (basically due to the opposite signs of the frictional and conservative forces). If the initial value of the inflaton is too close to $\phi_\text{end}$, the number of inflationary e-folds cannot be guaranteed to be greater than $N_*$ as required. In the extreme case, when $\phi_B$ is less than $\phi_\text{end}$, there is no slow-roll phase at all. This extreme situation is also shown in Fig. \ref{fig5} with $\phi_B$  exactly set to zero and $\dot \phi_B<0$. In this case, the inflaton first climbs up the left wing of the potential, then rolls down and oscillates around the bottom of the potential. Further,  the minimum values of the inflaton at the bounce indicate that all the observation-relevant initial conditions are those in which the bounce is dominated by the KE  of the potential. As plotted in Fig. \ref{starpotential}, when $\phi$ takes the value around negative unity, the PE  is around the order of $10^{-10}$ of the Planck energy, besides, the PE  at the right wing of the potential is even smaller. As  a result, for the Starobinsky potential, as long as we are concerned with the occurrence of a desired slow-roll phase, only the kinetic-energy-dominated bounce are phenomenologically relevant.

Similar to the case of chaotic inflation, there also exist three different phases in the Starobinsky potential -- the bouncing,  transition and slow-roll inflation,  when the bounce is largely dominated by the KE  of the scalar field. 
As in the case of chaotic inflation, a comparison of mLQC-I and mLQC-II with same initial conditions for $\phi_B$ shows the number of e-folds are slightly higher for mLQC-I at the end of SI , but the total number of e-folds are larger for mLQC-II. This is evident from the first blocks in Tables \ref{star1} and \ref{star2}.

 \begin{widetext}
 
\begin{table}\scriptsize
\caption{In this table, we compare the analytic and numeric results in mLQC-I with the Starobinsky potential. The analytic and numeric values of several observables are listed in a sequence of events, including Bounce, End of SI (superinflation), Equilibrium point when KE  equals PE, the turnaround point when $\dot \phi =0$, the onset of the slow-roll inflation and EOI (end of the inflation). The subscript `A' denotes analytic results and subscript `N' denotes numeric ones. All the e-folds are counted starting from the bounce to the particular event labelled in the first column of the Table   The Planck mass $m_{\mathrm{Pl}} $ is set to unity for conciseness.}
\begin{center}
 \begin{tabular}{|c||c|c||c|c||c|c||c|c||c|c|} 
 \hline
 Event  &\bf{$t_A$}  &\bf{$t_N $} & \bf{$\phi_A$} & \bf{$\phi_N$} & \bf{$\dot \phi_A$} & \bf{$\dot \phi_N$}&\bf{$H_A$} &\bf{$H_N$}&$N_A$& $N_N$\\ 
 \hline
 
  Bounce &0&0& -1.39 &-1.39& 0.440  & 0.440& 0 &$1.84\times10^{-12}$& 0&0\\ 
 End of SI  &0.364 &0.363&-1.25 &-1.25&  0.311&0.312& 0.452 &0.453&0.115&0.114 \\
KE=PE& $2.74\times10^5$ &$2.41\times10^5$& 0.924&0.891 & $5.94\times10^{-7}$ &$5.92\times10^{-7}$ &$1.22\times10^{-6}$ &$1.71\times 10^{-6}$ &4.51&4.50 \\ 
$\dot \phi=0$& $1.68\times10^6$&$1.44\times10^6$&1.08 &1.03&$2.48\times10^{-24}$ &$2.33\times10^{-24}$&  $1.23\times10^{-6}$&$1.23\times10^{-6}$&6.23&6.03\\
Slow-Roll &$4.18\times10^6$&$3.11\times10^6$  &1.07 &1.02&$-3.71\times10^{-9}$&$-6.24\times10^{-9}$ &$1.23\times10^{-6}$ &$1.22\times10^{-6}$ &9.30&8.08\\
EOI&$4.23\times10^7$ & $4.20\times10^7$  & 0.202&0.123&$-2.20\times10^{-7}$ &$-1.70\times10^{-7}$&$7.00\times10^{-7}$ &$6.02\times10^{-7}$&54.6&53.6\\
 \hline

 Bounce & 0& 0&-1.35 &-1.35 & 0.440 &0.440& 0&$1.84\times10^{-12}$&0&0\\ 
 End of SI &0.364 &0.363& -1.21 &-1.21 & 0.311 &0.312& 0.452&0.453&0.115&0.114\\
KE=PE&$2.73\times10^5$ &$2.41\times10^5$& 0.964 &0.932 & $5.97\times10^{-7}$ &$5.95\times10^{-7}$ &$1.22\times10^{-6}$ &$1.72\times10^{-6}$ &4.50&4.50\\ 
$\dot \phi=0$&$1.72\times10^6$  &$1.48\times10^6$ &1.12 &1.07 &$-1.24\times10^{-24}$ &$-4.72\times10^{-23}$&$1.23\times10^{-6}$ &$1.23\times10^{-6}$ &6.29&6.08\\
Slow-Roll &$3.46\times10^6$  & $2.99\times10^6$  & 1.12 &1.06&$-2.99\times10^{-9}$&$-5.22\times10^{-9}$ &$1.23\times10^{-6}$&$1.23\times10^{-6}$ &8.43 &7.93\\
EOI&$4.95\times10^7$  & $4.95\times10^7$& 0.202  &0.123&$-2.20\times10^{-7}$ &$-1.70\times10^{-7}$&$7.00\times10^{-7}$ &$6.02\times10^{-7}$&63.5&62.9\\
 \hline

  Bounce &0&0& 3.49&3.49& -0.440 & -0.440& 0&$1.84\times10^{-12}$ &0&0\\ 
 End of SI  &0.364&0.363& 3.35&3.35 & - 0.311&-0.312& 0.452&0.453&0.115&0.114 \\
KE=PE& $2.70\times10^5$ &$2.38\times10^5$ & 1.18&1.22 & $-6.04\times10^{-7}$&$-6.04\times10^{-7}$&$1.24\times10^{-6}$&$1.75\times 10^{-6}$  &4.50&4.50 \\ 
Slow-Roll &$2.69\times10^6$ &$2.57\times10^6$ &1.01 &1.06&$-5.53 \times10^{-9}$&$-5.32\times10^{-9}$ &$1.22\times10^{-6}$  &$1.23\times10^{-6}$&7.47&7.43\\
EOI&$4.97\times10^7$ & $4.92\times10^7$  & 0.171&0.123 &$-1.55\times10^{-7}$&$-1.70\times10^{-7}$ &$6.26\times10^{-7}$ &$6.02\times10^{-7}$&62.7 &62.5\\
 \hline
 
  Bounce & 0 &0& 3.50&3.50& -0.440 &-0.440& 0&$1.84\times10^{-12}$&0&0\\
 End of SI & 0.364 &0.363& 3.36&3.36 & -0.311&-0.312& 0.452 &0.453&0.115&0.114\\
KE=PE&$2.70\times10^5$&$2.38\times10^5$& 1.19 &1.22& $-6.04\times10^{-7}$ &$-6.04\times10^{-7}$&$1.24\times10^{-6}$&$1.75\times10^{-6}$ &4.50&4.50\\ 
Slow-Roll &$2.70\times10^6$&$2.59\times10^6$ & 1.01&1.07&$-5.39\times10^{-9}$ &$-5.18\times10^{-9}$& $1.23\times10^{-6}$&$1.23\times10^{-6}$&7.48&7.46\\
EOI&$5.10\times10^7$ & $5.04\times10^7$   & 0.171 &0.123&$-1.55\times10^{-7}$ &$-1.70\times10^{-7}$&$6.26\times10^{-7}$&$6.02\times10^{-7}$ &64.2&64.0\\
 \hline
 \hline

\end{tabular}

\label{star1}
\end{center}
\end{table}

\begin{table}\scriptsize
\caption{Similar to Table \ref{star1}, the analytic and numeric results are compared in mLQC-II with the Starobinsky potential.  All the notations in this table are the same as in Table \ref{t1}.}
\begin{center}
 \begin{tabular}{|c||c|c||c|c||c|c||c|c||c|c|} 
 \hline
 Event  &\bf{$t_A$}  &\bf{$t_N $} & \bf{$\phi_A$} & \bf{$\phi_N$} & \bf{$\dot \phi_A$} & \bf{$\dot \phi_N$}&\bf{$H_A$} &\bf{$H_N$}&$N_A$& $N_N$\\ 
 \hline
 
   Bounce &0&0& -1.39 &-1.39& 1.86 & 1.86& 0 &$2.96\times10^{-16}$& 0&0\\ 
 End of SI  &$8.22\times10^{-2}$ &$8.17\times10^{-2}$& -1.25 &-1.26 &  1.33&1.33& 1.96 &1.96&0.112&0.111 \\
KE=PE& $2.70\times10^5$ &$2.38\times10^5$& 1.16 &1.12& $6.03\times10^{-7}$ &$6.02\times10^{-7}$&$1.23\times10^{-6}$  &$1.74\times 10^{-6}$&4.98&4.98 \\ 
$\dot \phi=0$& $1.94\times10^6$&$1.68\times10^6$&1.32&1.26&$-1.03\times10^{-24}$&$5.70\times10^{-25}$ &  $1.24\times10^{-6}$&$1.24\times10^{-6}$&7.04&6.82\\
Slow-Roll &$3.30\times10^6$ &$3.03\times10^6$ &1.32&1.25&$-1.26\times10^{-9}$ &$-2.37\times10^{-9}$&$1.24\times10^{-6}$ &$1.24\times10^{-6}$&8.74&8.49\\
EOI&$1.08\times10^8$& $1.06\times10^8$   & 0.202&0.123&$-2.20\times10^{-7}$&$-1.70\times10^{-7}$ &$7.00\times10^{-7}$ &$6.02\times10^{-7}$&138 &133\\
 \hline
 
 Bounce & 0&0& -1.58&-1.58 & 1.86 &1.86& 0&$2.96\times10^{-16}$&0&0\\ 
 End of SI &$8.22\times10^{-2}$&$8.17\times10^{-2}$ & -1.44&-1.44 & 1.33 &1.33& 1.96&1.96 &0.112 &0.111\\
KE=PE&$2.73\times10^5$ &$2.41\times10^5$& 0.971&0.932& $5.97\times10^{-7}$ &$5.95\times10^{-7}$&$1.22\times10^{-6}$ &$1.72\times10^{-6}$&4.98&4.98\\ 
$\dot \phi=0$&$1.73\times10^6$ &$1.48\times10^6$ &1.13 &1.07 &$-1.65\times10^{-24}$&$-3.96\times10^{-23}$&$1.23\times10^{-6}$&$1.23\times10^{-6}$ &6.78&6.57\\
Slow-Roll &$3.43\times10^6$ & $2.92\times10^6$   & 1.12 &1.06&$-2.89\times10^{-9}$ &$-5.20\times10^{-9}$&$1.23\times10^{-6}$&$1.23\times10^{-6}$ &8.87&8.34\\
EOI&$5.08\times10^7$ & $4.96\times10^7$  & 0.202 &0.123&$-2.20\times10^{-7}$&$-1.70\times10^{-7}$ &$7.00\times10^{-7}$ &$6.02\times10^{-7}$&66.9 &63.4\\
 \hline

  Bounce &0 &0& 3.72 &3.72& -1.86 & -1.86& 0 &$2.96\times10^{-16}$& 0&0\\ 
 End of SI  &$8.22\times10^{-2}$&$8.17\times10^{-2}$& 3.59&3.59&  -1.33&-1.33& 1.96 &1.96&0.112&0.111 \\
KE=PE& $2.70\times10^5$ &$2.38\times10^5$& 1.18 &1.22& $-6.03\times10^{-7}$ &$-6.04\times10^{-7}$&$1.23\times10^{-6}$&$1.75\times 10^{-6}$  &4.98&4.98 \\ 
Slow-Roll &$2.68\times10^6$ &$2.58\times10^6$&1.00&1.06 &$-5.68\times10^{-9}$ &$-5.31\times10^{-9}$&$1.22\times10^{-6}$  &$1.23\times10^{-6}$&7.94&7.93\\
EOI&$4.83\times10^7$  & $4.92\times10^7$ & 0.171 &0.123&$-1.55\times10^{-7}$ &$-1.70\times10^{-7}$&$6.26\times10^{-7}$ &$6.02\times10^{-7}$&62.8 &62.9\\
 \hline
 
  Bounce & 0 &0& 4.00&4.00& -1.86 &-1.86& 0&$4.44\times10^{-16}$&0&0\\
 End of SI &$8.22\times10^{-2}$&$8.17\times10^{-2}$& 3.86&3.87 & -1.33 &-1.33& 1.96 &1.96&0.112&0.111\\
KE=PE&$2.68\times10^5$&$2.37\times10^5$& 1.45 &1.49& $-6.07\times10^{-7}$ &$-6.07\times10^{-7}$&$1.24\times10^{-6}$ &$1.76\times10^{-6}$&4.98&4.98\\ 
Slow-Roll &$3.04\times10^6$ &$2.98\times10^6$& 1.29&1.35&$-1.79\times10^{-9}$&$-1.65\times10^{-9}$& $1.24\times10^{-6}$&$1.24\times10^{-6}$&8.41&8.45\\
EOI&$1.54\times10^8$& $1.53\times10^8$   & 0.171&0.123&$-1.55\times10^{-7}$&$-1.70\times10^{-7}$ &$6.26\times10^{-7}$ &$6.02\times10^{-7}$&194&191\\
 \hline
 \hline
\end{tabular}
 
\label{star2}
\end{center}
\end{table}

\begin{figure}[h!]  
{
\includegraphics[width=7cm]{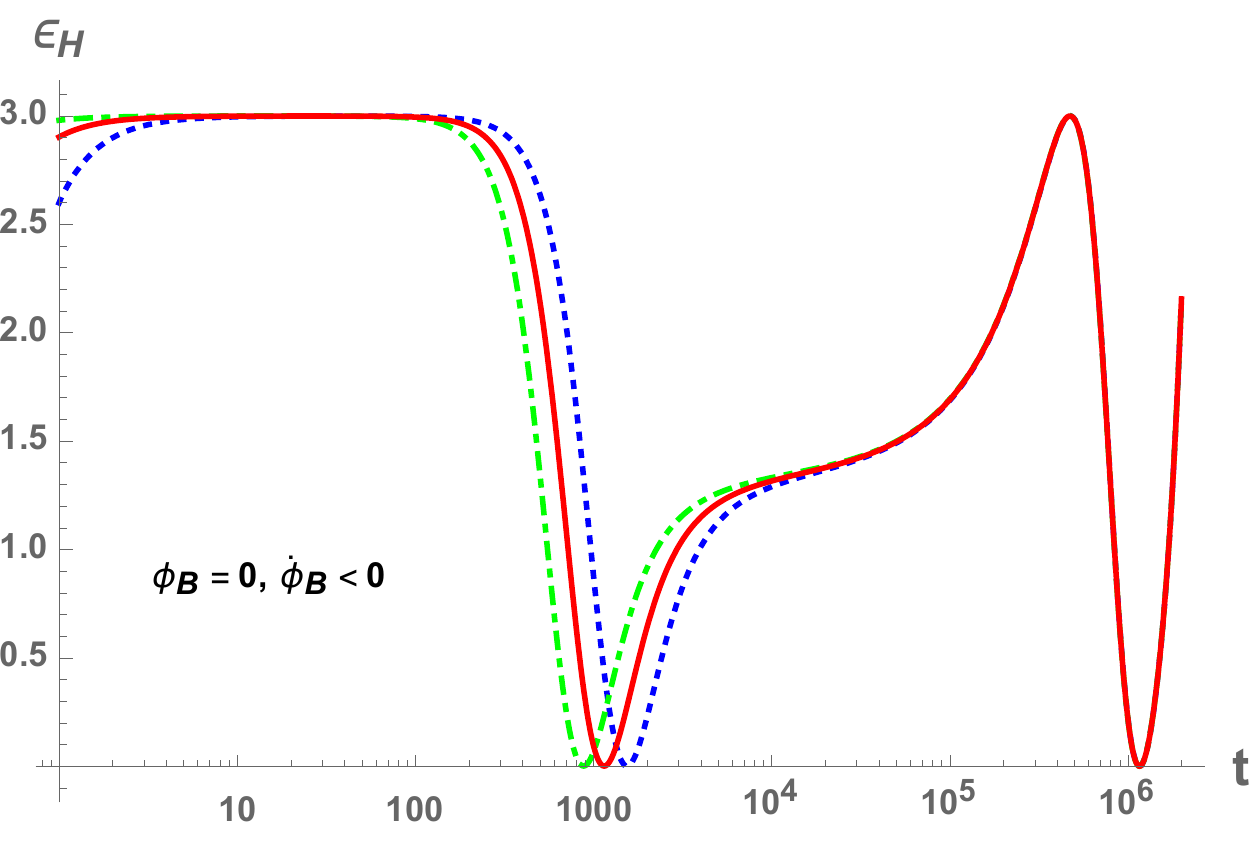}
\includegraphics[width=7cm]{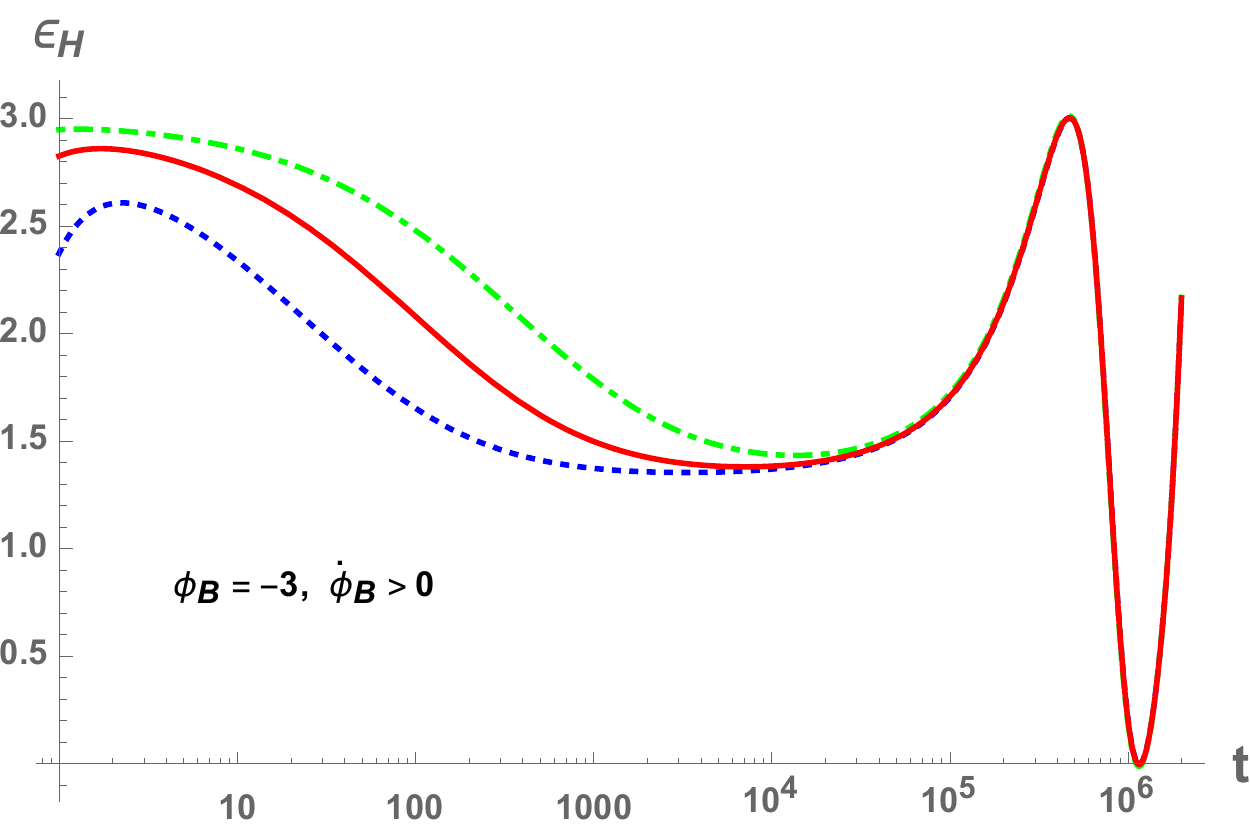}
\includegraphics[width=7cm]{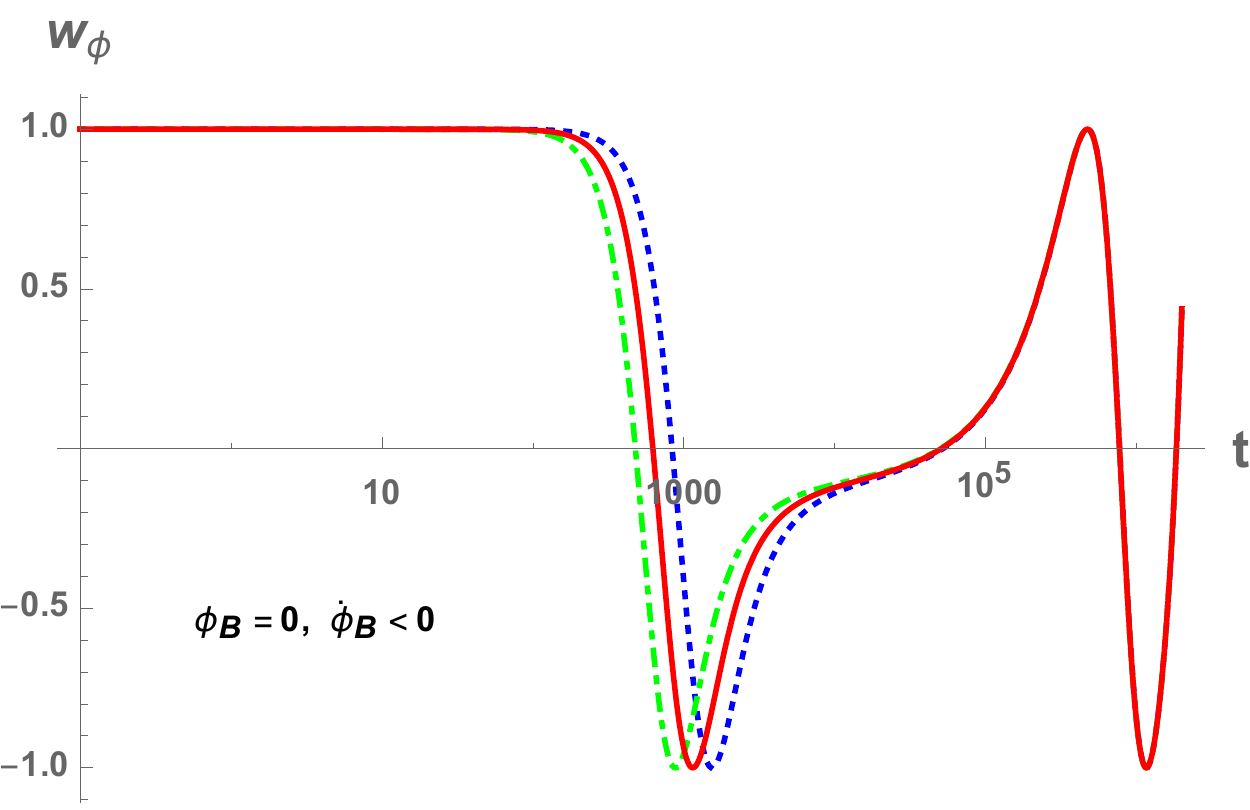}
\includegraphics[width=7cm]{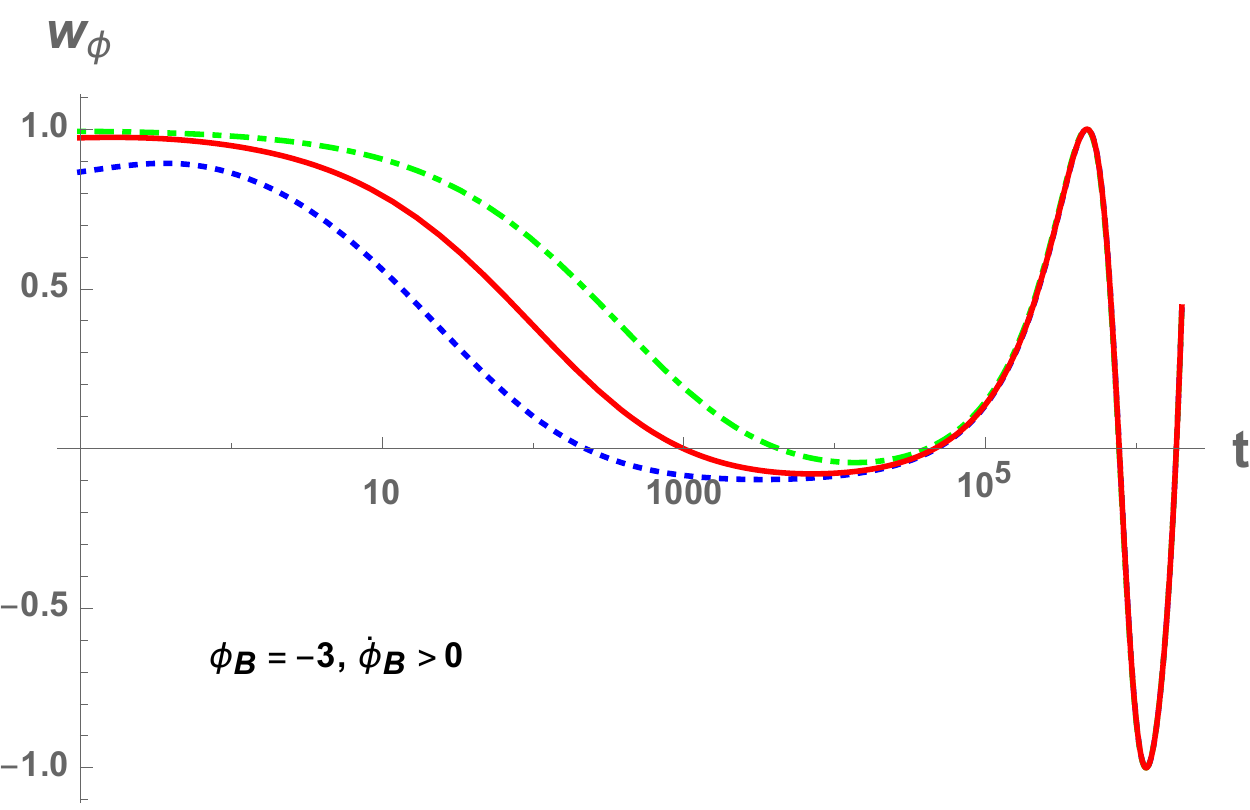}
\includegraphics[width=7cm]{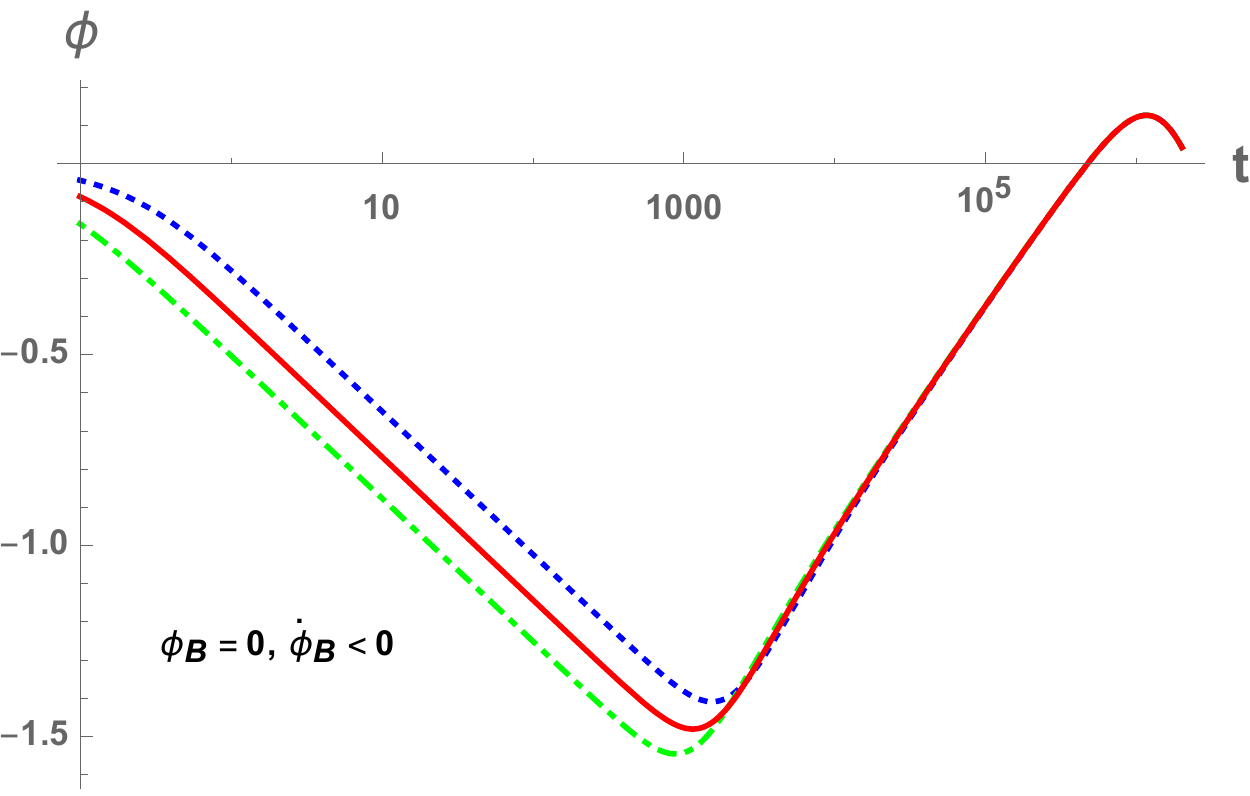}
\includegraphics[width=7cm]{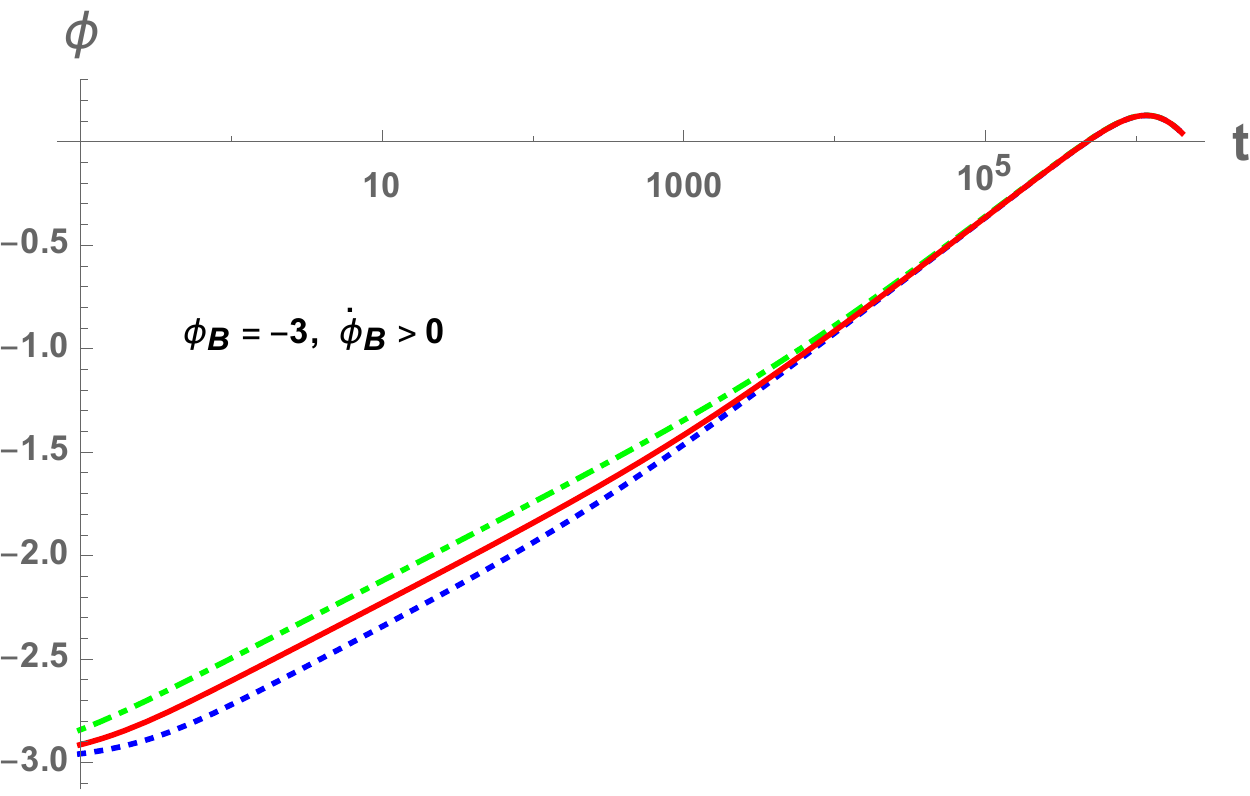}
}
\caption{This figure shows explicitly the situation  when the slow-roll inflation cannot be realized in mLQC (red solid curves), mLQC-I (blue dotted curves) and mLQC-II (green dot-dashed curves) with the Starobinsky potential. The Planck mass is set to unity.  }
\label{fig5}
\end{figure}
\end{widetext}

\section{Analytic Approximations for Kinetic-Energy Dominated Cases}
\label{Section3}
\renewcommand{\theequation}{3.\arabic{equation}}\setcounter{equation}{0}

As shown in the last section, the evolution of the universe before reheating has certain generic features for mLQC-I and mLQC-II when the KE  of the inflaton largely dominates the PE  at the quantum bounce. 
These generic features do not depend on the types of the potentials and the initial conditions of $a(t_B), \phi(t_B)$ and $\dot\phi(t_B)$, as long as the condition 
\bq
\lb{3.0}
\frac{1}{2}\dot\phi^2(t_B) \gg V(\phi_B),
\eq
where $\phi_B \equiv \phi(t_B)$. This is quite similar to the case of LQC \cite{ZWCKS17,ZWCKS16,SSWW17,SSW18a,Sha18,SSWW18b,bcl2018}. In this section, we shall find the analytical solutions of $a(t)$ and $\phi(t)$
in each of these three phases. In particular, we shall assume that the condition (\ref{3.0}) always holds. 

 As to be shown below, by simply replacing $\rho_c$ by $\rho^{A}_c$ in the analytical solutions found in LQC in \cite{ZWCKS17,ZWCKS16} will result in about $7 \sim 10 \%$ errors during the SI phase, 
  which are too large to be comparable with current  observations \cite{Planck2015}. To improve the accuracy, in the Subsection A we shall present a detailed analysis. In addition, to avoid errors that can be caused by the matching
  between the transition and slow-roll inflationary phases, as well as caused by the definition of the end of inflation, a detailed analysis is also provided in the Subsections B and C.

  \begin{figure}
{
\includegraphics[width=8cm]{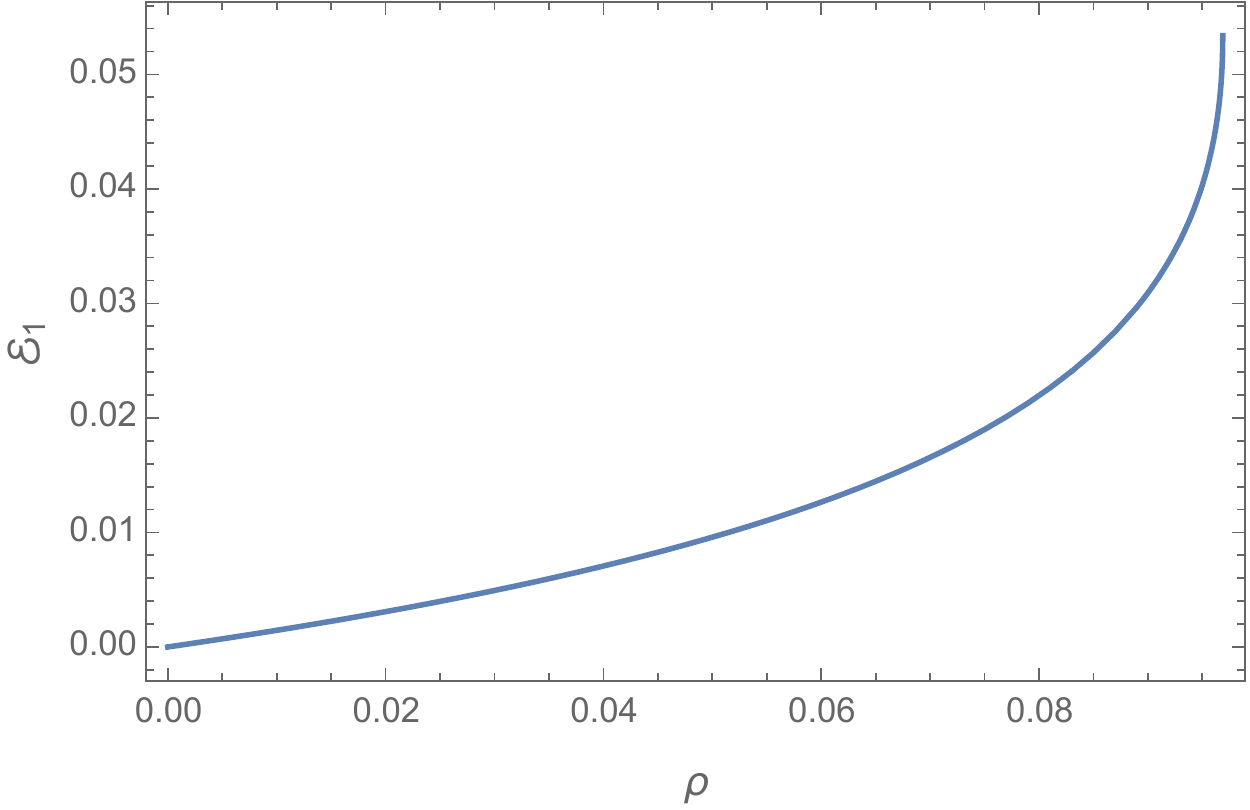}
}
\caption{In this figure,  $\mathcal E_1$ vs $\rho$ is plotted in mLQC-I, from which it can be seen that $\mathcal E_1$ given by
Eq.(\ref{2.6a})  is a monotonically increasing function of the energy density and it reaches its maximum value at the critical energy density $\rho=\rho_c^{{\scriptscriptstyle{\mathrm{I}}}} \simeq 0.097 \rho_{\mathrm{Pl}}$. 
In plotting this figure, we have set $\rho_{\mathrm{Pl}} =1$.}
\label{fig2.1}
\end{figure}

 \subsection {Analytic Solution in the Bouncing Phase}
 
 The main difference during the bouncing regime between mLQC-I and mLQC-II lies in the SI   phase in which the high-order terms in the energy density become important. After SI  ends, the energy density drops drastically, and the differences between mLQC-I/II and LQC become negligible. Therefore, it is safe to assume LQC to be valid towards the end of bounce regime, but there are notable differences near the bounce. We now discuss an appropriate approximation for mLQC-I/II which  keeps the relative error between analytic and numeric results comparatively small in the whole bouncing phase.

 \subsubsection{Bouncing Phase in mLQC-I}
 
In mLQC-I, the effective Hamiltonian gives rise to an asymmetric bounce with the modified FR equations given explicitly by Eqs.(\ref{1.4a})-(\ref{1.4b}) after the bounce and Eqs.(\ref{1.6a})-(\ref{1.6b}) before the bounce.   In the following, we shall focus 
  on the Klein-Gordon equation (\ref{ecl}) and the modified Friedmann equations (\ref{1.4a}) and (\ref{1.6a}), which can be cast in the form,   
 \bqn
\lb{2.5}
H^2 &=&
\frac{8\pi G \rho}{3}\left(1-\frac{\rho}{\rho^{\scriptscriptstyle{\mathrm{I}}}_c}\right)\nb\\&&
\times  \left\{\begin{array}{lr} \left(1  + \mathcal E_1\right), & t \ge t_B,\\
\frac{\alpha \rho_\Lambda}{\rho} \left(1+\mathcal E_2\right), & t \le t_B,\\
\end{array}\right.
\eqn
where   
\bqn
\lb{2.6a}
\mathcal E_1&\equiv&\frac{\gamma^2}{\gamma^2+1}\left(\frac{\sqrt{\rho/\rho^{\scriptscriptstyle{\mathrm{I}}}_c}}{1 +\sqrt{1-\rho/\rho^{\scriptscriptstyle{\mathrm{I}}}_c}}\right)^2,\\
\lb{2.6b}
\mathcal E_2&\equiv&\left(\frac{1-2\gamma^2+\sqrt{1-\rho/\rho^{\scriptscriptstyle{\mathrm{I}}}_c}}{4\gamma^2\left(1+\sqrt{1-\rho/\rho^{\scriptscriptstyle{\mathrm{I}}}_c}\right)}\right)\frac{\rho}{\rho^{\scriptscriptstyle{\mathrm{I}}}_c}.
\eqn
In the post-bounce phase when the bounce is dominated by the KE of the scalar field, it shall dominate the evolution of the universe in the whole bouncing phase, as shown explicitly in the last section. Then, we have $\rho \simeq P$, and 
Eq.(\ref{ecl}) yields,  
\bq
\lb{2.6c}
\rho \simeq \frac{\rho^{\scriptscriptstyle{\mathrm{I}}}_c}{a^6},
\eq
where the scale factor at the bounce is set to unity ($a_B = 1$). Now plugging the above relation into Eq. (\ref{2.5}), one can integrate out the modified  Friedmann equation explicitly to get
\bqn
\lb{a2}
&&\sqrt{96\pi G  \rho^{\scriptscriptstyle{\mathrm{I}}}_c(\gamma^2+1)}t=\xi -\sqrt{2\gamma^2+1}\nb\\
&&~~~~~~~+\frac{1+\gamma^2}{2\gamma}\ln\left[\frac{\left(\xi-\gamma\right)\left(\sqrt{2\gamma^2+1}+\gamma\right)}{\left(\xi+\gamma\right)\left(\sqrt{2\gamma^2+1}-\gamma\right)}\right],
\eqn
where the scale factor is related to $\xi$ via
\bq
\lb{a3}
a(\xi)=\left[\frac{\left(1+\xi^2\right)^2}{4\left(\gamma^2+1\right)\left(\xi^2-\gamma^2\right)}\right]^{1/6}.
\eq
Unfortunately, with this exact solution, there does not exist  a closed form of $\xi$ in terms of time $t$,  and hence no closed form of the scale factor could be found. 

In search of an approximate solution, we can consider the Taylor expansion of $\mathcal E_1$ in Eq. (\ref{2.6a}) around the bounce  as 
\bq
\lb{2.7}
\mathcal E_1=\frac{\gamma^2}{1+\gamma^2}-\frac{2\gamma^2}{1+\gamma^2}\left(1-\frac{\rho}{ \rho^{\scriptscriptstyle{\mathrm{I}}}_c}\right)^{1/2}+\mathcal O\left(\rho - \rho^{\scriptscriptstyle{\mathrm{I}}}_c\right),
\eq
where the leading term $\gamma^2/(\gamma^2+1) \approx 0.053$ for $\gamma = 0.2375$. Moreover, $\mathcal E_1$ turns out to be a monotonous function of the energy density as depicted in Fig. \ref{fig2.1}. At the bounce when $\rho =  \rho_{c}^{{\scriptscriptstyle{\mathrm{I}}}}$, it reaches the maximal value $\mathcal E^{\text{max}}_1 \simeq 0.053$.  As a result, we can first consider the approximation
 \bqn
\lb{2.5a}
H^2 \simeq
\frac{8\pi G \rho}{3}\left(1-\frac{\rho}{\rho^{\scriptscriptstyle{\mathrm{I}}}_c}\right),
\eqn
since it takes the same form as  the modified Friedmann equation in LQC with the substitution $\rho_c \rightarrow \rho^{\scriptscriptstyle{\mathrm{I}}}_c$. The solution of Eq. (\ref{2.5a}) for the KE dominated bounce is well-known and the scale factor and the scalar field in the bouncing phase are  given by 
\bqn
\lb{2.8}
a(t)&=&\left[1+ 24 \pi G \rho_{\text{c}}^{\scriptscriptstyle{\mathrm{I}}} t^2\right]^{1/6},\nb\\
\phi(t)&=& \phi_{\text{B}} \pm \frac{m_{\mathrm{Pl}} }{2\sqrt{3\pi}} \text{arcsinh}{\left(\sqrt{24\pi G  \rho_{\text{c}}^{\scriptscriptstyle{\mathrm{I}}}  }t\right)},
\eqn
where the scale factor at the bounce is again set to unity and $ \phi_{\text{B}}$ stands for the value of the scalar field at the bounce. The `$\pm$' sign corresponds to the positive/negative initial velocity respectively.  

\begin{figure}
{

\includegraphics[width=8cm]{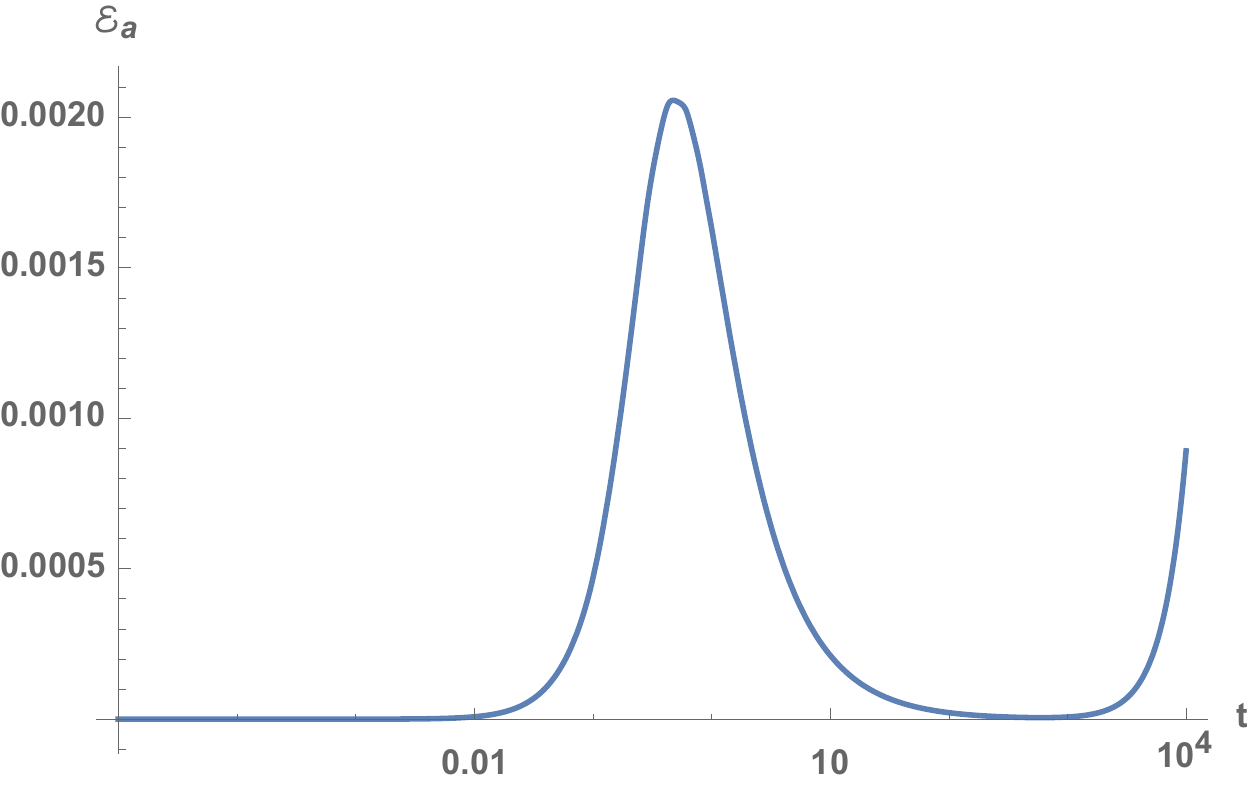}
\includegraphics[width=8cm]{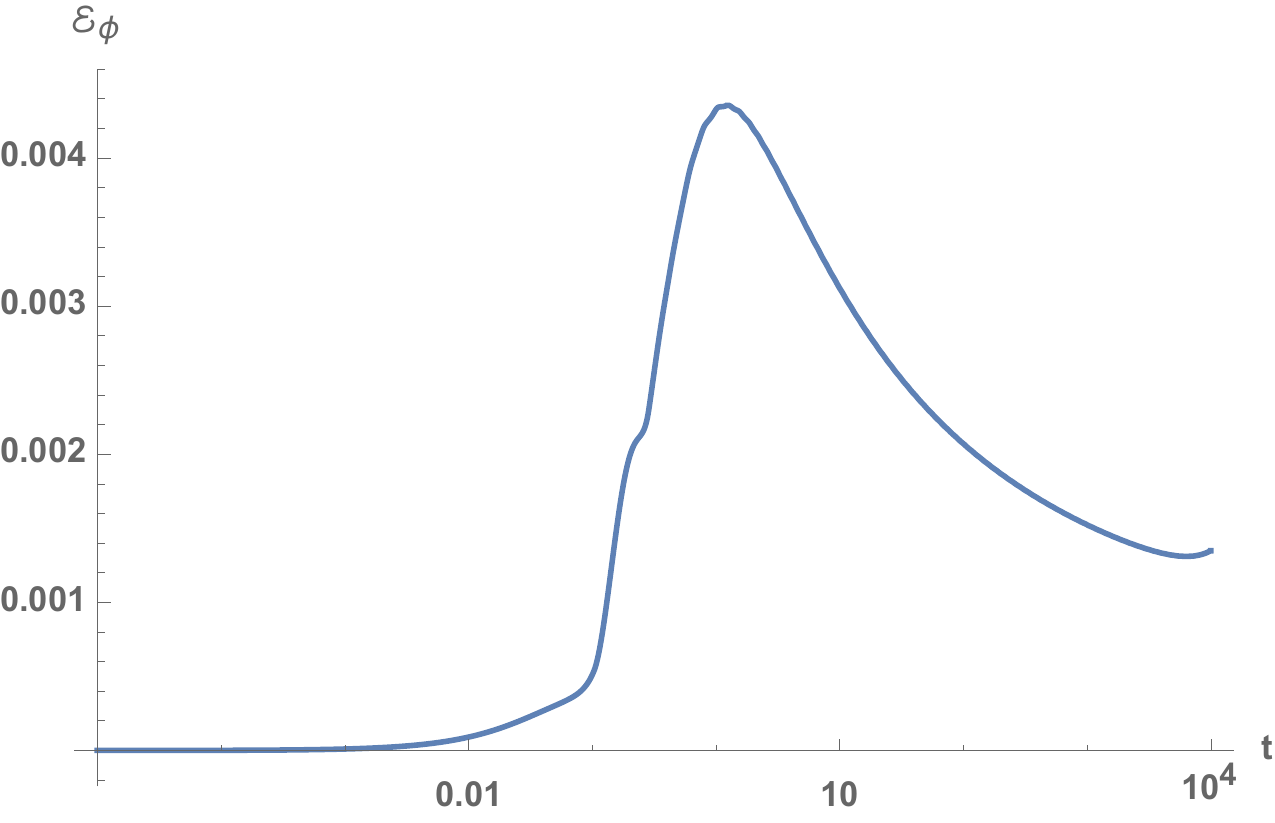}
\includegraphics[width=8cm]{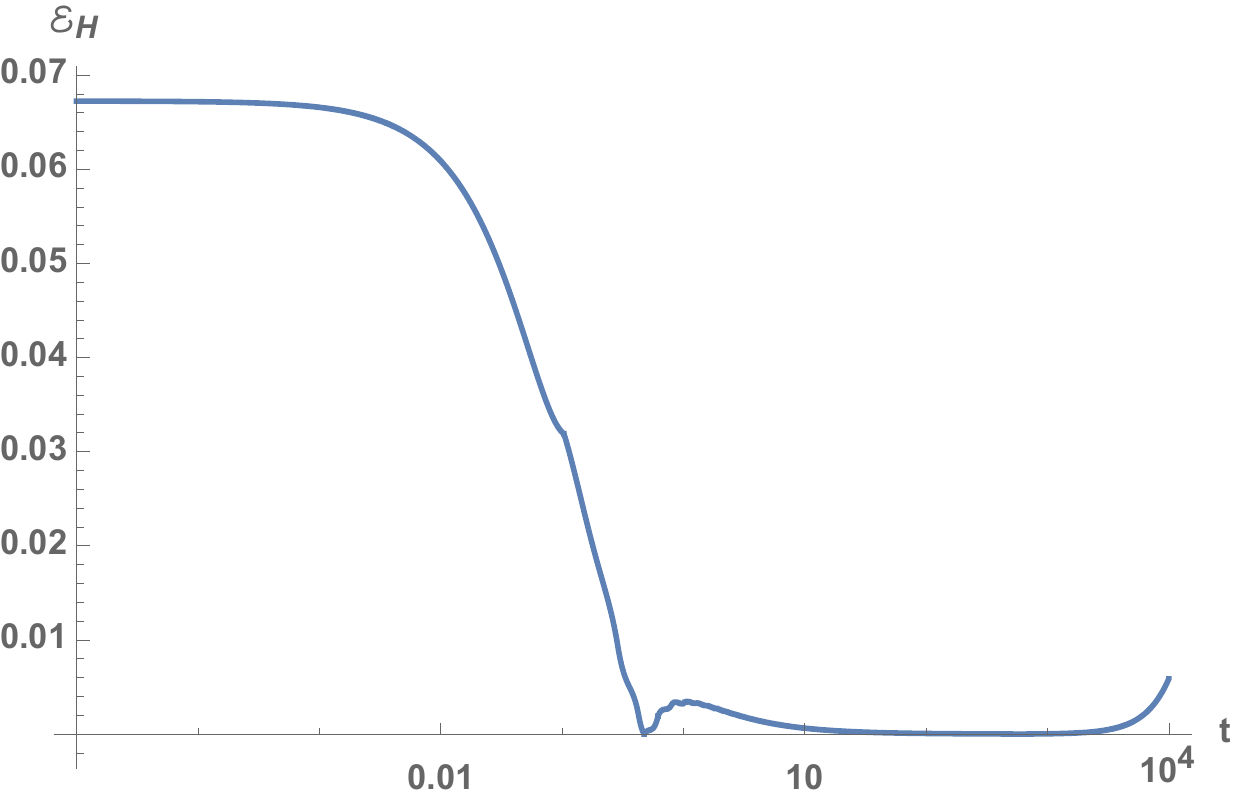}
\includegraphics[width=8cm]{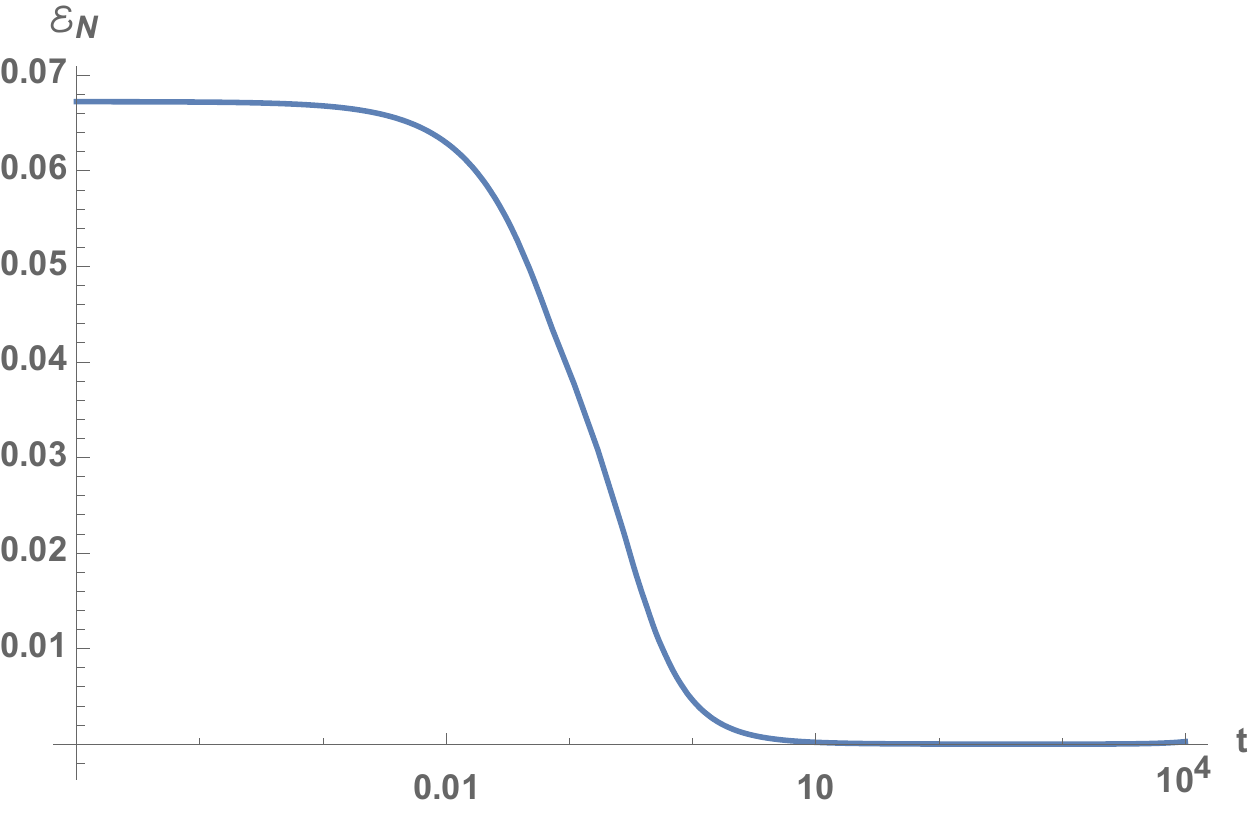}
}
\caption{Relative errors of the scale factor, scalar field,  Hubble rate, and e-folds  between numeric simulations and the approximation Eq. (\ref{2.8}) are computed until $t=10^4$ when the universe is filled with a single scalar field with the chaotic potential. The mass of the scalar field is set to $1.26\times10^{-6}$ and the initial conditions are chosen at the bounce with $\phi_B=0$ and $\dot \phi_B>0$.}
\label{f1}
\end{figure}

To understand the limitation of the approximation (\ref{2.5a}), we compare the numerical simulations with the analytic approximations (\ref{2.8}) in Fig. \ref{f1} where the relative errors for the scale factor, the Hubble rate, e-folds and the scalar field are shown explicitly. The relative error  $\mathcal E_{Q}$ between the exact  $Q_e$ and numerical $Q_n$ solutions  is defined by 
\bq
\lb{errors}
\mathcal E_{Q}\equiv 2\left|\frac{Q_e-Q_n}{Q_e +Q_n}\right|,
\eq
where $Q = \left\{a, \phi, H, N\right\}$.  The largest relative error of the Hubble rate and the e-folds occurs right at the bounce as shown explicitly in Fig. \ref{f1}, due to the   dropping of   the $\mathcal E_1$ term in the approximation (\ref{2.8}). The amount of the largest error which is about $7\%$  is independent of the form of the potential and the magnitude of the scalar field as long as the bounce is KE dominated. Besides, at a later time, say $t>10$, the approximation (\ref{2.8}) can be regarded as a very good approximation since the relative errors are less than $1\%$. Therefore, only some improvement near the bounce especially during the SI phase is required for a better accuracy. 

\begin{figure}
{
\includegraphics[width=8cm]{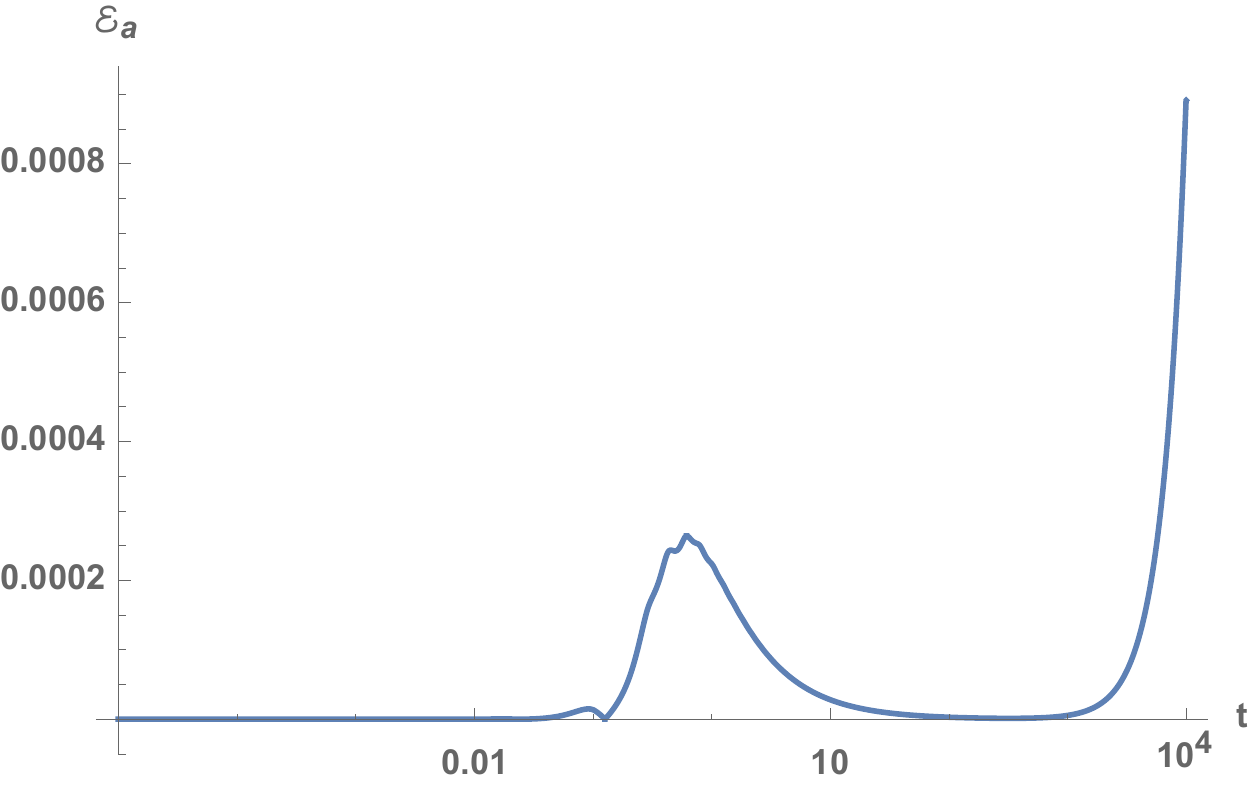}
\includegraphics[width=8cm]{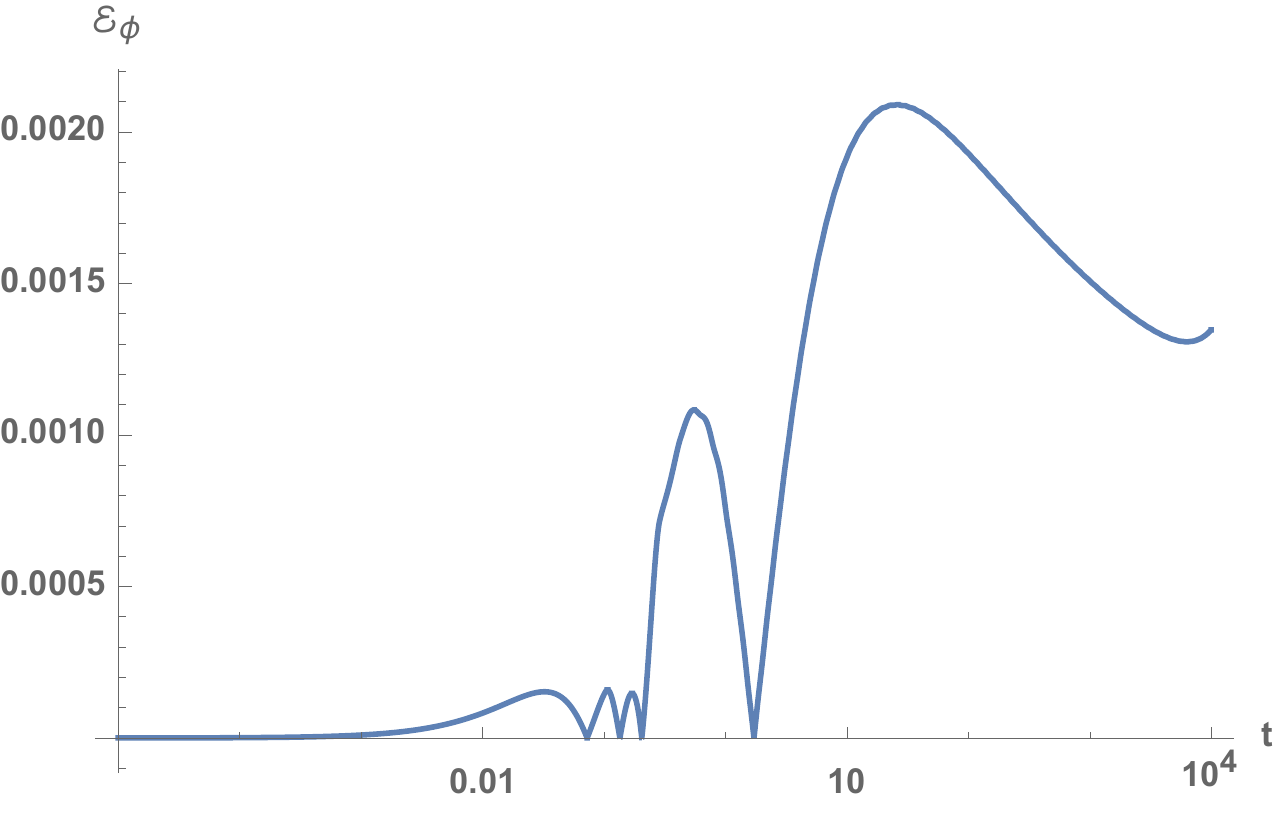}
\includegraphics[width=8cm]{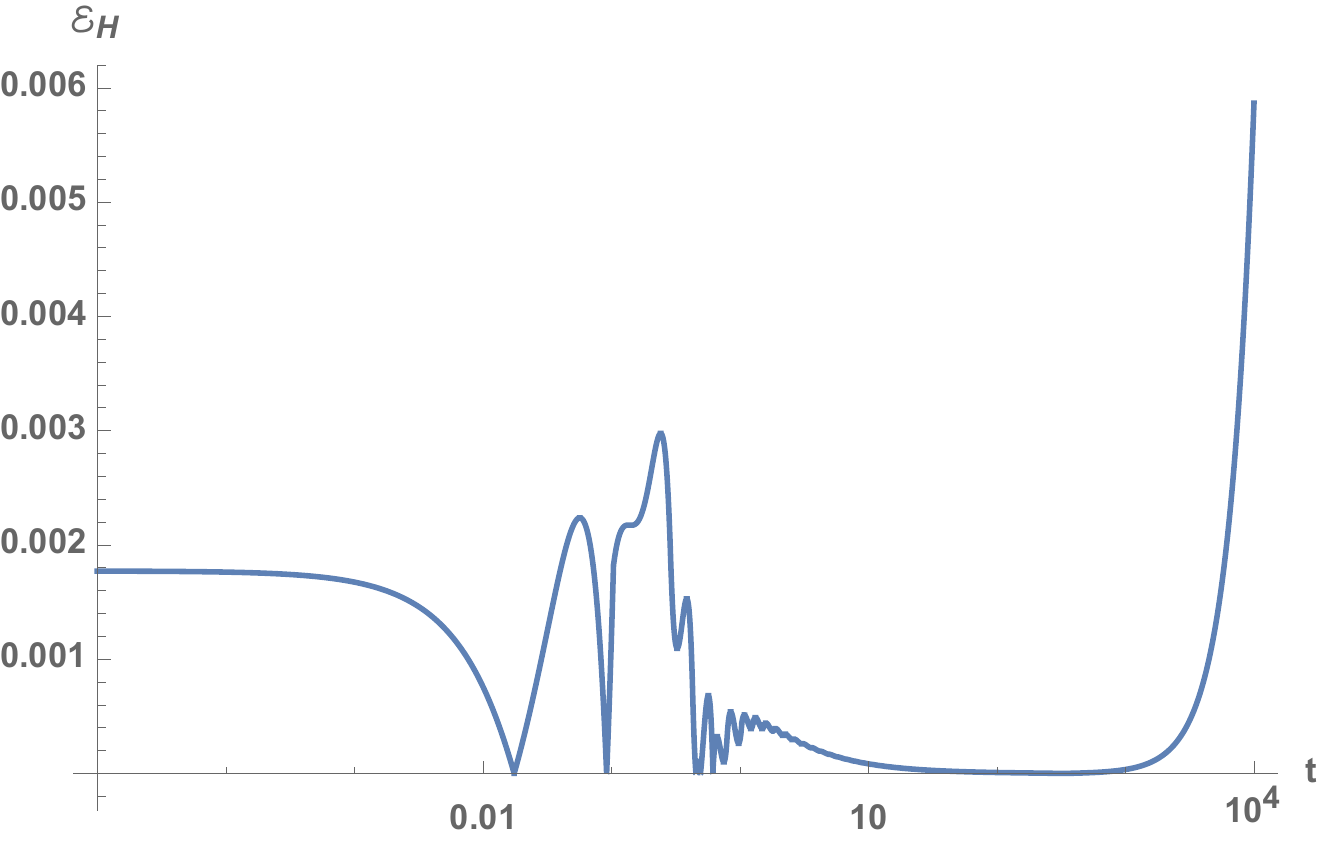}
\includegraphics[width=8cm]{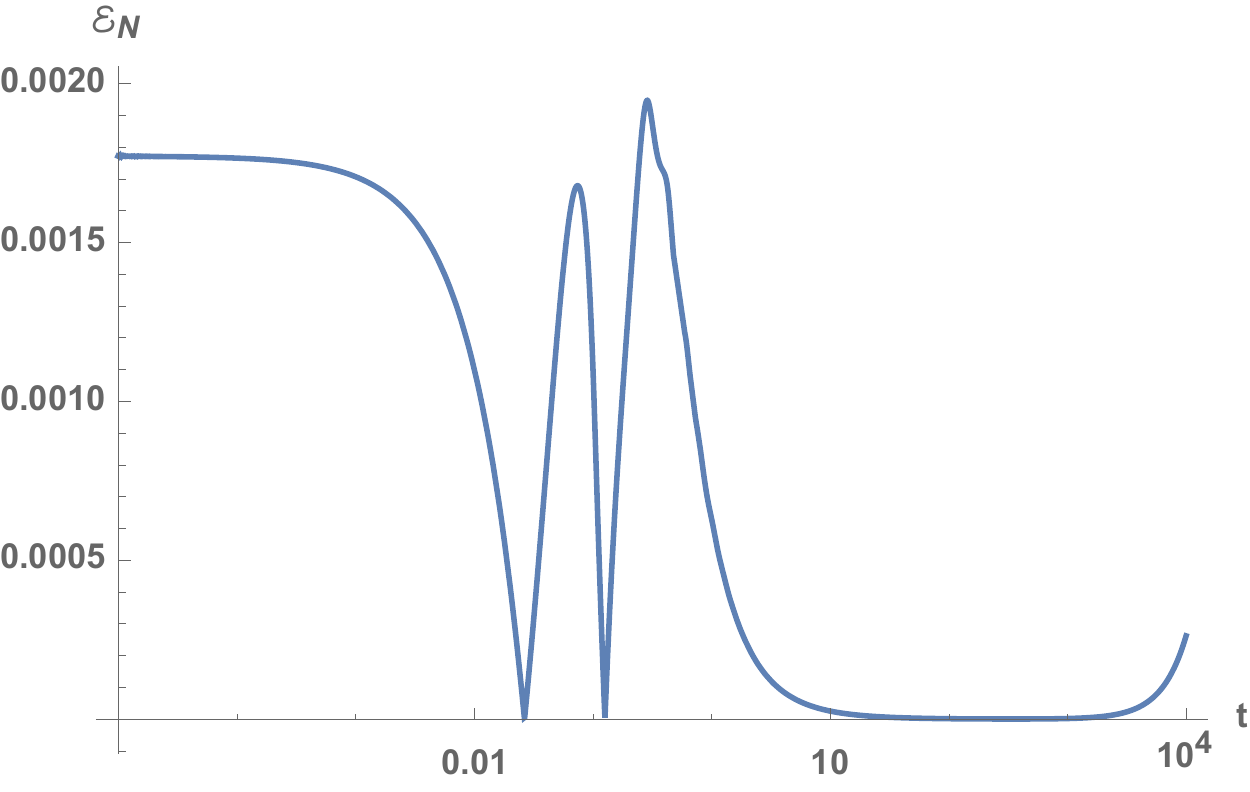}

}
\caption{The initial conditions of the numeric simulations are chosen the same as in Fig. \ref{f1}. The relative errors of the same quantities between the modified solutions (\ref{2.8m}) and numeric simulations are shown in the figure  for the chaotic potential.}
\label{f2}
\end{figure}

In order to reduce the error during the SI phase, one can approximate the RHS of Eq. (\ref{a2})  with a simpler function of $\xi$ to obtain a closed form of $\xi$. Another approach is to modify the approximate solution Eq. (\ref{2.8}) with reasonable corrections. Considering this approach, we find  that a better approximation can be achieved by the modified solution 
\bqn
\lb{2.8m}
a(t)&=&\left[1+ 24 \pi G \rho_{\text{c}}^{\scriptscriptstyle{\mathrm{I}}} \left(1+\frac{A \gamma^2}{1+B t}\right)t^2\right]^{1/6},\nb\\
\phi(t)&=& \phi_{\text{B}} \pm \frac{m_{\mathrm{Pl}}  \text{arcsinh}{\left(\sqrt{24\pi G  \rho_{\text{c}}^{\scriptscriptstyle{\mathrm{I}}} \left(1+\frac{C \gamma^2}{1+Dt}\right) }t\right)}}{\sqrt{12\pi G \left(1+\frac{C \gamma^2}{1+D t}\right)}},\nb\\
\eqn
where the parameters $A$, $B$, $C$ and $D$ are fixed through numerical simulations. 
The best fitting   is provided by 
\bqn
\lb{fittingA}
A=C=1.2, \quad B=6, \quad D=2. 
\eqn
In the simulations, we have found that the choices of $A$ and $C$ are closely related with  the values of $\mathcal E_H$ and $\mathcal E_\phi$ near the bounce, while $B$ and $D$ affect the height of the maximums of $\mathcal E_H$ and $\mathcal E_\phi$ in the SI phase. More specifically, in mLQC-I, we choose $A=1.2$ so that the initial error of $\mathcal E_H$ in Fig. \ref{f2} is less than $0.2\%$ which is minimized in comparison to other choices of $A$. Besides, $B=6$ also minimizes the values of two peaks in the interval  $t\in(0.01, 1)$ which turn out to be less than $0.3\%$. Similar considerations were applied to the scalar field and the choices of $C$ and $D$ as well.

With the best fitting parameters in Eq. (\ref{fittingA}), we compare the relative errors of our modified solution (\ref{2.8m}) with the numeric simulations in Fig. \ref{f2}.  As can been seen from this figure, a noticeable improvement is made in accuracy. The largest error of the Hubble rate/e-folds is now reduced to around $0.3\%$/$0.2\%$ in the SI phase. This improvement is due to the asymptotic behavior of solution (\ref{2.8m}): when $t\rightarrow 0$, $a\rightarrow \left[1+24\pi G \rho_{\text{c}}^{\scriptscriptstyle{\mathrm{I}}} \left(1+A\gamma^2\right)t^2 \right]^{1/6}$. The correction  $A \gamma^2$ is thus vital in reducing the error in the SI phase. Similar considerations also apply to the scalar field. 

In Tables \ref{t1} and  \ref{star1}, we compare our analytical solutions with numerical results for various initial conditions. From these tables it can be seen that our (approximate) analytical solutions track the (exact) numerical ones very well, as long as the condition (\ref{3.0}) is satisfied. 

\subsubsection{Bouncing Phase in mLQC-II}

In mLQC-II,  the  modified FR equations  are given explicitly by Eqs.(\ref{2.6aa}) and (\ref{2.6bb}), using which we find that the matter field  also satisfies the Klein-Gordon equation (\ref{ecl}).
The modified Friedmann equation can be cast in the form,  
\bqn
\lb{2.9}
H^2
=\frac{8\pi G \rho}{3}\left(1-\frac{\rho}{\rho_c^{{\scriptscriptstyle{\mathrm{II}}}}}\right)\left(1+\mathcal E_3\right),
\eqn
where 
\bqn
\lb{2.10}
\mathcal E_3(\rho)&\equiv&\frac{2+8\gamma^2(\gamma^2+1)\rho/\rho_c^{{\scriptscriptstyle{\mathrm{II}}}}}{1+2\gamma^2 \rho/\rho_c^{{\scriptscriptstyle{\mathrm{II}}}}+\sqrt{4\gamma^2(\gamma^2+1)\rho/\rho_c^{{\scriptscriptstyle{\mathrm{II}}}}+1}} -1\nb\\
&=& 2\gamma^2-\frac{\gamma^2(2+\gamma^2)}{1+2\gamma^2}\left(1-\frac{\rho}{\rho_c^{{\scriptscriptstyle{\mathrm{II}}}}}\right)+\mathcal O \left( (\rho-\rho_c^{{\scriptscriptstyle{\mathrm{II}}}} )^2\right).\nb \\
\eqn
Similar to mLQC-I, when the bounce is KE dominated, with the help of Eq. (\ref{2.6c}), the modified Friedmann equation (\ref{2.9}) can be integrated out explicitly and in the current case it yields
\bq
\lb{a5}
\xi-\frac{\gamma^2}{\sqrt{\gamma^2+1}}\arctan \frac{\xi}{\sqrt{1+\gamma^2}}=\sqrt{\frac{24\pi G   \rho_c^{{\scriptscriptstyle{\mathrm{II}}}}}{ 1+\gamma^2}}t .
\eq
But now  the scale factor is given by 
\bq
\lb{a6}
a(\xi)=\left(\frac{(\gamma^2+1)(\xi^2+1)^2}{\xi^2+\gamma^2+1}\right)^{1/6}.
\eq
As in the case of mLQC-I, from the above two relations, no closed form of the scale factor can be found. 

\begin{figure}
{
\includegraphics[width=8cm]{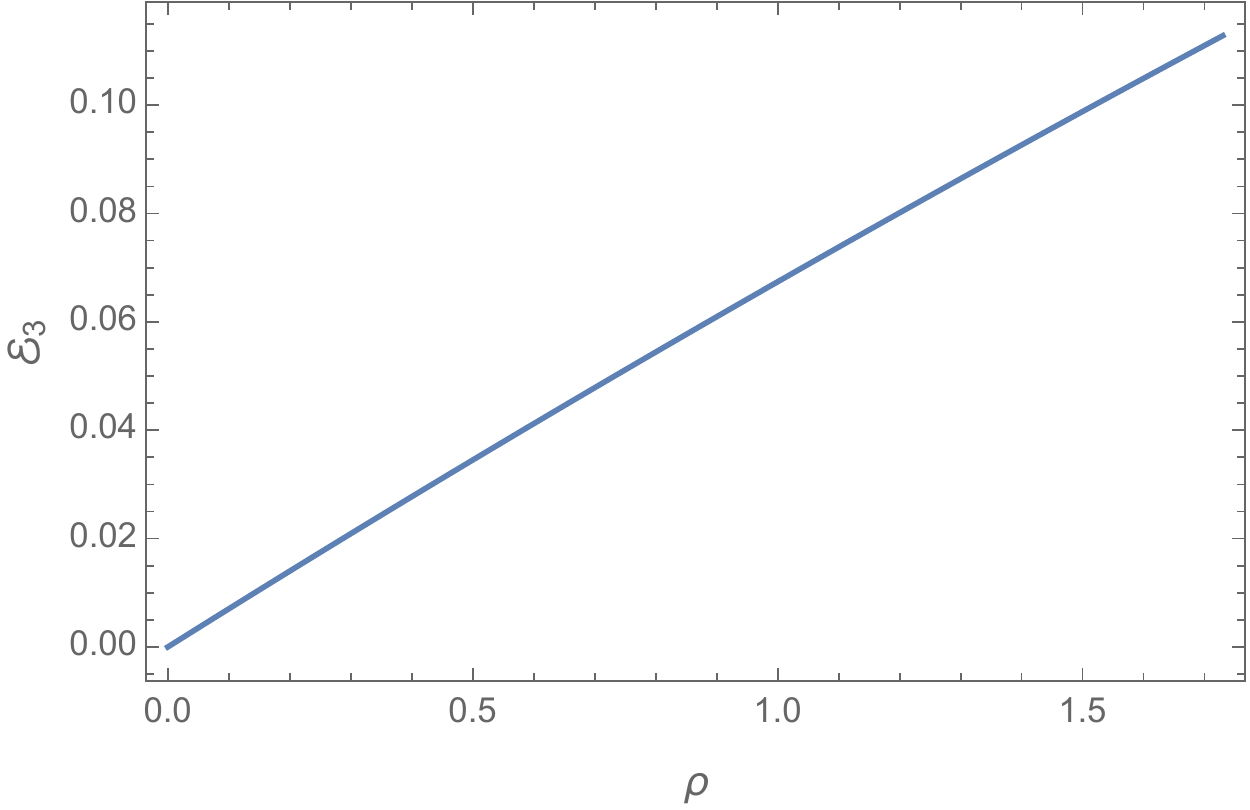}
}
\caption{This figure is for  mLQC-II, from which it can be seen  that $\mathcal E_3$ defined by Eq. (\ref{2.10}) is monotonically increasing function of the energy density, and reaches its maximum value at the critical energy density $\rho=\rho_c^{{\scriptscriptstyle{\mathrm{II}}}}\approx 1.73  m^4_\text{Pl}$. }
\label{fig2.2}
\end{figure}

Similar to mLQC-I, we can first ignore the $\mathcal E_3$ term, such that the solutions are given by,
\bqn
\lb{2.8aa}
a(t)&=&\left[1+ 24 \pi G \rho_{\text{c}}^{\scriptscriptstyle{\mathrm{II}}} t^2\right]^{1/6},\nb\\
\phi(t)&=& \phi_{\text{B}} \pm \frac{m_{\mathrm{Pl}} }{2\sqrt{3\pi}} \text{arcsinh}{\left(\sqrt{24\pi G  \rho_{\text{c}}^{\scriptscriptstyle{\mathrm{II}}}  }t\right)}.
\eqn
As depicted in Fig. \ref{fig2.2},  $\mathcal E_3$ is a monotonous function of the energy density and it takes the maximum value ($\approx 0.11$) right at the bounce.  As a result, simply disregarding $\mathcal E_3$ in the analytic approximation would generate a large  relative error of the Hubble rate in the SI phase (more than $10\%$) which will in turn cause errors in the intermediate regime of the power spectra if we use  the above analytic solution directly. 

To reduce the error in the SI phase, one can use the modified solution given by Eq. (\ref{2.8m}) with the substitution $\rho_c^{{\scriptscriptstyle{\mathrm{I}}}}\rightarrow \rho_c^{{\scriptscriptstyle{\mathrm{II}}}}$, i.e., 
\bqn
\lb{2.8mA}
a(t)&=&\left[1+ 24 \pi G \rho_{\text{c}}^{\scriptscriptstyle{\mathrm{II}}} \left(1+\frac{A \gamma^2}{1+B t}\right)t^2\right]^{1/6},\nb\\
\phi(t)&=& \phi_{\text{B}} \pm \frac{m_{\mathrm{Pl}}  \text{arcsinh}{\left(\sqrt{24\pi G  \rho_{\text{c}}^{\scriptscriptstyle{\mathrm{II}}} \left(1+\frac{C \gamma^2}{1+Dt}\right) }t\right)}}{\sqrt{12\pi G \left(1+\frac{C \gamma^2}{1+D t}\right)}}.\nb\\
\eqn
 Then, we find that the best fitting is given by,  
\bq
\lb{parameter2}
A=2.5, \quad B=7, \quad C=D=2.
\eq 
These parameters can be fixed in the same way as in mLQC-I. The values of $A$ and $B$ affect the initial values of $\mathcal E_H$ and $\mathcal E_\phi$   near the bounce while $C$ and $D$ change the maximums of $\mathcal E_H$ and  $\mathcal E_\phi$ in the SI phase. In mLQC-I, we choose $A=2.5$ so that the initial $\mathcal E_H$ is less than $0.1\%$ while $C=2$ keeps the maximum error of the Hubble rate in the SI phase less than $0.3\%$. Similar considerations are applied to the scalar field parameters.

\begin{figure}
{

\includegraphics[width=8cm]{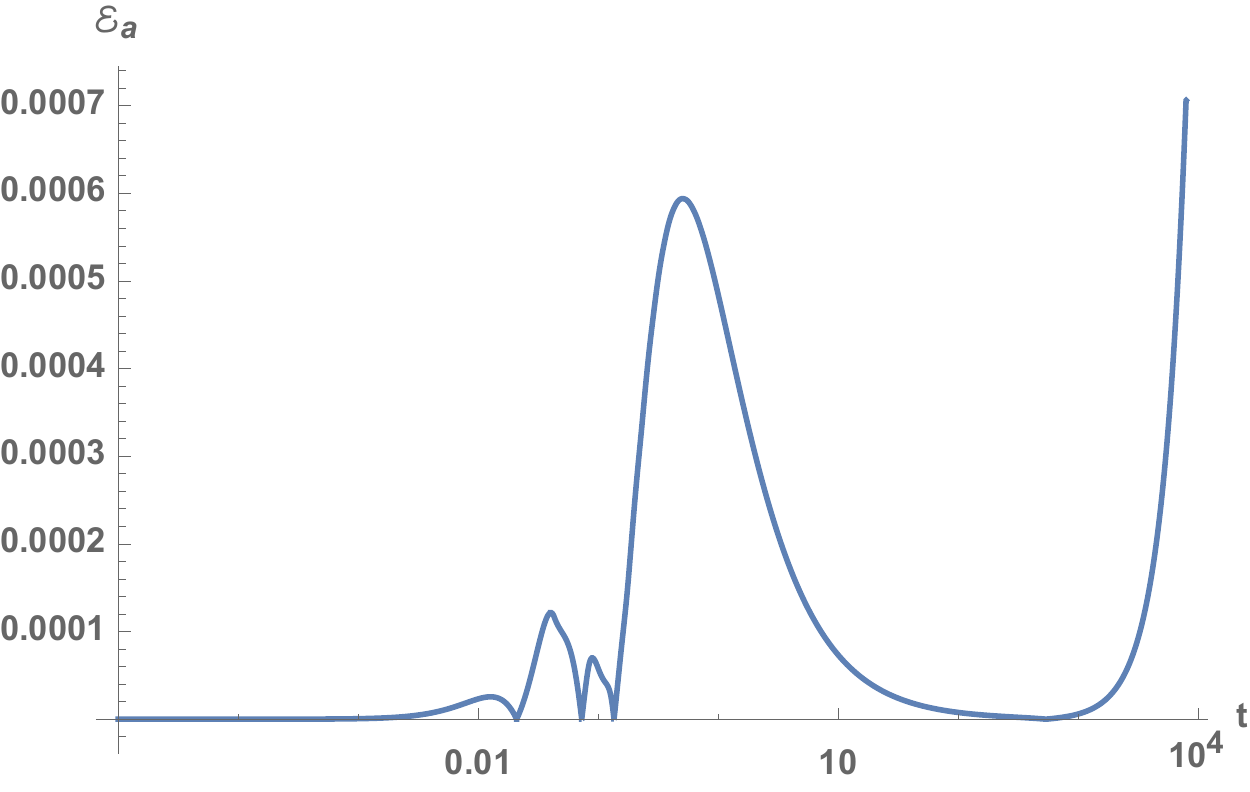}
\includegraphics[width=8cm]{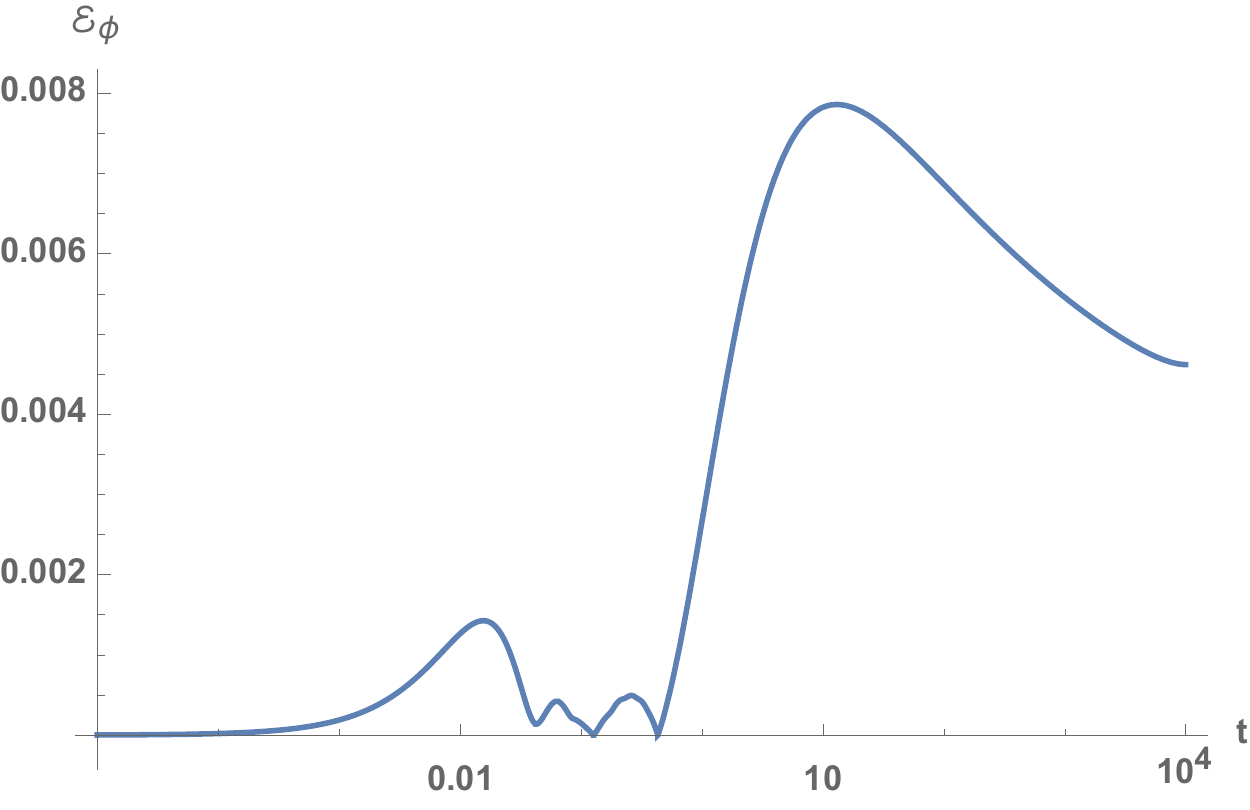}
\includegraphics[width=8cm]{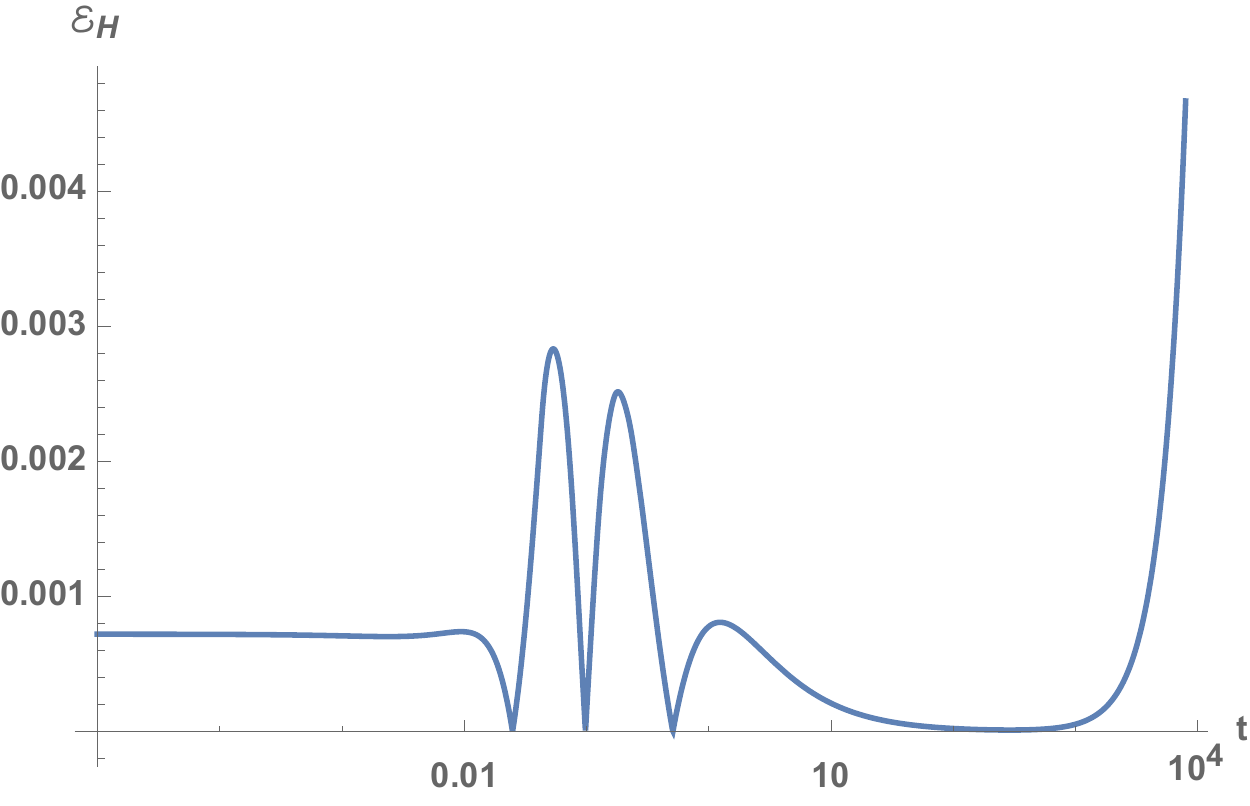}
\includegraphics[width=8cm]{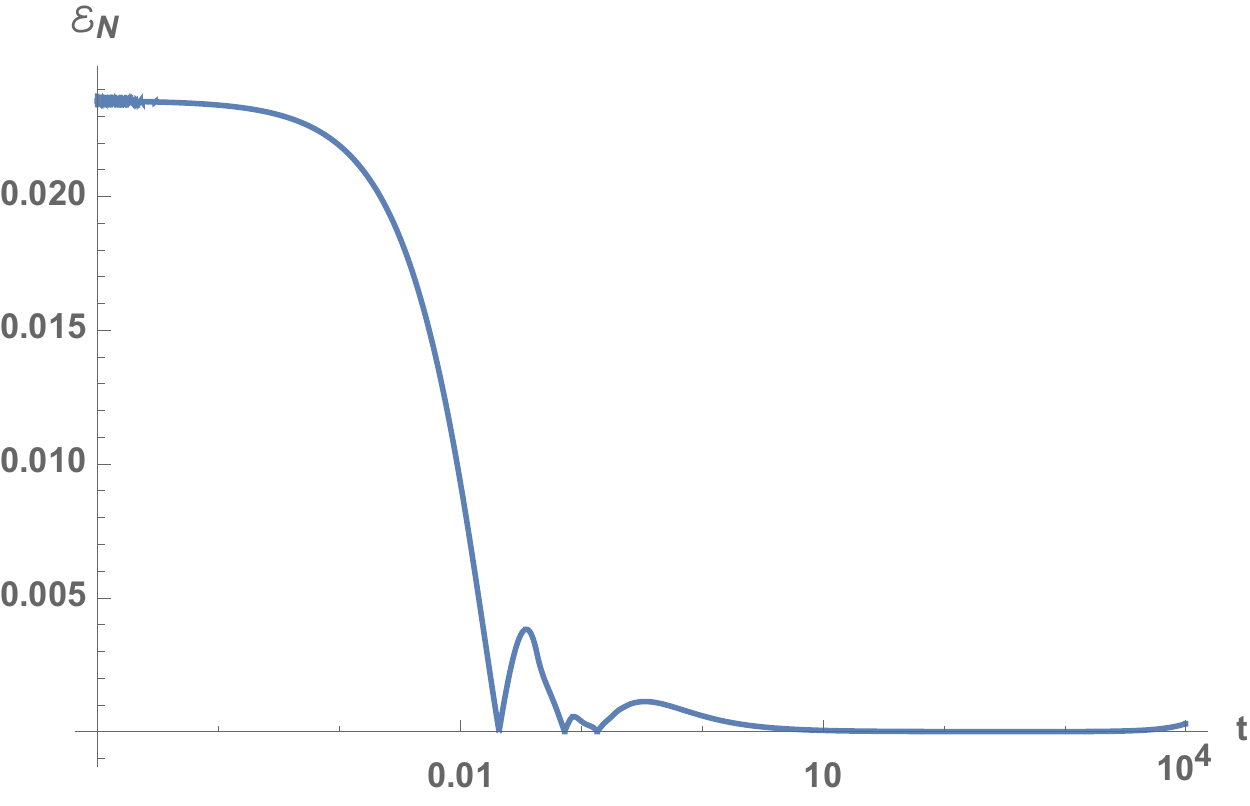}

}
\caption{The initial conditions of the numeric simulations are chosen the same as in Fig. \ref{f1}. The relative errors  between the modified solutions (\ref{2.8mA}) and numeric simulations are shown in the figure for the chaotic potential. In mLQC-II, the parameters are chosen as in Eq. (\ref{parameter2}).}
\label{f3}
\end{figure}

In Fig. \ref{f3},  the relative errors of the scale factor,  Hubble rate, e-folds and  scalar field are depicted until $t=10^4$. The accuracy of our analytic solution (\ref{2.8mA}) is again improved due to the correction term $A\gamma^2$ as $t\rightarrow 0$.

As in mLQC-I, in  Tables \ref{t2} and  \ref{star2}, we compare the numerical and our analytical solutions 
with  various initial conditions. From these tables it can be seen that the (approximate) analytical solutions track the numerical (exact) ones extremely well, provided that   the condition (\ref{3.0}) is satisfied. 

As explained above, this is mainly because that once KE   dominates at the bounce, then it dominates during the entire pre-inflationary phase as it can be seen from Fig. \ref{fig4}. This also happens in LQC \cite{ZWCKS17,ZWCKS16}, and explains well the fact that the evolution of the background of the universe in pre-inflationary regime is generic and independent of the inflationary potentials, as long as KE  dominates at the quantum bounce. This feature is shared by LQC and mLQCs.

 \subsection { Analytic Solutions in Transition Phase }
 
At the  point   $\frac{1}{2}\dot \phi^2_\text{eq}=V(\phi_\text{eq})$, the universe is in the transition phase, in which  the KE  starts to lose its dominant role to the PE   of the inflaton. 
As the typical energy scale in this phase  is around $10^{-12} \rho_\text{Pl}$, the modified  Friedmann equations in mLQC-I/II can be approximated by its classical counterpart, which is,  
\bq
\lb{transition1}
H^2=\frac{8 \pi G}{3} V(\phi).
\eq 
Note  that in the transition phase the change in the magnitude of the scalar field is very small  \cite{bcl2018}. Hence, the potential terms in the Klein-Gordon equation,
\bq
\lb{kleingordon}
\ddot \phi+3 H \dot \phi+V_{,\phi}=0,
\eq
can be Taylor expanded around the equilibrium point $\phi=\phi_\text{eq}$. To get the the approximate solution, only the first-order terms in $(\phi-\phi_\text{eq})$ are considered, where the subscript `eq' denotes quantities at the  point KE = PE. In this way, the scalar field  can be described effectively by a damped harmonic oscillator.   
In particular, the analytic solution of $\phi$ can now be expressed as 
\bq
\lb{transition2}
\phi(t)=\psi(t)-  \left(\frac{V'_\text{eq}}{\beta_\text{eq}} - \phi_\text{eq}\right),
\eq
where $\psi(t)$ is given by 
\bq
\psi= A_+ e^{-\lambda_+ \left(t-t_\text{eq}\right)}+A_- e^{-\lambda_- \left(t-t_\text{eq}\right)},
\eq
and 
\bqn
&&\lambda_\pm=\frac{1}{2}\left(\alpha_\text{eq} \pm\sqrt{\alpha_\text{eq}^2-4\beta_\text{eq}}\right), \quad A_\pm=\pm\frac{\lambda_\mp \psi_\text{eq}+\dot \psi_\text{eq}}{\lambda_--\lambda_+}, \nb\\  
&& \alpha_\text{eq}=\sqrt{24\pi V_\text{eq}/m^2_{\mathrm{Pl}}}, \quad \quad \quad \quad \beta_\text{eq}=V^{''}_\text{eq}.
\eqn
Soon after the transition phase, the universe enters into the slow-roll inflationary phase when the acceleration of the scalar field becomes very small.

\subsection { Analytic Solutions in slow-roll Inflationary phase}

In this phase, the typical energy scale of the scalar field is below $10^{-12} \rho_\text{Pl}$. As a result, the analytic approximations of physical observables in mLQC-I/II are the same as in GR 
 in which the well known slow-roll phase is characterized by two conditions,
\bq
(i) ~~\frac{1}{2}\dot \phi^2 \ll V(\phi), \quad \quad  (ii) ~~ |\ddot \phi|\ll |H\dot \phi |.
\eq
Therefore, the evolution equations in this regime can be well-approximated by 
\bqn
\lb{slowroll1}
H^2&=&\frac{8\pi G}{3} V(\phi), \\
\lb{slowroll2}
3 H \dot \phi&+&\frac{d V(\phi)}{d\phi}=0.
\eqn
Depending on the specific form of the inflationary potential, one can solve for $\phi$ and the Hubble rate with the help of two slow-roll conditions. In particular, it can be easily shown that in the chaotic inflation, the approximate solution of the scalar field turns out to be 
\bq
\lb{chaoticinf}
\phi^{\text{chaotic}}_\text{inf}=\phi_i\pm\frac{m}{\sqrt{12\pi G}}\left(t-t_i\right),
\eq
here $t_i$, $\phi_i$ represent the time and the value of the scalar field at the onset of the slow roll when $|\eta_H|=0.03$, and `$\pm$' denotes the sign of $\dot \phi_i$. As the $\ddot \phi$ term is ignored in the slow-roll approximation,  there is only one free parameter in the solution (\ref{chaoticinf}) that can be fixed by the value of the scalar field at the onset of inflation, namely, $\phi(t_i)$. Besides, $\phi(t_i)$ is determined by  the analytic solution  (\ref{transition2}) in the transition phase. 
Comparing these solutions one finds a discontinuity in the first derivative with respect to $\phi$. 
To make the first derivative of our piecewise solution also continuous at $t_i$, we can add to Eq. (\ref{chaoticinf}) one exponential term, that is, 
\bq
\phi^{\text{chaotic, new}}_\text{inf}=F e^{- \chi \left(t-t_i\right)},
\eq
with $\chi$ set to a large number, say 100. The value of $F$ can be fixed by requiring the first derivative of $\left( \phi_\text{inf}+\phi^{\text{new}}_\text{inf}\right)$ and Eq. (\ref{transition2}) with respect to $t$ are also continuous at the onset of the inflation. In mLQC-I,  if we choose $\chi=100$, $\phi_B=-0.200, \dot \phi_B=0.440$, then $F=6.50\times10^{-11}$ which implies $\phi^{\text{new}}_\text{inf}$ quickly decays away right after the onset of the slow-roll inflation and thus its impact on the slow-roll inflation is negligible.

For the Starobinsky potential, after a straightforward integration of Eqs. (\ref{slowroll1})-(\ref{slowroll2}), one finds
\bq 
\lb{starinf}
\phi^{\text{star}}_\text{inf}(t)=\sqrt{\frac{3}{16\pi G}}\ln\Big[e^{\sqrt{\frac{16\pi G}{3}}\phi_i}-\frac{2m\left(t-t_i\right)}{3}\Big].
\eq

Let us note some non-trivialities arising from comparison of our approximate solutions with the numerical evolution during the inflationary phase for both of the potentials.  For the chaotic potential, from the last row (labelled by `EOI') of each block in Tables \ref{t1}-\ref{t2}, one can find that the numeric and analytic solutions are close to one another, in particular, the difference of the e-folds is less than three in all the cases shown in the tables. This difference can be understood due to the fact that the inflation always lasts for a  longer period than the slow-roll. Since we have set the moment for the end of inflation at $\epsilon_H=1$, there is always a period of non-slow-roll inflation when $\eta_H\gg 0.03$ right before $\epsilon_H=1$.  In chaotic inflation, the error induced by this non-slow-roll inflationary phase is small as this phase is comparably short lived. For example, we find for Table \ref{t1}, when $\phi_N=-0.200$ and $\dot \phi=0.440$, after the universe enters into the inflationary phase, $\eta_H$ grows up again to $0.03$ at $N_N=28.8$, while the end of inflation (EOI) occurs at $N_N=29.9$. During the last $1.1$ e-folds before the EOI, the slow-roll approximations become invalid. In other words, the $\ddot \phi$ term in the Klein-Gordon equation has to be taken into account in the non-slow-roll inflationary evolution of the universe. The effects of the non-slow-roll inflation do not manifest  with large relative errors for the chaotic potential as $\eta_H\approx \left(\eta_V-\epsilon_V\right)=0$ which implies that the non-slow-roll inflation only lasts for a very short time. 

\begin{figure}
{
\includegraphics[width=8cm]{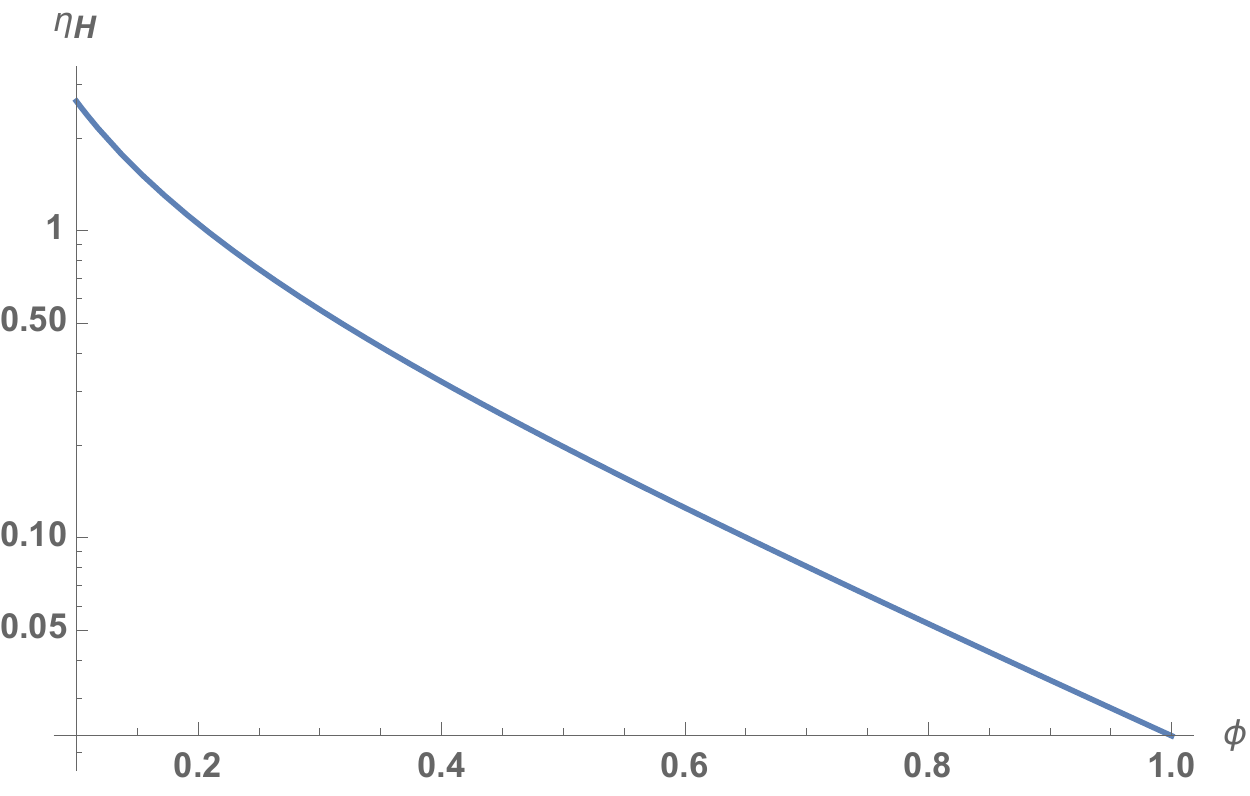}
}
\caption{For Starobinsky potential, the slow-roll inflation ends much earlier that the inflation itself. This figure shows how $\eta_H$ changes when the value of the scalar field is close to the one ($\phi=0.123$) at which the inflation ends. }
\label{etah}
\end{figure}

\begin{figure} 
{
\includegraphics[width=8cm]{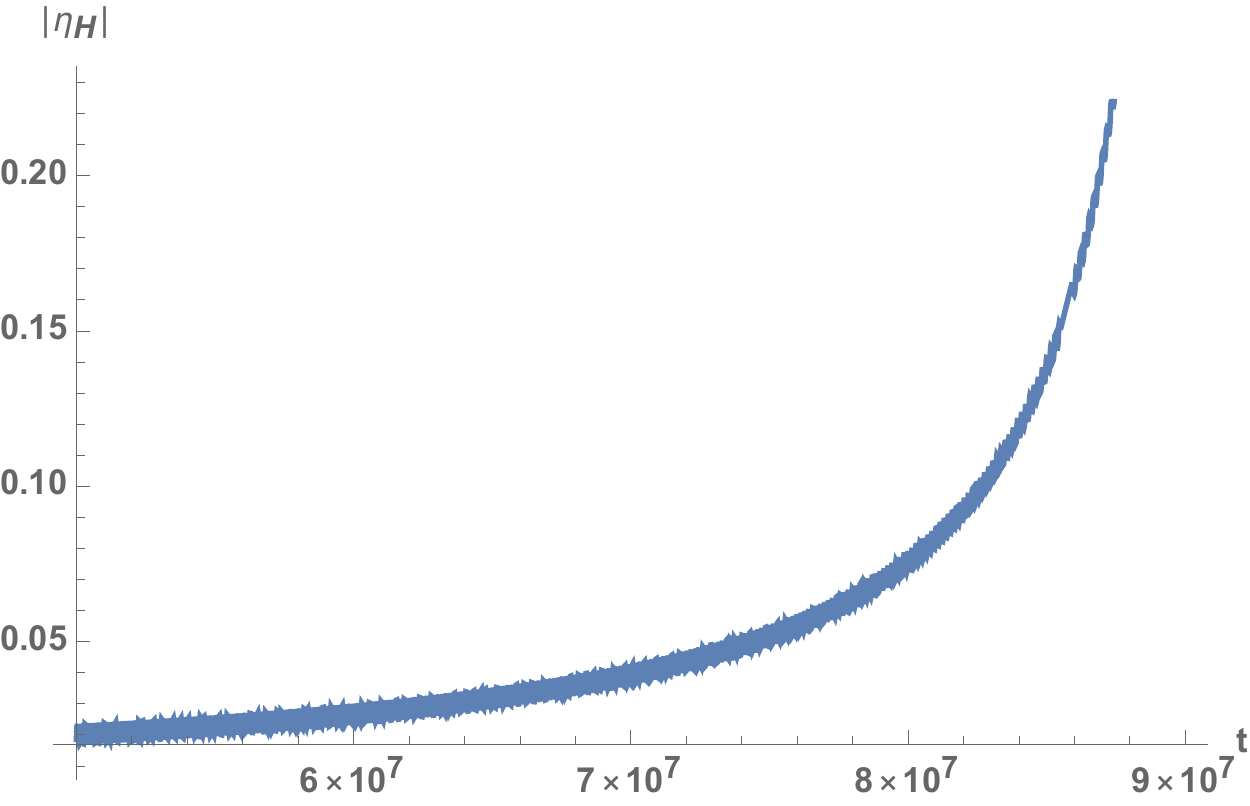}
\includegraphics[width=8cm]{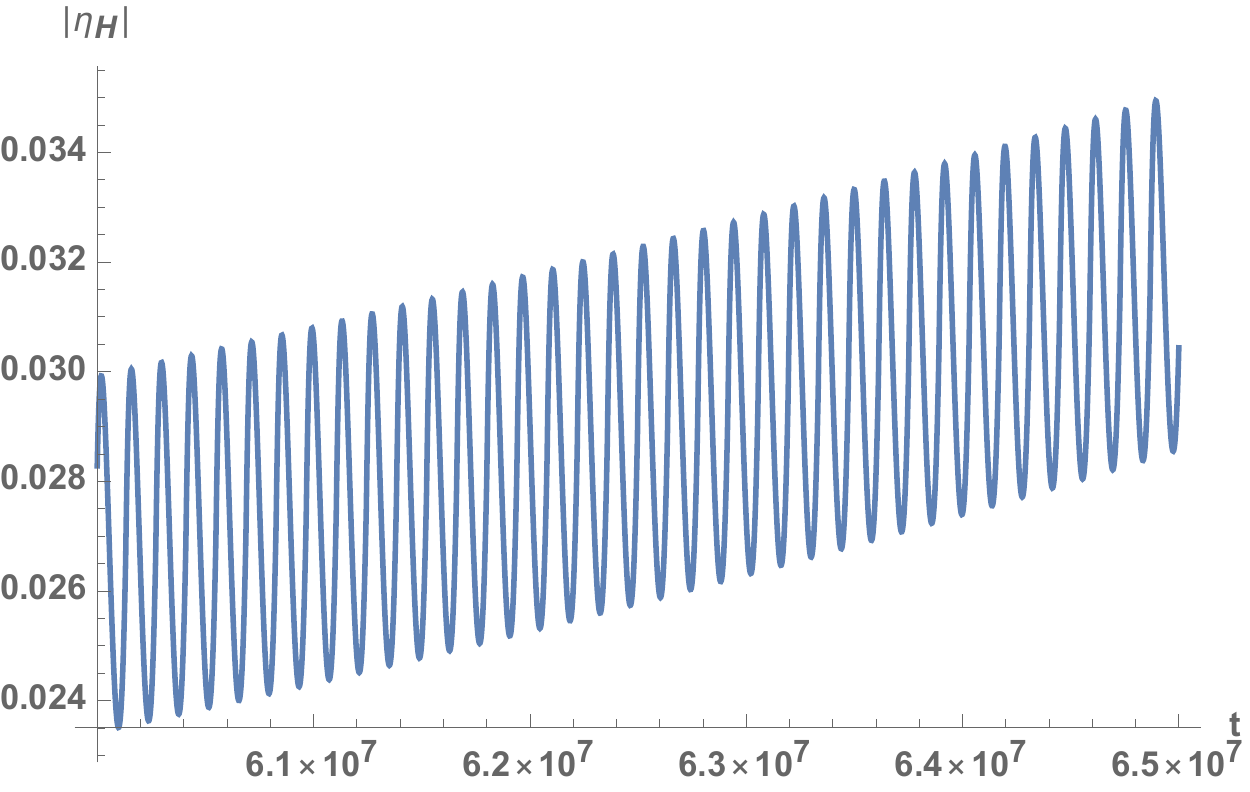}
}
\caption{In mLQC-I,  the initial conditions are set to $\phi_B=-1.20m_\mathrm{Pl}$ and $\dot \phi_B=0.440m_\mathrm{Pl}$. The top panel shows how $|\eta_H|$ changes in the interval $t\in\left(5\times10^7,9\times10^7\right)m^{-1}_\mathrm{Pl}$, while the bottom panel depicts a zoom-in view of $|\eta_H|$ in the interval $t\in\left(6\times10^7,6.5\times10^7\right)m^{-1}_\mathrm{Pl}$.   }
\label{eta1}
\end{figure}

However, the situation is quite different in the Starobinsky potential. In this case there is a considerable amount of e-folds in which the non-slow-roll inflation takes place, as can be seen from 
\bq
\eta_H\approx \left(\eta_V-\epsilon_V\right)=\frac{4}{3-3e^{\sqrt{\frac{16\pi G}{3}}\phi}},
\eq
which is a monotonic function of $\phi$. In this case, $|\eta_H|\le 0.03$ is satisfied only in the interval $\phi\ge 0.932m_\text{Pl}$. Near the end of inflation when $\phi=0.123m_\text{Pl}$ (this number comes from numeric simulations in Tables \ref{star1}-\ref{star2}), $\eta_H$ quickly grows up to unity  as shown in Fig. \ref{etah} and thus the slow-roll condition (\ref{slowroll2}) is violated and the approximate solution (\ref{starinf}) becomes invalid. This is found to be true in numerical simulations. For example,  in mLQC-I, if $\phi_B=-1.20$ and $\dot \phi_B=0.440$, we find that the slow-roll ends at $N_N=78.5$ when $t=6.03\times10^7$, $|\eta_H|=0.03$, $\phi=0.956$ and $H=1.22\times 10^{-6}$. Therefore, in the last $34.5$ e-folds (the total e-folds in this case is $113$), the slow-roll approximation we have used in deriving the solution (\ref{starinf}) results in relative errors to accumulate at an increasing rate. In Fig. \ref{eta1}, we show the behavior of $\eta_H$. This figure shows that there is only one single phase of slow-roll inflation since the averaged value of $\eta_H$ is monotonically increasing even though $\eta_H$ itself is highly oscillating. It turns out that if the analytic solution (\ref{starinf}) is used, then the total e-folds at the end of inflation would be  $139$, much larger than the numerical result $113$. The situation remains similar for the other initial conditions in Tables \ref{star1}-\ref{star2}.  As a result, the analytic solution (\ref{starinf}) must be modified in order to make it fit in the entire inflationary phase, not only the slow-roll part.  This means the acceleration term in the Klein-Gordon equation has to be kept during the non-slow-roll inflationary phase. However, to solve the coupled system of Eqs.  (\ref{transition1})-(\ref{kleingordon}) without any approximations is very difficult. Recall that in the transition phase, we used that $\Delta \phi$ is small so that the Taylor expansion of the equations around the equilibrium point to the linear order in $\phi$ becomes a good approximation. However, we can not apply the same technique for the inflationary phase as the change in $\phi$ can be large. As a result,  many higher-order terms in the powers of $\phi$ would be required to ensure a better accuracy which can in turn make the system rather involved and unsolvable.  Instead we follow a different strategy by appealing to the empirical solution as we have done to find the analytical solution in the bouncing phase. 

\begin{figure}
{
\includegraphics[width=8cm]{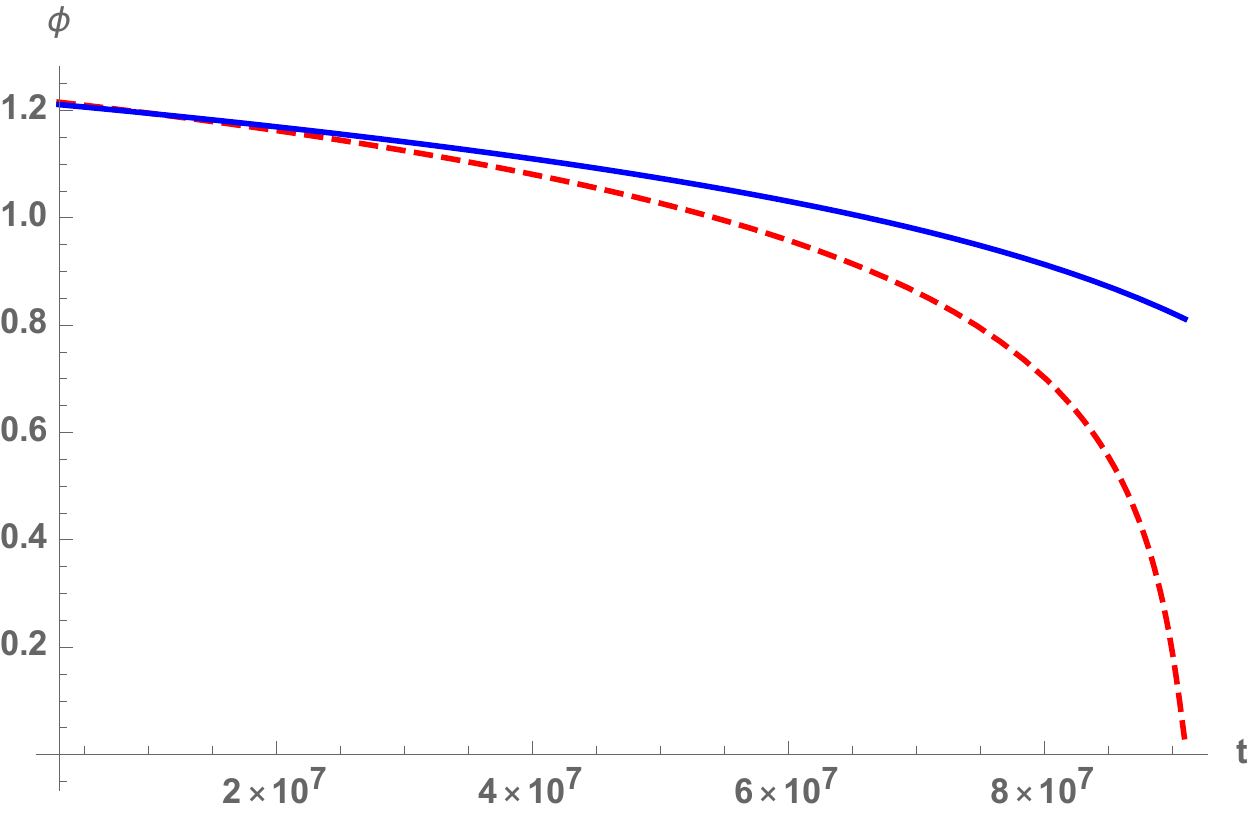}
\includegraphics[width=8cm]{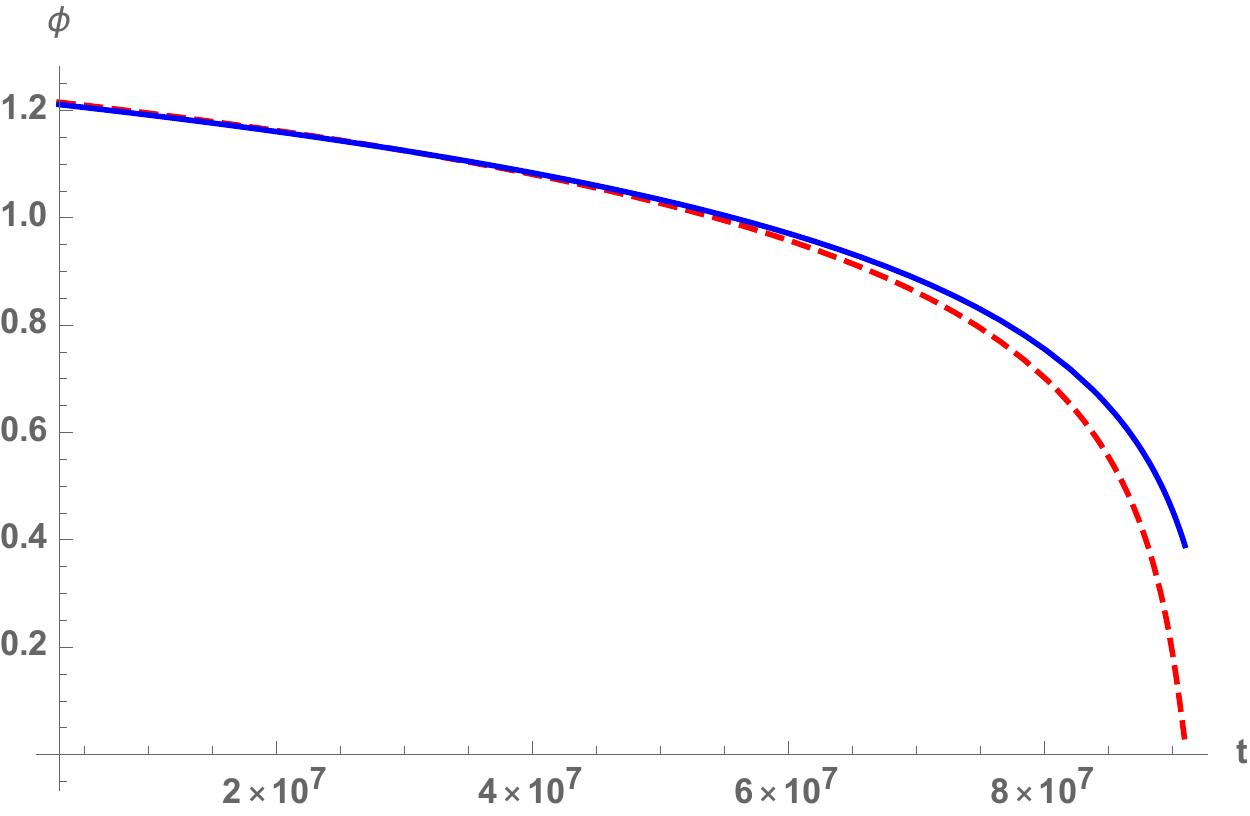}
\includegraphics[width=8cm]{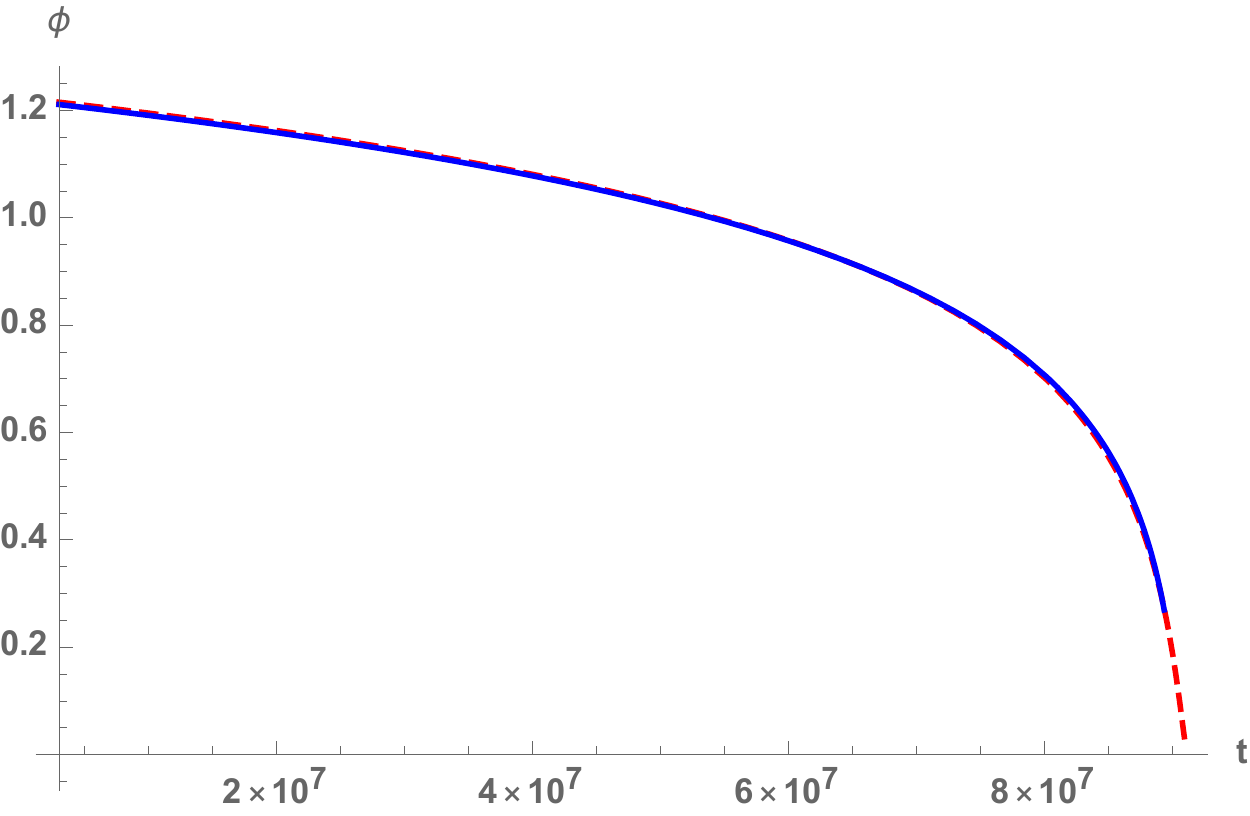}
}
\caption{In mLQC-I,  the initial conditions are set to $\phi_B=-1.20m_\mathrm{Pl}$ and $\dot \phi_B=0.440m_\mathrm{Pl}$. The red dashed curves are results on the value of the scalar field from the numeric simulations while blue solid curves are from the analytic solution (\ref{starinf2})  with $\alpha=1.5$ (top panel), $\alpha=1.25$ (middle panel) and $\alpha=1.21$ (bottom panel). To directly compare the results from different choices of $\alpha$, the inherited error at the onset of the inflation due to the transition phase has already been deducted.}
\label{phi}
\end{figure}

\begin{figure}
{
\includegraphics[width=8cm]{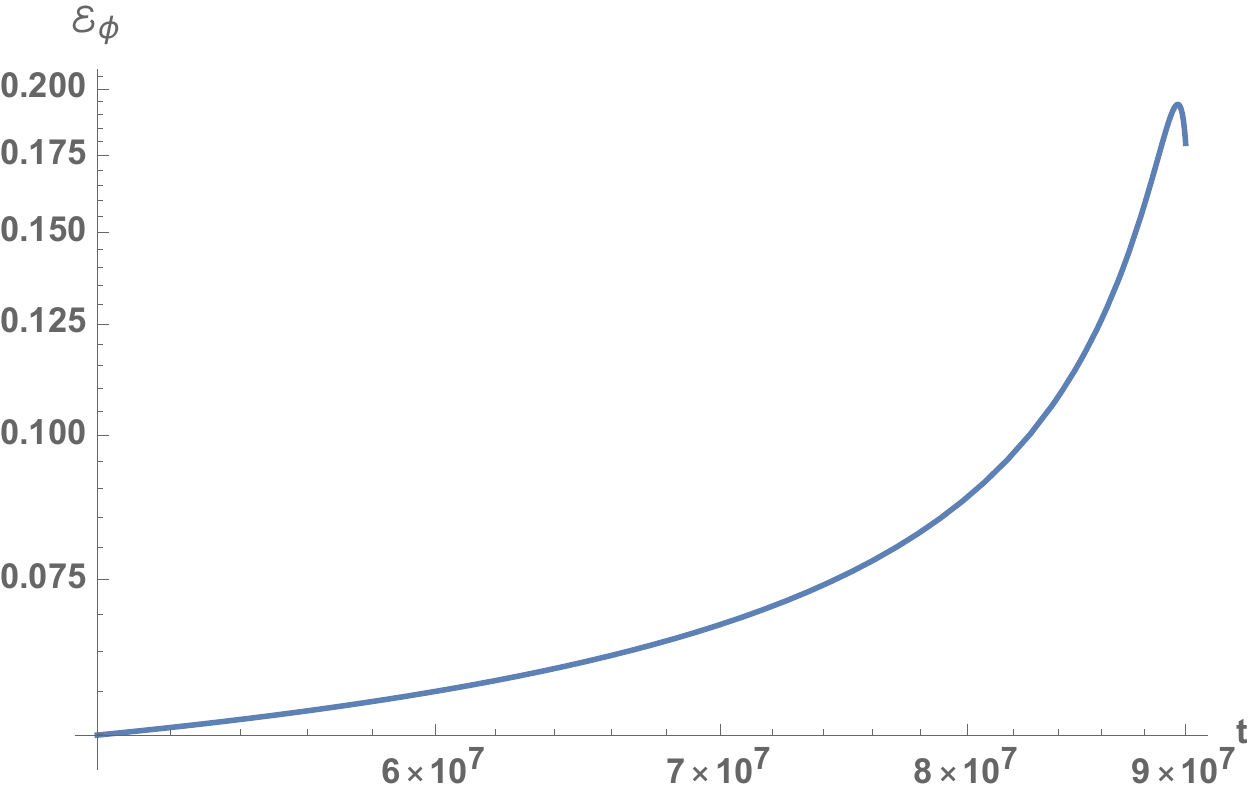}
\includegraphics[width=8cm]{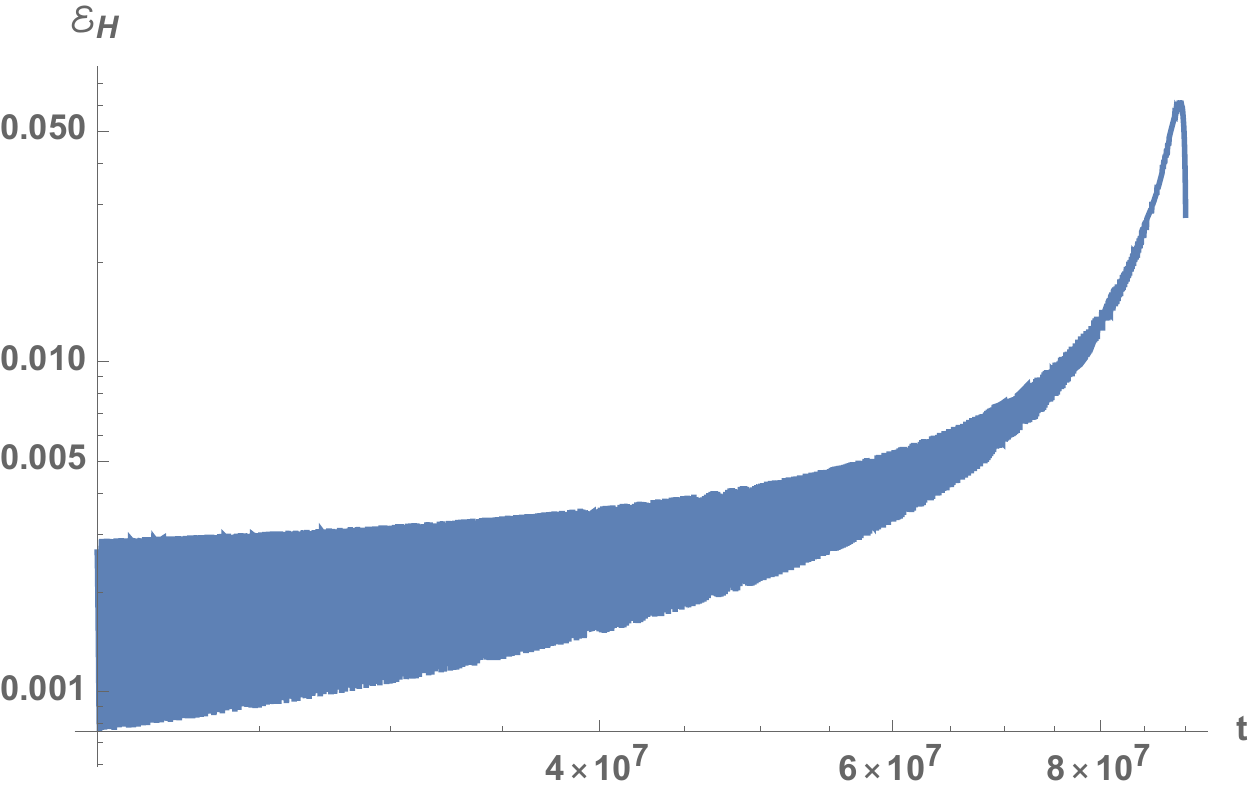}
\includegraphics[width=8cm]{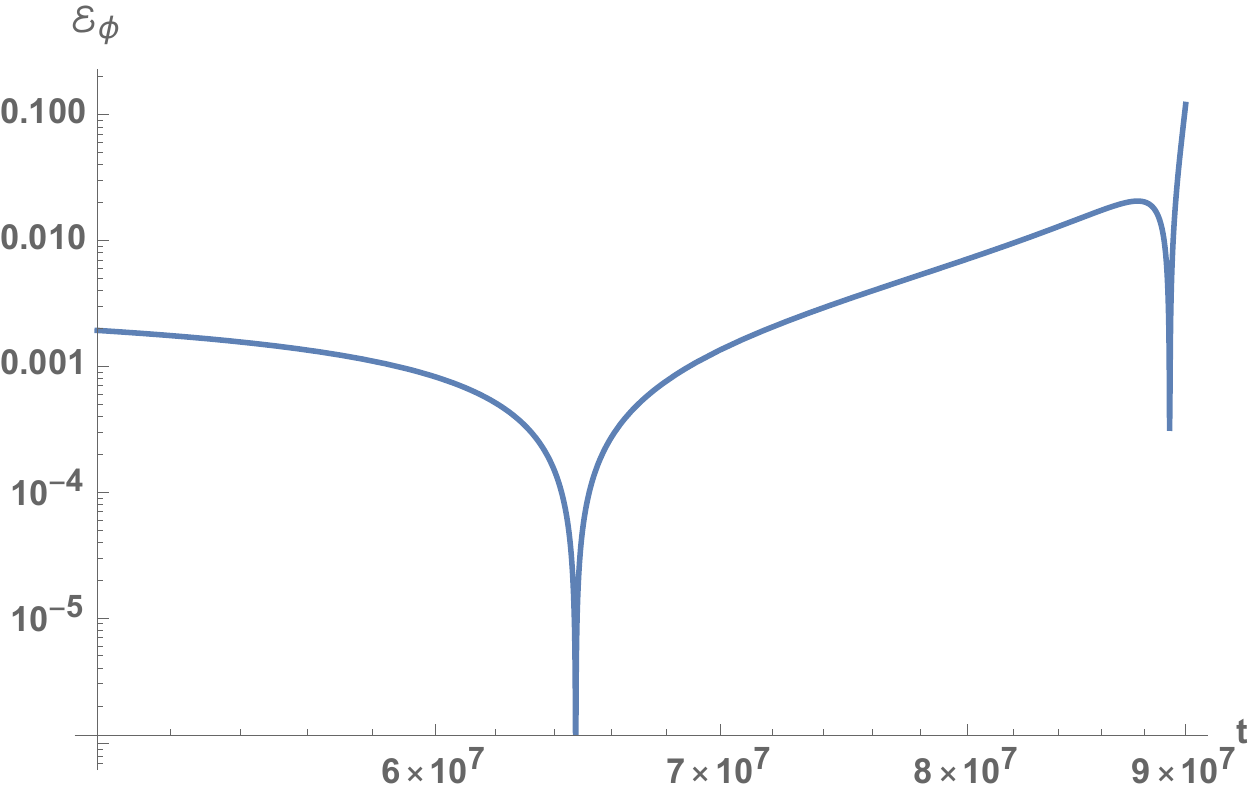}
\includegraphics[width=8cm]{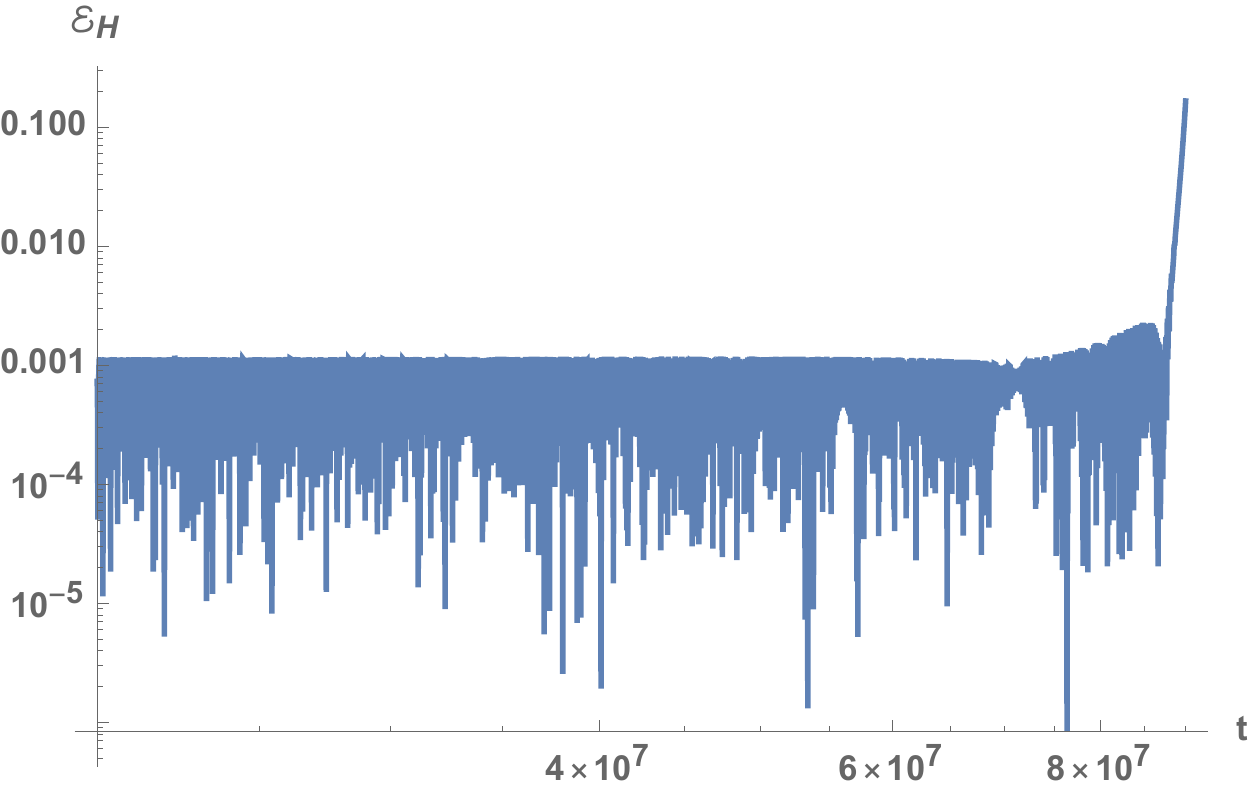}
}
\caption{With the same initial conditions as in Fig. \ref{phi}, the relative error between Eq. (\ref{starinf2}) and the numerics are depicted for $\alpha=1.21$. In the last two panels, we deduct the inherited errors from the transition  phase which is about $\Delta \phi\approx 0.06 m_\mathrm{Pl}$. This highlights the errors only in the inflationary phase.}
\label{starerror}
\end{figure}

As a first step to search for an empirical solution, it is important to understand the deviation of Eq. (\ref{starinf}) from numerical solution. In Fig. \ref{phi}, for mLQC-I, the initial conditions at the bounce are set to $\phi_B=-1.20$ and $\dot \phi=0.440$. The top panel compares the results from Eq. (\ref{starinf}) (solid curve) and numeric simulation (dashed curve), which shows the accumulated deviation between two curves in the inflationary phase. The solid curve decreases much more slowly than the dashed one, which actually confirms that, starting from  $t=4\times10^7$, simply ignoring the $\ddot \phi$ terms brings considerable errors. Therefore, the key task here is how to change the slope of the solid curve. It turns out that generally one can try the solution in the form 
\bq 
\lb{starinf2}
\phi_\text{inf}(t)=\sqrt{\frac{3}{16\pi G}}\ln\Big[e^{\sqrt{\frac{16\pi G}{3}}\phi_i}-\frac{ m\left(t-t_i\right)}{\alpha}\Big].
\eq
When $\alpha=1.5$, Eq. (\ref{starinf2}) reduces to Eq. (\ref{starinf}) while different choices of $\alpha$ can effectively change the slope of the solid curve as depicted in the middle and bottom panels of Fig. \ref{phi}. We find that the best fit between Eq. (\ref{starinf2}) and the numeric simulation can be achieved by choosing $\alpha=1.21$. In Fig. \ref{starerror}, the first two panels show the relative error for the solution (\ref{starinf2}) when $\alpha=1.21$. The big relative error near the end of inflation is due to the smallness of the magnitude of the scalar field ($\approx 0.123m_\text{Pl}$), and comparatively large inherited error from the transition phase $\Delta \phi \approx 0.06$ at $t_i$. Once this inherited error is deducted, the relative errors resemble the two bottom panels in Fig. \ref{starerror}. As the choice $\alpha=1.21$  helps reduce the error in a considerable amount, we have used this value of $\alpha$ in Eq. (\ref{starinf2}) and perfomed more simulations and comparisons with different initial conditions. The results are listed in Tables. \ref{star1}-\ref{star2}.

\begin{figure}[h!]  
{
\includegraphics[width=7cm]{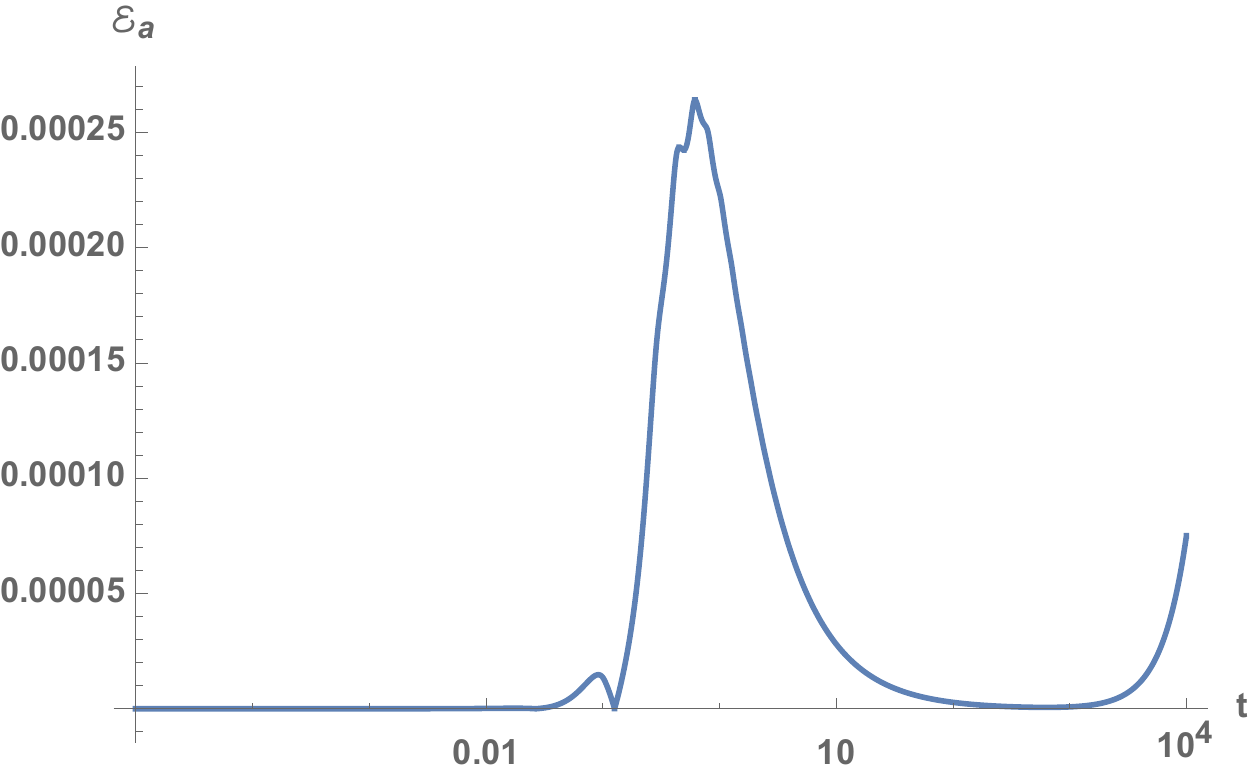}
\includegraphics[width=7cm]{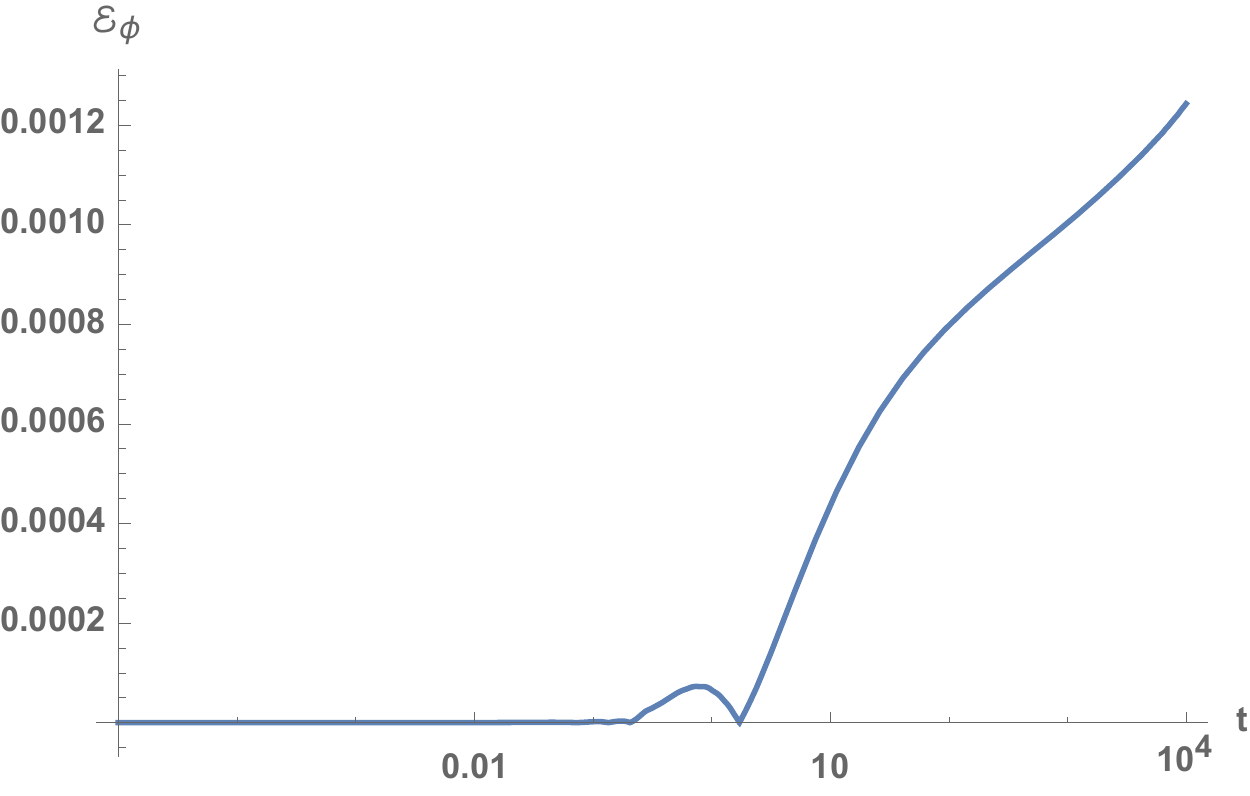}
\includegraphics[width=7cm]{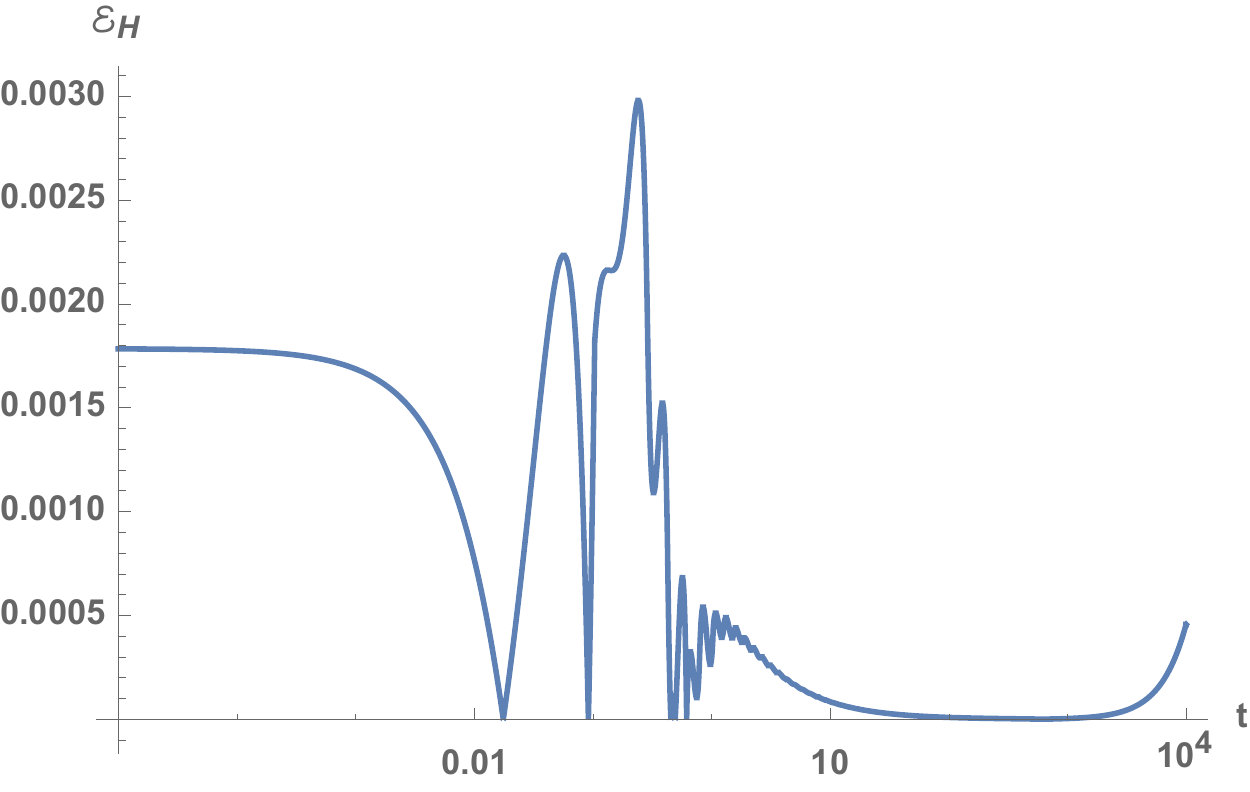}
\includegraphics[width=7cm]{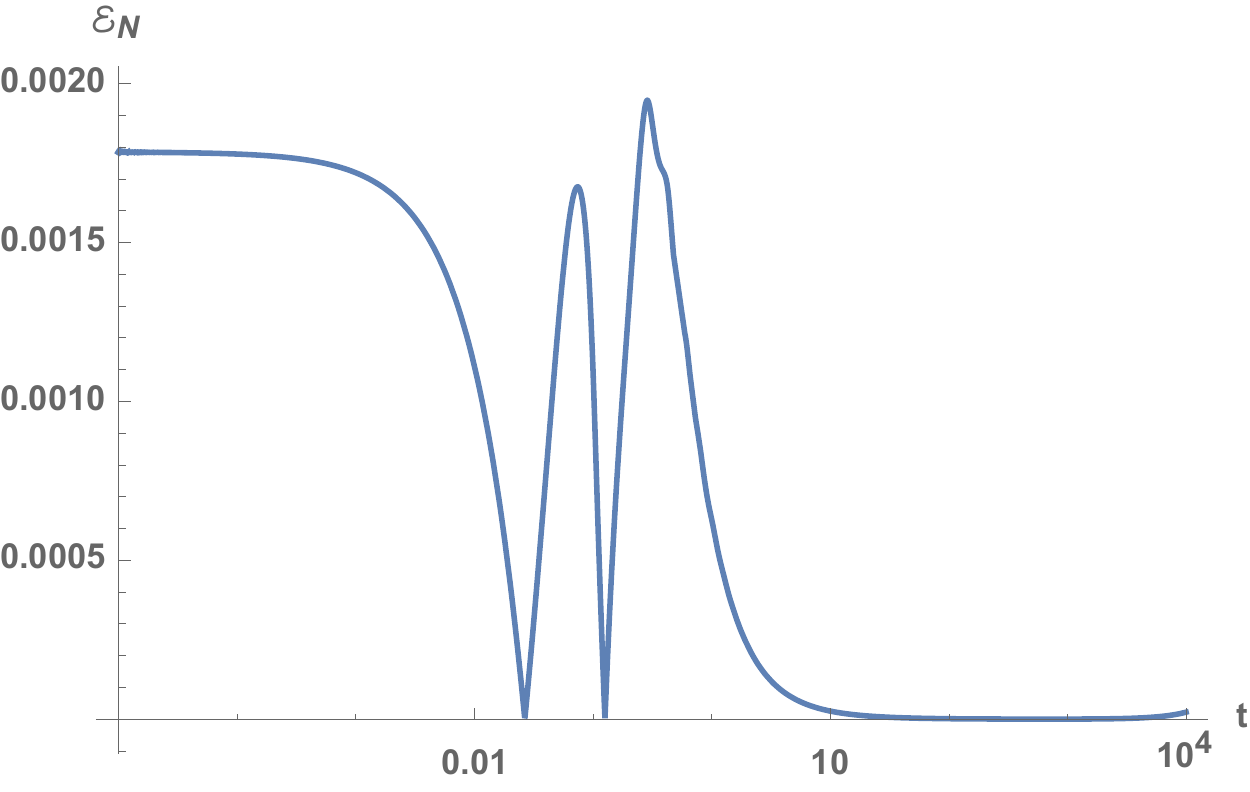}

}
\caption{In this figure, we show the relative error between the numeric and analytic results for the Hubble rate, the scale factor, the e-folds and the scalar field in mLQC-I for the Starobinsky potential with $\phi_B=3.50m_{\mathrm{Pl}} , \dot \phi_B<0$.}
\label{fig11}
\end{figure}

\begin{figure}[h!]  
{

\includegraphics[width=7cm]{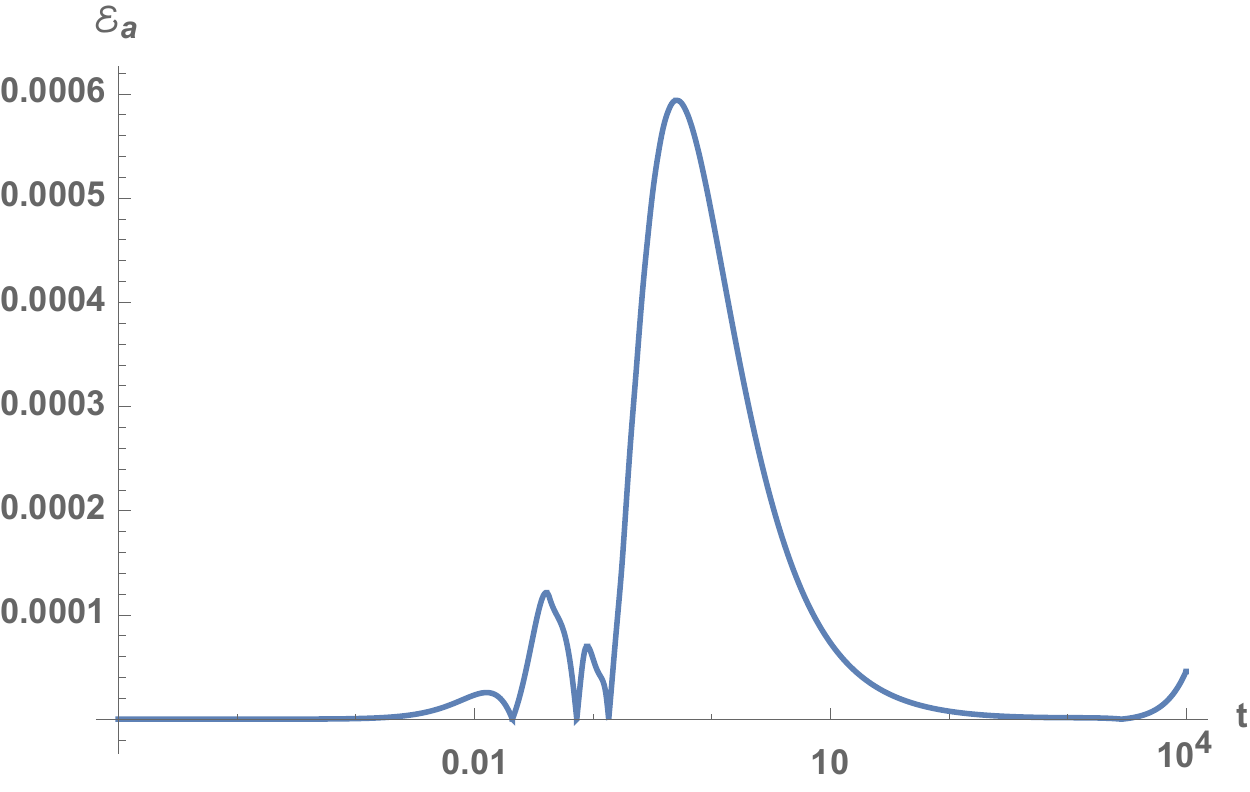}
\includegraphics[width=7cm]{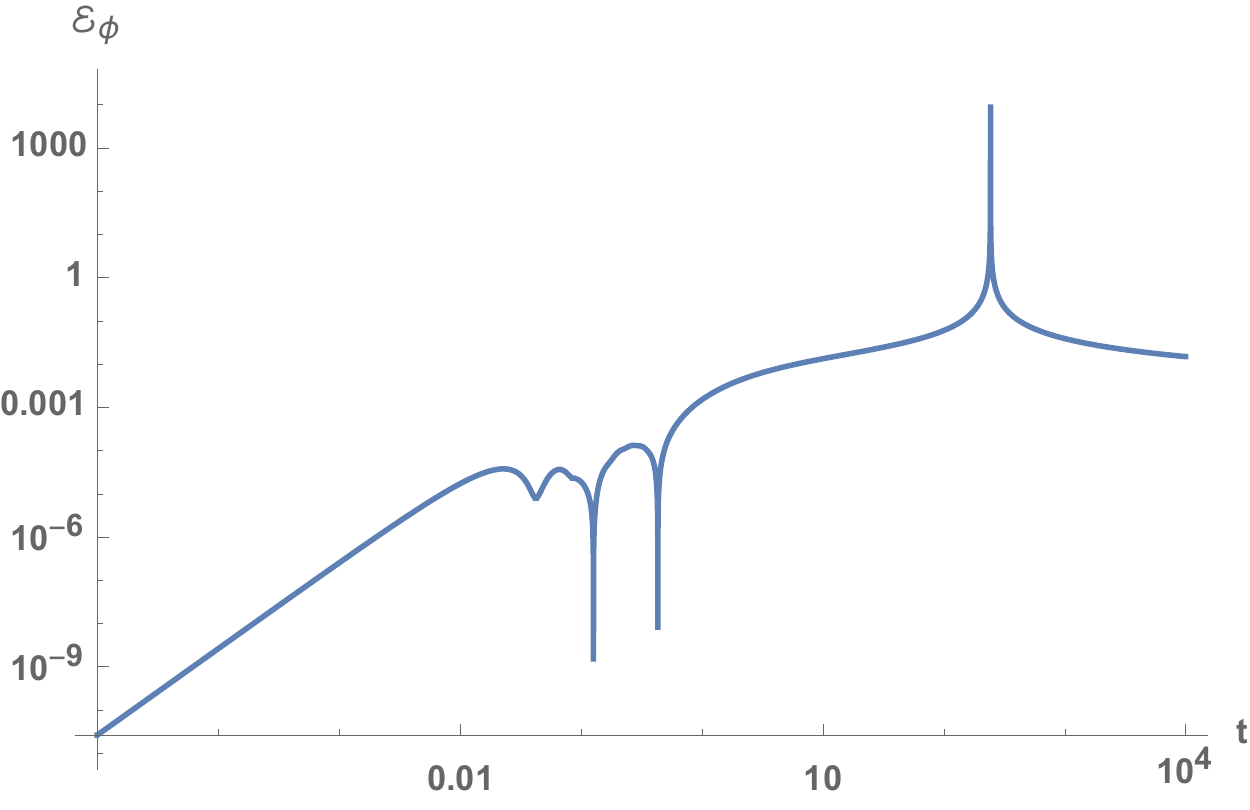}
\includegraphics[width=7cm]{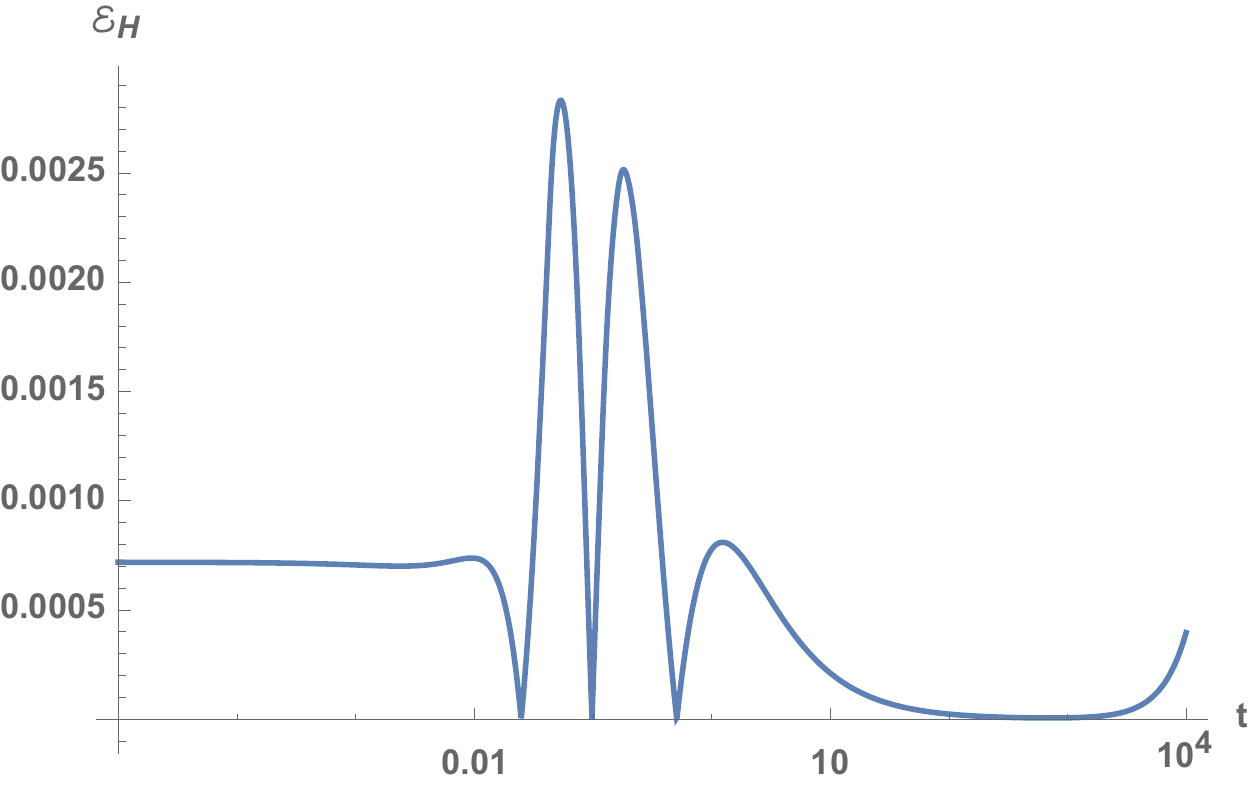}
\includegraphics[width=7cm]{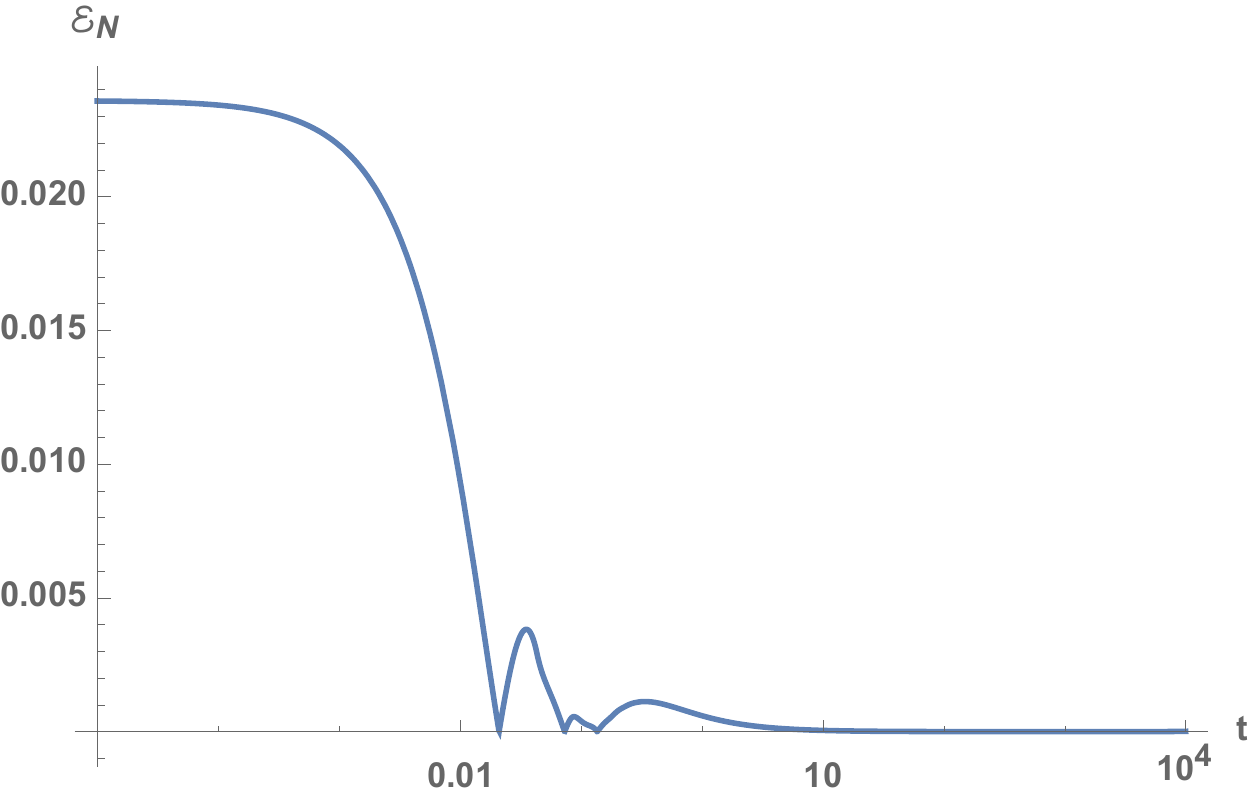}

}
\caption{In this figure, we show the relative error between the numeric and analytic results for  the Hubble rate, the scale factor, the e-folds and the scalar field in mLQC-II  for the Starobinsky potential with $\phi_B=-1.40m_{\mathrm{Pl}} , \dot \phi_B>0$.}
\label{fig12}
\end{figure}

To check the validity of our analytic solutions, especially, in the bouncing phase, relative errors of a few important observables are further studied with the Starobinsky potential, and shown in Fig. \ref{fig11}-\ref{fig12}. Compared with Fig. \ref{f2}-\ref{f3} for the chaotic potential,  one can find the errors of the Hubble rate, the scale factor as well as the e-folds are all insensitive to the form of the potentials as long as the KE  of the scalar field is dominant at the bounce, such that Eq. (\ref{2.6c}) amounts to a good approximation of the behavior of energy density. In this sense, the form of our modified analytic solutions (\ref{2.8m}) and (\ref{2.8mA}), respectively,  for mLQC-I and mLQC-II, are universal. 

\section{Probability of Slow-Roll Inflation in loop cosmology}
\renewcommand{\theequation}{4.\arabic{equation}}\setcounter{equation}{0}

In this section, we  compute the probability for inflation following the analysis in Ref. \cite{as2011} (which we refer the reader for details for LQC). 
 To consider the probability of the slow-roll inflation in the modified LQC models,   let us first consider  the phase space $\mathbb{S} $ of  the modified Friedmann equation and  Klein-Gordon equations,
 which consists of four variables,  $v, b$  from the gravitational sector and $\phi, p_\phi$ from the matter sector. Using the canonical communication relations, the symplectic form on the four-dimensional phase space is 
\bq
\Omega=d p_\phi \wedge d\phi+ \frac{d v \wedge d b}{4 \pi G \gamma}.
\eq 
The vanishing of the effective Hamiltonian constraint reduces the four-dimensional phase space to the hypersurface $\bar\Gamma$, on which we have 
\bq
\lb{4.2}
\mathcal{C}= 16\pi G \left\{{\cal{H}}_{grav}(v, b)  + \frac{p_{\phi}^2}{2v} + v V(\phi) \right\} \simeq 0,
\eq
where $``\simeq"$ means that the equality  holds only on $\bar\Gamma$. On the other hand, the phase space $\mathbb{S}$ is isomorphic to a 2-dimensional gauge-fixed surface $\hat\Gamma$ of $\bar\Gamma$, which is intersected by each dynamical trajectory once and only once \cite{as2011}. Since $b$ satisfies (\ref{1.3}), it is monotonically decreasing, as long as the matter field satisfies the weak energy condition. Thus, a natural parameterization of this 2-dimensional surface is $b = $ constant, say, $b_0$. Then, using the above constraint we find
\bq
\lb{4.3}
p^{\mathrm{A}}_{\phi} = v \left\{-2\left[\hat{\cal{H}}_{grav}^{\mathrm{A}}   +  V(\phi)\right]\right\}^{1/2},
\eq
where $\mathrm{A} = \mathrm{I}, \mathrm{II}$, and 
\bq
\lb{4.4}
\hat{\cal{H}}_{grav}^{\mathrm{A}} \equiv v^{-1} {\cal{H}}_{grav}^{\mathrm{A}}(v, b_0) .
\eq
 Note that in writing down the above expression, we have chosen only the ``+" sign for $p_{\phi}$, and $\hat{\cal{H}}_{grav}^{\mathrm{A}}   +  V(\phi)$ is negative definite.  From Eqs.(\ref{ham}) and (\ref{2.1}) we can see that
$\hat{\cal{H}}_{grav}^{\mathrm{A}} = \hat{\cal{H}}_{grav}^{\mathrm{A}}(b_0) = $ constant on $\hat\Gamma$. Hence, we obtain 
\bqn
\lb{4.5}
\left. dp^{\mathrm{A}}_{\phi}\right|_{\hat\Gamma} =  \frac{p^{\mathrm{A}}_{\phi}}{v} dv - \frac{v^2 V_{,\phi}}{p_{\phi}} d\phi.  
\eqn
 Thus, the pulled-back symplectic structure $\hat\Omega$ reads
 \bqn
\lb{4.6}
\left. \hat\Omega^{\mathrm{A}}\right|_{\hat\Gamma} =   \left\{-2\left[\hat{\cal{H}}_{grav}^{\mathrm{A}}(b_0)   +  V(\phi)\right]\right\}^{1/2} d\phi \wedge dv.  
\eqn
Then, the Liouville measure $d\hat\mu_L$ on $\hat\Gamma$ is given by
\bqn
\lb{4.7}
d\hat\mu^{\mathrm{A}}_L =   \left\{-2\left[\hat{\cal{H}}_{grav}^{\mathrm{A}}(b_0)   +  V(\phi)\right]\right\}^{1/2} d\phi dv.  
\eqn

Now the key observation is that  $d\hat\mu_L$   does not depend on $v$. As a result, although the integral $\int dv$ is infinite, it will get cancelled in the probability calculations as it shows up both in the denominator and the numerator. This occurs because the rescaling of $v$ amounts to a gauge transformation and the integral $\int dv$ simply gives the infinite length of the gauge orbits.  Therefore, the measure for the space of physically distinct solutions can be taken as 
\bq
\lb{measure1}
d\omega^{\scriptscriptstyle{\mathrm{A}}}=\left\{-2\left[\hat{\cal{H}}_{grav}^{\mathrm{A}}(b_0)   +  V(\phi)\right]\right\}^{1/2} d\phi .
\eq
The 2-dimensional phase space $\hat\Gamma$ is reduced further to an interval $\mathbb{S}_0 = \left\{\phi: \phi\in \left(\phi_{\mathrm{min}}, \phi_{\mathrm{max}}\right)\right\}$.  It should be noted that such a defined measure depends explicitly on the choice of $b_0$, a choice that is arbitrary. However, in loop cosmology there exists a preferred one, that is, the choice of $b(t)$ at the quantum bounce \cite{as2011},
\bq
\lb{4.8}
b_0 = b\left(t_B\right).
\eq
Thus, the probability of the occurrence of an event $E$ becomes
\bq
\lb{4.9}
P(E) = \frac{1}{{\cal{D}}} \int_{\mathcal{I}(E)}{\left\{-2\left[\hat{\cal{H}}_{grav}^{\mathrm{A}}(b_0)   +  V(\phi)\right]\right\}^{1/2} d\phi},
\eq
where $\mathcal{I}(E)$ is the interval on the $\phi_B$-axis, which corresponds to the physically distinct initial conditions in which the event $E$ happens, and ${\cal{D}}$ is the total measure 
\bq
\lb{4.10}
 {\cal{D}} \equiv  \int_{\phi_{\mathrm{min}}}^{\phi_{\mathrm{max}}} {\left\{-2\left[\hat{\cal{H}}_{grav}^{\mathrm{A}}(b_0)   +  V(\phi)\right]\right\}^{1/2} d\phi}.
\eq

Now, we are ready to apply the above formulas to the modified LQC models. Let us start with mLQC-I. In order to compare with the results obtained in LQC \cite{as2011}, we further assume that
 $V(\phi)$ is given by the chaotic potential. Then, we find that 
\bq
\lb{4.11}
\sin\left(\lambda b_B\right)=\sqrt{\frac{1}{2\gamma^2+2}}, \quad \sin\left(2\lambda b_B\right)=\frac{\sqrt{2\gamma^2+1}}{\gamma^2+1}.
\eq
Substituting them into Eqs.(\ref{4.3}) and (\ref{4.4}),  we get
\bq
\lb{4.12}
p^{{\scriptscriptstyle{\mathrm{I}}}}_\phi=v\left(2\rho^{{\scriptscriptstyle{\mathrm{I}}}}_c-2 V\right)^{\frac{1}{2}}.
\eq
Therefore, the measure for the space of physically distinct solutions is given by 
\bq
\lb{measure1}
d\omega^{\scriptscriptstyle{\mathrm{I}}}=\left(2\rho^{{\scriptscriptstyle{\mathrm{I}}}}_c-2 V\right)^{\frac{1}{2}} d\phi,
\eq
and the probability for the desired slow-roll not to happen is 
\bq
\lb{4.13}
P^{\scriptscriptstyle{\mathrm{I}}}(\text{not realized})\lesssim\frac{\int^{0.917}_{-5.158} d \omega^{\mathrm{I}} }
{\int^{\phi^{\mathrm{I}}_{\text{max}}}_{-\phi^{\mathrm{I}}_{\text{max}}}{ d \omega^{\mathrm{I}}}} \simeq 1.12\times 10^{-5},
\eq
where $\phi^{\mathrm{I}}_{\text{max}}$ is given by Eq.(\ref{Max}).

In mLQC-II, following a similar analysis, it can be shown that the measure for the physically distinct solutions is given by,
\bq
\lb{measure2}
d\omega^{\scriptscriptstyle{\mathrm{II}}}=\left(2\rho^{{\scriptscriptstyle{\mathrm{II}}}}_c-2 V\right)^{\frac{1}{2}} d\phi,
\eq
and the probability for the desired slow-roll to not happen in mLQC-II is 
\bq
\lb{4.14}
P^{\scriptscriptstyle{\mathrm{II}}}(\text{not realized})\lesssim 2.62\times 10^{-6}.
\eq

It is interesting to note that in LQC, the probability for desired slow roll inflation to not occur turned out to be \cite{as2011},
\bq
\lb{4.15}
P^{\scriptscriptstyle{\mathrm{LQC}}}(\text{not realized})\lesssim2.74\times 10^{-6},
\eq
which is smaller  than the result for mLQC-I and slightly larger than the one for mLQC-II. 
Therefore, we find that the desired slow-roll inflation is very  favorable in modified loop quantum cosmological models.

\section{Conclusions}
\renewcommand{\theequation}{5.\arabic{equation}}\setcounter{equation}{0}

Let us summarize the main findings of this manuscript. Our goal was to investigate two major issues: the generic properties of the background evolution of the universe in the pre-inflationary regime and the probability of occurrence of a desired slow-roll inflation in the framework of the modified loop 
quantum cosmologies. This was studied for both cases in which either KE or PE of the inflaton dominates at the quantum bounce. The bounce was chosen as the initial moment for numerical evolution. 
It was found that once KE dominated initially, it dominates the evolution of the universe for a long period, which is of the order of $10^{4} \; t_{\mathrm{Pl}}$. The exact time  depends on the specific potential, as it can be seen from
Fig. \ref{fig4}.  In fact, for such KE-dominated case, the evolution of the universe during the whole period before  reheating can be divided into three different phases: {\em bouncing, transition, and slow-roll inflation}. 
This division is generic in the sense that it is independent of the initial conditions at the bounce, the potentials of the scalar field and the models of mLQCs, as long as the condition 
$\mathrm{KE} (t_B) \gg \mathrm{PE} (t_B)$,
 holds at the bounce. During each of the three phases, the expansion factor $a(t)$ and the scalar field $\phi(t)$ can be obtained analytically. In particular, during the bouncing phase, they are
 given by  Eqs.(\ref{2.8m})-(\ref{fittingA}) for mLQC-I, and Eqs.(\ref{2.8mA})-(\ref{parameter2}) for mLQC-II. These explicit solutions trace the numerical (exact) ones extremely well, and  the relative errors in the SI phase are less than 
 $0.3\%$ for both chaotic and Starobinsky potentials (except for the e-folds in the Starobinsky potential). We expect that the same holds for other potentials, too, given the genericness features we uncovered in our study and negligible role played by potential in a kinetic-dominated bounce. 
  This is a robust result, and similar conclusions were also reached in LQC \cite{ZWCKS16,ZWCKS17}. This result primarily occurs because the corrections 
 appearing in mLQCs are of high-orders and their differences are significant only in the bouncing regime.  On the other hand, in the chaotic inflation, when the bounce is dominated by the PE of the inflaton, the slow-roll inflation now takes place shortly after the bounce when the universe is still in the SI phase. As a result, the universe expands with extremely fast speed during the early stage of the slow-roll which makes the inflationary e-folds a colossal number and imposes difficulties on the simulations. However, by concentrating on a small interval close to the threshold value of the scalar field which leads to the desired slow-roll, we evade simulating the whole process from the bounce to the end of the slow-roll and prove that all the PE dominated initial conditions allow for the desired slow-roll as well. As a result, the part of the parameter space consistent with the observations in the chaotic inflation is given by Eqs. (\ref{infefold1})-(\ref{infefold2}) while for the Starobinsky potential, only the KE dominated bounce is phenomenologically relevant. 
 
 To define the probability of the occurrence of a desired slow-roll inflation, we first generalize the analysis in \cite{as2011} for LQC to mLQCs. 
  We define the probability of the occurrence of an event $E$   by Eq.(\ref{4.9}), and  find that for the chaotic potential the probability is $ P^{{\scriptscriptstyle{\mathrm{I}}}}(\text{not realized})\lesssim  1.12\times 10^{-5}$ 
for mLQC-I and $P^{{\scriptscriptstyle{\mathrm{II}}}}(\text{not realized})\lesssim 2.62\times 10^{-6}$ for mLQC-II, respectively.  This is comparable to  $P (\text{not realized})\lesssim2.74\times 10^{-6}$  obtained for LQC \cite{as2011}. These results show that even though there are non-trivial differences in the modified Friedmann dynamics of studied models in the Planck regime, qualitative predictions  about probability do not get affected. 

Finally, we note that the linear perturbations of mLQCs were recently studied numerically in \cite{IA19}. It would be very interesting to study the same problem analytically along the lines
presented in \cite{ZWCKS16,ZWCKS17,Zhu1,BFL1,wuqiang}, by using the analytical solutions obtained in our analysis along with the method recently developed   in \cite{Zhu2,Zhu3,Zhu5}. This exercise will be performed in a future publication. 

\section*{Acknowledgements} 
A.W. is supported  by the National Natural Science Foundation of China (NNSFC) 
with the Grants Nos. 11375153, 11675145. B.F.L. is supported by NNSFC
with the Grants No. 11847216 and the Fundamental Research Funds for the Provincial Universities of Zhejiang in China with Grants No. RF-A2019015.
P.S. is supported by NSF grant PHY-1454832.

\section*{Appendix A: Fixing parameters and the e-folds for desired slow-roll inflation}
\renewcommand{\theequation}{A.\arabic{equation}}\setcounter{equation}{0}

For single field slow-roll inflation, the scalar power spectrum amplitude and the scalar spectral index for the pivot mode $k_*=0.05~ \text{Mpc}^{-1}$ is given by \cite{Planck2015}
\bqn
\lb{app1}
A^*_s&=&\frac{8V(\phi_*)}{3\epsilon_V(\phi_*)m^4_{\mathrm{Pl}}}=2.21\times10^{-9}, \\
 n^*_s&=&1+2\eta_V(\phi_*)-6\epsilon_V(\phi_*)=0.965,
\eqn
where $\phi_*$ denotes the value of the scalar field at the moment when the pivot mode exits the horizon during inflation. As a result, the parameter in the potential and $\phi_*$ can be found from $A_s$ and $n_s$. Up to first-order terms in the slow-roll parameters, the observable e-folds from the horizon crossing to the end of slow-roll inflation are given by 
\bq
\lb{app2}
N_*=\int^{\phi_*}_{\phi_\text{end}}d\phi \frac{8 \pi V}{m^2_{\mathrm{Pl}}V_{,\phi}} .
\eq 
Here $\phi_\text{end}$ is the value of the scalar field at the end of the slow-roll inflation whose value can be determined from the condition $\epsilon_V(\phi_\text{end})=1$. In this way, once $\phi_\text{end}$ and $\phi_*$ are known, the e-folds for the desired slow-roll inflation can be fixed as well. Finally, we list our results in Table \ref{appb}.
\begin{widetext}

\begin{table}[H]
\caption{In this table, we list the parameters in the inflationary potentials, the e-folds for the desired slow-roll inflation, the predicted tensor-to-scalar ratio in each potential as well as the values of the scalar field at the horizon crossing and the end of inflation. }
\begin{center}
 \begin{tabular}{||c|c|c|c|c|c|c||} 
 \hline
 \bf{Inflationary Potentials}  & \bf{$V(\phi)$} & \bf{$\phi_\text{end}/m_{\mathrm{Pl}} $} & \bf{$\phi_*/m_{\mathrm{Pl}} $}  & \bf{Parameter} & \bf{$N_*$}&\bf{r}\\ [1ex] 
 \hline
 
 Quadratic Chaotic & $\frac{1}{2}m^2\phi^2$ & $\pm\frac{1}{2\sqrt{\pi}}\approx 0.282$ & $\pm3.02$ & $m=1.26\times 10^{-6} m_{\mathrm{Pl}} $ &56.6&0.14\\ [1.2ex] 
 \hline
  Starobinsky & $\frac{3m^2}{32 \pi G}\left(1-e^{-\sqrt{\frac{16 \pi}{3m^2_{\mathrm{Pl}}}}\phi}\right)^2$ & $\frac{1}{4}\sqrt{\frac{3}{\pi}}\ln\left(1+\frac{2}{\sqrt{3}}\right)\approx 0.188$ &$1.07$ & $m=2.49\times 10^{-6}~m_{\mathrm{Pl}} $ &55.0&$3.49\times10^{-3}$\\ [1.2ex] 
 \hline 
\end{tabular}
\label{appb}
\end{center}
\end{table}
\end{widetext}

\end{document}